\pdfoutput=1
\documentclass[12pt,a4paper,aps,preprint,preprintnumbers,superscriptaddress,nofootinbib]{revtex4-1}
\usepackage[english]{babel}
\usepackage[utf8]{inputenc}
\usepackage{graphicx}   % need for figures
\usepackage{slashed}
\usepackage{epstopdf}
\usepackage{verbatim}   % useful for program listings
\usepackage{color}      % use if color is used in text
\usepackage{subfigure}  % use for side-by-side figures
\usepackage{subcaption}
\usepackage{multirow}

\usepackage[colorlinks=true, pdfstartview=FitV, linkcolor=blue, citecolor=blue, urlcolor=blue]{hyperref}
\usepackage{float}
\restylefloat{table}
\usepackage{bm}
\usepackage[normalem]{ulem}
\usepackage{amsmath,amssymb}
\def\addresses#1#2{\hbox to \hsize{\@tablebox{#1}\hfil\@tablebox{#2}}}
\def\@tablebox#1{\vtop{\hsize=5in \begin{flushleft} #1 \end{flushleft}}}

\def\beq{\begin{equation}}
\def\eeq{\end{equation}}
\def\bit{\begin{itemize}}
\def\eit{\end{itemize}}
\def\be{\begin{eqnarray}}
\def\ee{\end{eqnarray}}
\def\bray{\begin{array}}
\def\eray{\end{array}}

\definecolor{darkgreen}{rgb}{0,0.6,0}

\definecolor{orange}{rgb}{1,0.5,0}
\definecolor{blue}{rgb}{0,0,1}                          

\begin{document}

\title{Numerical study on the gauge symmetry of electroweak amplitudes}

\author{Wang-Fa Li}
\affiliation{School of Physics, Sun Yat-Sen University, Guangzhou 510275, China}

\author{Junmou Chen}
\email[Corresponding author. ]{chenjm@jnu.edu.cn}

\affiliation{Jinan University, Guangzhou, Guangdong, China}

\author{Qian-Jiu Wang}
\affiliation{School of Physics, Sun Yat-Sen University, Guangzhou 510275, China}

\author{Zhao-Huan Yu}
\email[Corresponding author. ]{yuzhaoh5@mail.sysu.edu.cn}
\affiliation{School of Physics, Sun Yat-Sen University, Guangzhou 510275, China}

\begin{abstract}
%\JM{5-component Goldstone equivalence Feynman rules -- vertices comprise of gauge and Goldstone components -- precise relation between couplings of sub-vertices guarranteed by MWI --  1. direct test by applying MWI; 2. modify one of the sub-couplings to see if and how MWI changee. 3. Add SMEFT operators, then apply method 2.  }

%\JM{Electroweak (EW) amplitudes in the gauge-Goldstone 5-component formalism have no gauge cancellation, greatly simplifying numerical computations.  A vertex in the 5-component formalism comprises both gauge and Goldstone components,  thus has multiple couplings, between which there are precise relations required by gauge symmetry.}

Electroweak (EW) amplitudes in the gauge-Goldstone five-component formalism have a distinctive property: gauge symmetry is imprinted in the amplitudes, manifested as the massive Ward identity (MWI) $k^M\mathcal M_M=0$.  In this study, we used the HELAS package to numerically study gauge symmetry in EW amplitudes. First, we directly tested gauge symmetry by examining the MWI of amplitudes. Second, we modified the couplings within a vertex and among vertices to check if and how the MWI changes.
Third, we tested gauge symmetry by considering the couplings modified by operators from the standard model effective field theory (SMEFT). Similar to the standard model, there are relations between different couplings that are protected by gauge symmetry.  We observed that, if we modify the couplings to deviate from  the relations, the MWI is violated. In contrast, the MWI is restored when the relations between couplings reduce to those in the SMEFT.

%\JM{write MWI first, then ask how MWI gauge symmetry is reflected on amplitudes, then explain it's the precise relations between couplings within and without vertices. } 

\end{abstract}

\maketitle

% \titlepage \maketitle
%\newpage
%\flushbottom

\section{Introduction}

%\JM{modifying intro: 
%1. intro new feynman rules with Goldstone equivalence
%2. gauge symmetry in amplitudes -- compared to massless
%3. compared to the standard formalism in computing massive amplitudes:  precise relation between couplings from no where v.s. from gauge symmetry 
%4. arrangement of the whole paper}

The gauge cancellation of electroweak (EW) scattering amplitudes in the standard model (SM)~\cite{Glashow:1961tr,Weinberg:1967tq,Salam:1968rm} has been both a theoretical problem hindering physical analysis and a practical problem in numerical calculations. In recent years, a new framework for computing EW amplitudes based on Goldstone equivalence has been proposed and implemented in the software package HELAS~\cite{Chen:2022gxv}. The main ingredients of this framework are as follows: 1) taking Goldstone equivalence~\cite{Cornwall:1974km,Chanowitz:1985hj,Gounaris:1986cr}, so that the polarization vector of a massive vector boson has no $k^\mu$ term and is instead composed of its Goldstone component and a remnant gauge  term; 2) combining the gauge components and their Goldstone components of fields or polarization vectors into single five-component objects at the level of  polarization vectors, propagators, and vertices; and 3)  imposing a special light-cone gauge defined by the gauge direction $n^\mu=(1,-{\vec k}/{|\vec k|})$, dubbed the Feynman diagram gauge~\cite{Chen:2022gxv} or Goldstone  equivalence gauge~\cite{Chen:2016wkt}. It has been demonstrated in Ref.~\cite{Chen:2022gxv}  with many examples that EW amplitudes in this new scheme do not have gauge cancellation, thereby solving a long-standing problem.  Further studies on this topic can be found in Refs.~\cite{Hagiwara:2024ron, Chen:2022xlg, Furusato:2024uli, Hagiwara:2024xdh, Zheng:2024lwl, Furusato:2024uli, Furusato:2024ghr, Jeong:2025hwj}.  Earlier but incomplete treatment of the five-component formalism can be found in \cite{Kunszt:1987tk,Wulzer:2013mza,Cuomo:2019siu}. %Related works on the Feynman diagram gauge in massless theories can be seen in Refs.~\cite{}. 

The new framework in Ref.~\cite{Chen:2022gxv} is essentially a reorganization and combination of ``pieces" from existing Feynman rules.  Although ideal to be combined with  the Feynman diagram gauge, it can also be applied to other gauges, such as the Feynman gauge. Therefore, it is necessary to distinguish the five-component framework from any specific choice of gauge. Additionally, since the framework relies on the Goldstone equivalence theorem~\cite{Cornwall:1974km, Chanowitz:1985hj,Gounaris:1986cr} in an essential way, it is appropriate to refer to it as the Goldstone equivalence (GE) representation of Feynman rules, as we will do from now on. Correspondingly, the standard Feynman rules will be called the gauge representation, as they exclusively treat the  physical content of massive vector bosons as quanta of gauge fields. %The jargons of  GE representation and gauge representation will be used in the remaining of this paper. 

Although the absence of gauge cancellation is crucial for practical applications, it is not the only important property of the GE representation of EW Feynman rules. Another intriguing property is that the GE representation directly imprints gauge symmetry at the level of amplitudes, as expressed by the massive Ward identity (MWI), which is named in analogue to its massless counterpart.
Specifically, for a scattering amplitude $\varepsilon^{\mu(*)}(k) \mathcal M_\mu $ involving an external vector boson $V$ with mass $m_V$ and four-momentum $k^\mu$, the MWI is~\cite{Cornwall:1974km, Chanowitz:1985hj, Gounaris:1986cr, Bagger:1989fc}
\begin{equation}\label{eq:MWI}
  k^\mu \mathcal M_\mu = \mp im_V \mathcal{M}(\varphi),
\end{equation}
where the $-$ ($+$) sign corresponds to the case where the vector boson is in the initial (final) state, and $\mathcal M(\varphi)$ represents the amplitude obtained by replacing the vector boson with the corresponding Goldstone boson $\varphi$.
The MWI \eqref{eq:MWI} is a fundamental identity for a spontaneously broken gauge theory, e.g., the EW gauge theory. This study aimed to investigate the gauge symmetry of EW amplitudes using the numerical tool HELAS~\cite{Hagiwara:1990dw, Murayama:1992gi, Chen:2022gxv}.

Similar to the Ward identity in gauge  theories with massless gauge bosons, the MWI guarantees precise relations among different diagrams of an amplitude, which further imply exact relations among different vertices or, equivalently, different couplings within the same theory. Considering the example of $W^+ W^- \rightarrow W^+ W^-$ with the helicity combination of $\mathrm{TTTT}$, where $\mathrm{T}$ represents a transverse polarization, the amplitude includes three vertices $WWZ$, $WWA$, and $WWWW$ with couplings $g_{WWZ}$, $g_{WWA}$, and $g_{WWWW}$. The Ward identity requires that
\begin{equation}
    g_{WWWW}=g_{WWZ}^2+g_{WWA}^2.
\end{equation}
Conversely, any deviation from this relation would result in violations of the gauge symmetry and the Ward identity.
%\JM{gneralize to massive case -- our study: directly test; anomoloug coupilng in SM; SMEFT}
When the helicity combination involves longitudinal components, the precise relations the MWI brings about   not only involve couplings of gauge bosons, but also those of Goldstone bosons and the Higgs boson. Thus, the structure of gauge symmetry is much richer in massive amplitudes, which we study in detail. 
Moreover, the gauge symmetry and MWI are not limited to the SM and can be applied to other theories. One theory of particular interest is the standard model effective field theory (SMEFT)~\cite{Coleman:1969sm, Callan:1969sn, Weinberg:1980wa, Leung:1984ni, Buchmuller:1985jz, Grzadkowski:2010es}, which is a direct extension of the SM.

The remainder of this paper is organized as follows. In Section II, we provide a brief introduction to the GE representation of EW interactions, emphasizing the relations among different couplings and parameters. In Section III, we directly test the EW gauge symmetry with the MWI~\eqref{eq:MWI} across multiple processes and helicity combinations. In Section IV, we examine the EW gauge symmetry through anomalous couplings, including both within individual vertices and among multiple vertices. Finally, in Section V, we study the connection between anomalous couplings (Higgs self-couplings and Yukawa couplings) and certain SMEFT operators using the MWI. We demonstrate that, when the anomalous couplings are adjusted according to the SMEFT operators, the MWI is satisfied, ensuring the gauge invariance  and self-consistency of the  theory.

%Importantly, on-shell gauge symmetry $k^M \mathcal{M}_M =0$ implies that Goldstone contribution should also be included in the vertices, exactly as is done in our new scheme.  In this paper we will confirm 

% The following is the main message of our paper: (on-shell) gauge symmetry for massive theory means  $k^\mu/m_v$ term in the longitudinal polarization vector is equivalent to its Goldstone mode in amplitudes, thus there are infinite many equivalent forms of the longitudinal polarization vector by making an arbitrary shift of $k^M$: $\epsilon^M_L\rightarrow \epsilon^M_L+\xi k^M$;  similarly  all the vertices should include 

%\newpage

\section{GE representation of EW interactions }
\label{sec:GE_int}

%\JM{Goldstone equivalence -- GE representation: gauge-Goldstone 5-components formalism -- double line diagram - Goldstone equivalence gauge  }

In this section, we describe key ingredients of the GE representation, including polarization vectors, gauge choice, propagator, vertices, and Feynman rules.

\subsection{Polarization Vectors in the Five-Component Formalism}

The central identity of our program is the MWI \eqref{eq:MWI}, which essentially states the equivalence between the $k^\mu$ terms in gauge fields and the corresponding Goldstone fields. Moreover, the longitudinal polarization vector of a massive vector boson $V$ with four-momentum $k^\mu$ can be decomposed into 
\begin{equation}\label{eq:long_pol_vec:4dim}
\epsilon^\mu_\mathrm{L}(k) =\frac{k^\mu}{m_V}-\frac{m_V}{n\cdot k}\, n^\mu    
\end{equation}
with $n^\mu=(1, -{\vec k}/{|\vec k|})$, and the subscript $\mathrm{L}$ representing a longitudinal polarization. We can eliminate the $k^\mu$ term in $\epsilon_\mathrm{L}^\mu(k)$ by using the MWI, resulting in the amplitude with a longitudinal vector boson:
\begin{equation}
    \mathcal M(V_\mathrm{L}) \equiv \epsilon_\mathrm{L}^\mu(k) \mathcal M_\mu =  -\frac{m_V}{n\cdot k}\, n^\mu \mathcal M_\mu - i\mathcal M(\varphi),
\end{equation}
where the MWI \eqref{eq:MWI} is used in the second step.

The generalization to multiple vector bosons is straightforward, but it leads to an excessive number of terms, making calculations cumbersome. To address this issue, we define five-component longitudinal polarization vectors in the GE representation by combining the gauge boson wave function and the Goldstone boson ``wave function'' as 
\begin{equation}\label{eq:pol_vec_5dim}
\begin{aligned}
    \epsilon^M_\mathrm{L}(k) &\equiv \left(-\frac{m_V n^\mu}{n\cdot k},~ i \right)  \ \ \ \text{for initial state,} \\
    \epsilon^{*M}_\mathrm{L}(k) &\equiv \left(-\frac{m_V n^\mu}{n\cdot k},~ -i\right) \ \ \  \text{for final state,}
\end{aligned}
\end{equation}
with $\mu = 0,1,2,3$ and $M = 0,1,2,3,4$.
In addition, the five-component transverse polarization vectors are defined as $ \epsilon^M_\pm(k) = \big(\epsilon^\mu_\pm(k),~0\big)$.
Then, the MWI \eqref{eq:MWI} can be rewritten as 
\begin{equation}\label{eq:MWI:alt}
    k^M\mathcal{M}_M=0 ~~\text{for incoming,}\qquad
    k^{*M}\mathcal{M}_M=0 ~~\text{for outgoing,}
\end{equation}
with $\mathcal{M}^M \equiv (\mathcal{M}^\mu, \mathcal{M}(\varphi))$, and the five-component ``momentum'' $k^M \equiv (k^\mu, -im_V)$. The five-dimensional ``metric'' for index contraction is $g_{MN} \equiv \operatorname{diag}(+1, -1, -1, -1, -1)$.
The spin sum of the polarization vectors becomes
\begin{equation}\label{eq:spin_sum}
   \sum_{s = \pm,\mathrm{L}}\epsilon^M_s(k) \epsilon^{*N}_s(k)= -g^{MN}+\frac{k^Mn^N+n^Mk^{*N}}{n\cdot k},
\end{equation}
with $g^{MN} = \operatorname{diag}(+1, -1, -1, -1, -1)$ and $n^M \equiv (n^\mu, 0)$.
%\JM{The purpose of choosing $g_{44}=-1$ is to be accordance with spin-sum tensor as below.}
Writing the gauge and Goldstone components separately, Eq.~\eqref{eq:spin_sum} becomes
\begin{equation}\setlength{\arraycolsep}{.5em}
    \sum_{s = \pm,\mathrm{L}} \epsilon^M_s(k) \epsilon^{*N}_s(k) = 
   \begin{pmatrix}
       -g^{\mu\nu}+\dfrac{k^\mu n^\nu+n^\mu k^\nu}{n\cdot k}  &  i\dfrac{m_Vn^\nu}{n\cdot k}\\[1em]
       -i\dfrac{m_Vn^\mu}{n\cdot k}  & 1 
    \end{pmatrix}.
\end{equation}
The mixing between gauge and Goldstone degrees of freedom is manifest here. When $m_V\rightarrow 0$, they decouple from each other.

Because of the MWI \eqref{eq:MWI:alt}, we can go further to state that the longitudinal polarization vector can be equivalently defined by an arbitrary shift:
\begin{equation}\label{eq:long_pol_vec:generic}
\epsilon_\mathrm{L}^M(k,\lambda) \equiv \left(-\frac{m_V n^\mu}{n\cdot k}, ~ i \right) + \lambda \,\frac{k^M}{m_V}
\end{equation}
which gives the same amplitudes for any value of $\lambda$. Choosing $\lambda = 1$ yields $\epsilon_\mathrm{L}^M(k,1) = (\epsilon_\mathrm{L}^\mu(k), 0)$, which corresponds to the conventional case using Eq.~\eqref{eq:long_pol_vec:4dim}. We refer to the case of $\lambda =1$ as the gauge form of the longitudinal polarization.  The case of $\lambda =0$ corresponds to Eq.~\eqref{eq:pol_vec_5dim}, which we call the GE form.  %with $\lambda=0$ and the gauge form with $\lambda=1$ are two special cases of the generic from \eqref{eq:long_pol_vec:generic}.

%\JM{propagator? -- vertices and relations of couplings }

\subsection{Gauge Choice and Propagator}

The five-component formalism can also be naturally derived from the Lagrangian, which includes a term $m_V V_\mu \partial^\mu\varphi$ that mixes the gauge field with the Goldstone field after gauge symmetry breaking~\cite{Chen:2022gxv}. Therefore, it is natural to combine gauge and Goldstone components into one five-component object.

The GE representation, implemented through the five-component formalism, is compatible with any gauge except the unitary gauge, which eliminates Goldstone bosons from the Feynman rules. In the $R_\xi$ gauges, the gauge-fixing Lagrangian term  $-(\partial^\mu V_\mu -m_V \varphi)^2/(2\xi)$ is introduced to cancel the gauge-Goldstone mixing term. This results in the propagator for a massive vector boson in the five-component formalism taking the form of
%For example,  Feynman gauge takes a simple form in the 5-component form:
%\begin{equation}
%    D^{MN}=\frac{-ig^{MN}}{k^2-m_V^2}
%\end{equation}
\begin{equation}
    D^{MN}(k) = \frac{-ig^{MN}}{k^2-m_V^2+i\varepsilon} +\frac{i(1-\xi)}{(k^2-m_V^2)(k^2-\xi m_V^2)}   
       \begin{pmatrix}
       k^\mu k^\nu  &  0\\
       0            & m_V^2
    \end{pmatrix}.
\end{equation}
Setting $\xi = 1$ yields the Feynman gauge. 

%$R_\xi$ gauge including Feynman gauge, however, still hasn't completely eliminated $k^\mu$ terms, which are also found in $g^{\mu\nu}$. Another way to see the fact is the numerator does not equal to the spin sum in Eq.~\eqref{eq:spin_sum} similar to the fermion.  \JM{irrelevant and too long?}
%\ZH{It is difficult to understand this paragraph.}

To completely eliminate the $k^\mu$ terms, we can choose the gauge condition $n^\mu(x) V_\mu(x) =0$, where $n^\mu(x)$ is the Fourier transformation of $n^\mu(k) = (1,-{\vec{k}}/{|\vec{k}|})$, by adding the Lagrangian term $-[n^\mu(x) V_\mu(x)]^2/(2\alpha)$. Its specific form is irrelevant here. Applying this to a massive gauge theory, the five-dimensional vector boson propagator becomes
\begin{equation}\label{eq_prop_5c}
    D^{MN}(k) = \frac{-i}{k^2-m_V^2}\left(g^{MN}-\frac{k^Mn^N+n^Mk^{*N}}{n\cdot k}\right)+\alpha \,\frac{k^Mk^{*N}}{(n\cdot k)^2}.
\end{equation}
The numerator of the gauge independent part is precisely the spin sum of polarization vectors in Eq.~\eqref{eq:spin_sum} for an on-shell momentum. Taking $\alpha \to 0$, we  obtain the polarization vector in Eq.~\eqref{eq:long_pol_vec:generic} with $\lambda=0$. This is known as the Goldstone equivalence gauge or the Feynman diagram gauge~\cite{Chen:2022gxv}.

\subsection{Vertices and Feynman Rules}

%\JM{a vertex composed of many sub-vertices, also many parameters in one vertex, but there is relation between them ;double line notes -- VVh, ffV, VVV, VVVV\&VVhh\&VVVh: couplings and relations}

\begin{figure}[!t]
    \centering
    \includegraphics[width=0.8\linewidth]{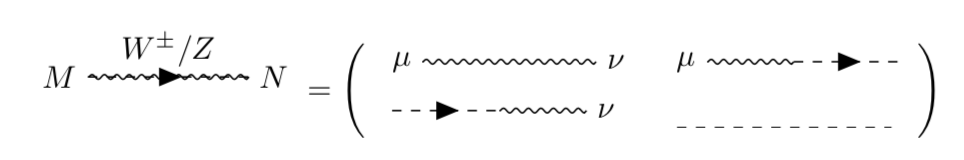}
    \caption{Five-component vector boson propagator in the double-line notation. }
    \label{fig:dbl-line_prop}
\end{figure}

\begin{figure}[!t]
    \centering
    \includegraphics[width=1\linewidth]{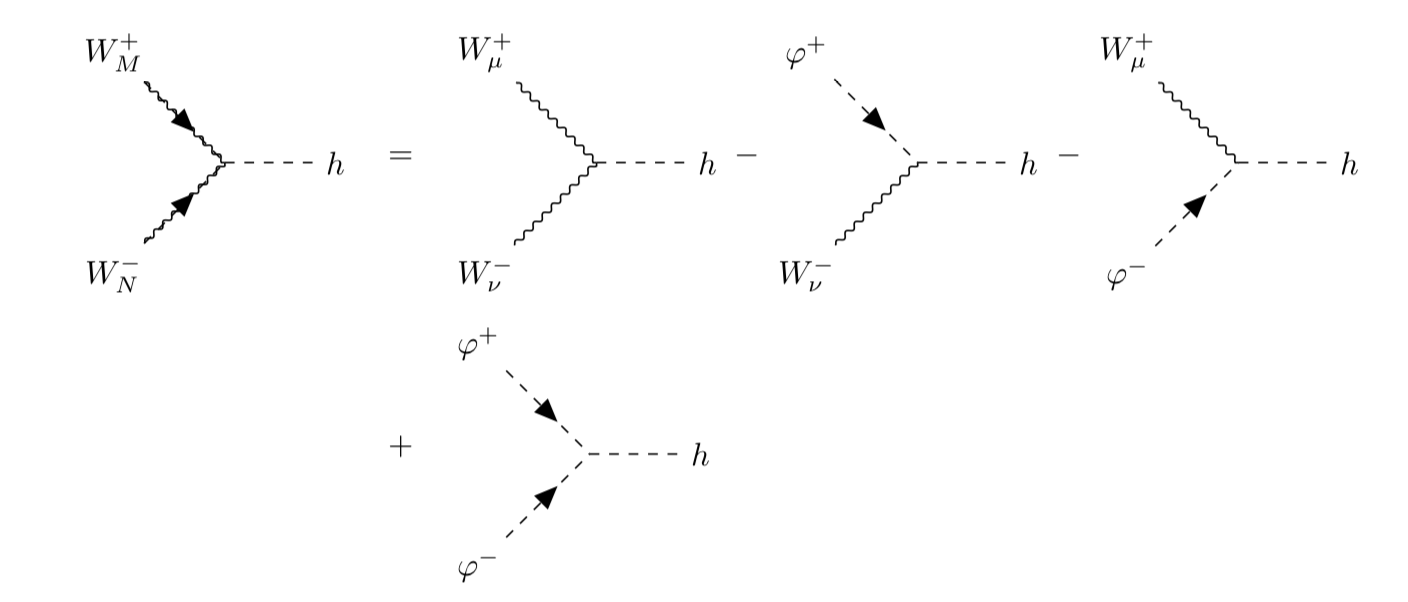}
    \caption{Five-component vertex of $WWh$ in the double-line notation. The minus sign before $\varphi W h$ comes from the $g_{44}=-1$ component of the ``metric" $g_{MN}$.  }
    \label{fig:dbl-line_wwh}
\end{figure}

With the polarization vectors and propagator expressed in the five-component formalism, this framework can be naturally extended to vertices involving massive vector bosons. Accordingly, we introduce a double-line notation by overlapping a wavy line with a dashed line to represent the massive vector boson in Feynman rules: the wavy line represents the gauge components, whereas the dashed line represents the Goldstone component. As examples of this double-line notation, we illustrate the decomposition of the vector boson propagator in Fig.~\ref{fig:dbl-line_prop} and that of the $WWh$ vertex in Fig.~\ref{fig:dbl-line_wwh}.

The new form of vertices brings additional subtleties.  For example, we denote the $WWh$ vertex as $V_{WWh}^{MN}$, where the subscript indicates the vertex and the superscript represents indices. In the GE representation, a vertex involving massive vector bosons has multiple parameters.  For instance, $V^{MN}_{WWh}$ not only includes the pure gauge components $V^{\mu\nu}_{WWh}$ with coupling $g_{WWh}$ but also involves the gauge-Goldstone components $V^{\mu 4}_{WWh} = V^{\mu}_{W\varphi h}$ with coupling $g_{W\varphi h}=g_{\varphi W h}$ and the pure Goldstone component $V_{WWh}^{44} = V_{\varphi\varphi h}$ with coupling $\lambda_{\varphi\varphi h}$. We can further express the vertex as $V^{MN}_{WWh}(g_{WWh}, g_{\varphi Wh}, \lambda_{\varphi\varphi h}; m_h, m_W)$.  However, these couplings are not independent in the SM.  They are all determined by the gauge coupling $g$, excluding the mass parameters.  These relations originate from the spontaneous symmetry breaking of the EW theory, reflecting of the underlying gauge symmetry.  Similar relations also exist for other vertices such as $ff'W$ and $WWZ$.  Further details are discussed in Sec.~\ref{sec:anom_coup}. 

Finally, let us examine the free parameters in the SM. The EW gauge and Higgs sectors of the SM contain only four free parameters, which we choose to be $m_W$, $g$, $\theta_\mathrm{W}$, and $m_h$, representing  the $W$ boson mass, $\mathrm{SU}(2)_\mathrm{L}$ gauge coupling, weak mixing angle, and Higgs boson mass, respectively. All the other parameters in these sectors are derived from them. For example, the $Z$ boson mass is given by $m_Z=m_W/\cos\theta_\mathrm{W}$, and the Higgs self-coupling is $\lambda_h = 2m_h/v^2=g^2 m_h/(2m_W^2)$. In the fermion sector, each Yukawa coupling is defined as $\lambda_f = g m_f/(\sqrt{2}m_W)$, introducing an additional parameter, the corresponding fermion mass $m_f$. %When those parameters deviate from SM, they will be label as ``an".
If we assume that the particle masses have been precisely measured and are treated as inputs, the free parameters in the EW sectors of the SM reduce to just two: $g$ and $\theta_\mathrm{W}$.

%The vertex $V_{WWZ}^{MNP}$ in GE representation comprises of 3 types of sub-vertices: $V^{\mu\nu\rho}_{WWZ}$, $V_{\varphi\varphi V}^{\mu}$ and $V_{\varphi VV}^{\nu\rho}$. Those couplings  are related to each other by 
%\begin{eqnarray}
%    WWZ: \ \  &&  g\equiv 2 g_{\varphi^+w^-\varphi^0}= 2g_{w^+\varphi^-\varphi^0}  \ \   c_w\equiv \frac{g_{wwz}}{g} \nonumber\\
%          \ \  &&  g_{\varphi^+\varphi^- z} = g\frac{c_{2w}}{2c_w} \ \ \  g_{w^+\varphi^- z}=-g_{\varphi^+ w^- z}= gm_W\frac{s_w}{c_w}\ \ \  g_{w^+w^-\varphi^0}=0
%\end{eqnarray}
%with $c_w\equiv \cos\theta_W, c_{2w}=\cos 2\theta_W, s_w=\sin\theta_W$.  All couplings are determined by 3 parameters: $g, c_w, m_W$. 

%Similar to $WWZ$, the vertex of $WWA$ includes different sub-vertices, the couplings of which are related as 
%\begin{eqnarray}
%    WWA: \ \  && 
%\end{eqnarray}

%Similarly,  related vertices also include both types of components,  thus will be represented by double line notation too.  Taking the example of $WWZ$ , $WWh$ and $WWZZ$  in  Fig.().  

\section{Direct Test of Gauge Symmetry }

%Eq.(\ref{eq:pol_vec}) is in contrast with  the  longitudinal polarization vector   in gauge representation, which in 5-component  can be  written as
%$\epsilon^M_L=(\frac{k^\mu}{m_V}-\frac{m_V}{n\cdot k}n^\mu,0)=\epsilon^{*M}_L$ . Comparing the two,  there is a difference of $k^M/m_V$:
%\begin{equation}
%    \epsilon_L^M(\text {GE})=\epsilon_L^M(\text{Gauge})-\frac{k^M}{m_V}.
%\end{equation}
%The two different forms of  longitudinal polarization vector must give the same amplitudes, resulting in  massive Ward identity of Eq.(\ref{eq:MWI}):
%\begin{equation}
%    \text{incoming}:k^M\mathcal{M}_M=0 \ \ \ \  \ 
%    \text{outgoing}:k^{*M}\mathcal{M}_M=0 
%\end{equation}

In this section, we directly test the gauge symmetry in EW scattering amplitudes using the MWI~\eqref{eq:MWI:alt}. We consider the EW processes $W^+ W^-\rightarrow t \bar t$ and $W^+W^-\rightarrow W^+W^-$ as two examples, replacing the polarization vectors of $W$ bosons in the amplitudes with the five-component momenta $k^{(*)M}$ and computing $k^M\mathcal M_M$ utilizing HELAS in the GE representation.

\begin{figure}[!t]
\centering
\subfigure[]{\includegraphics[height=0.33\textwidth]{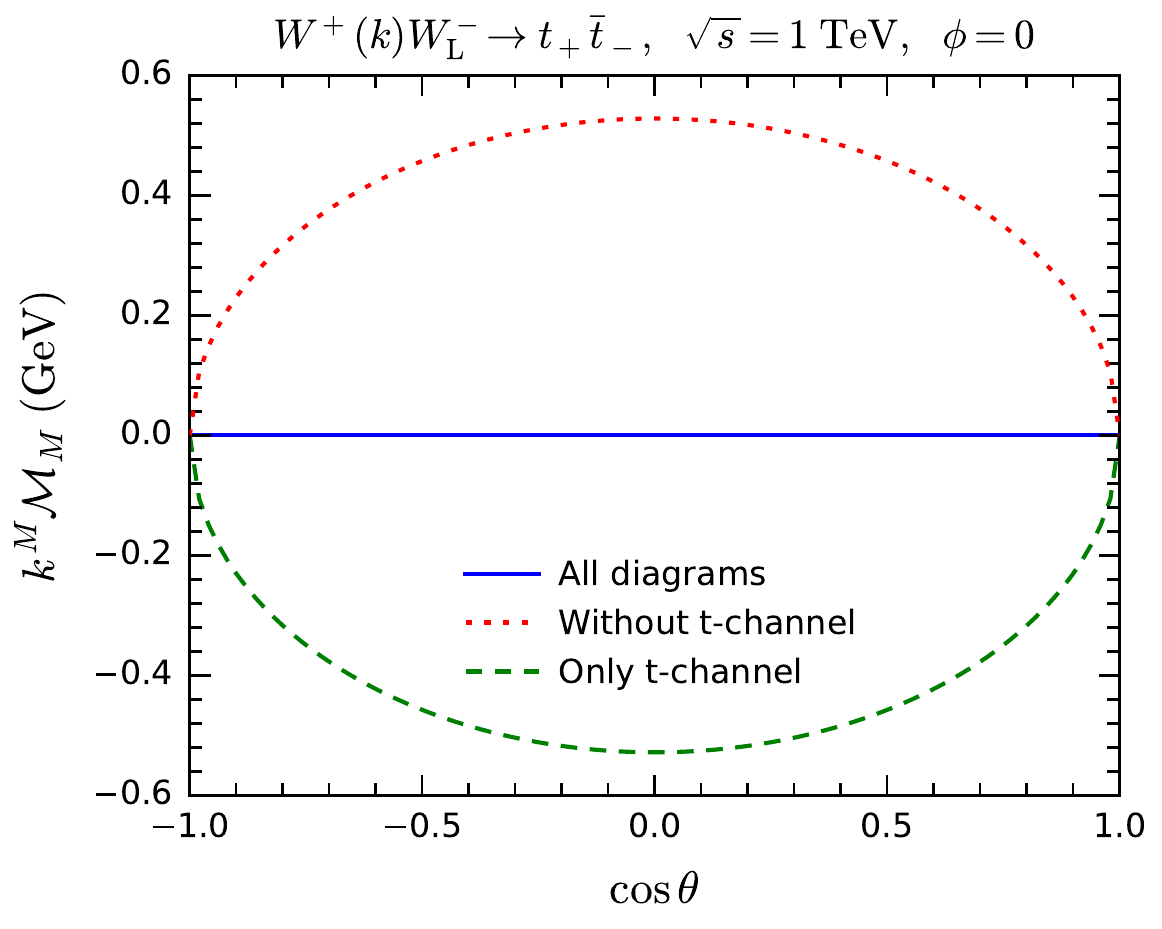}}
\subfigure[]{\includegraphics[height=0.33\textwidth]{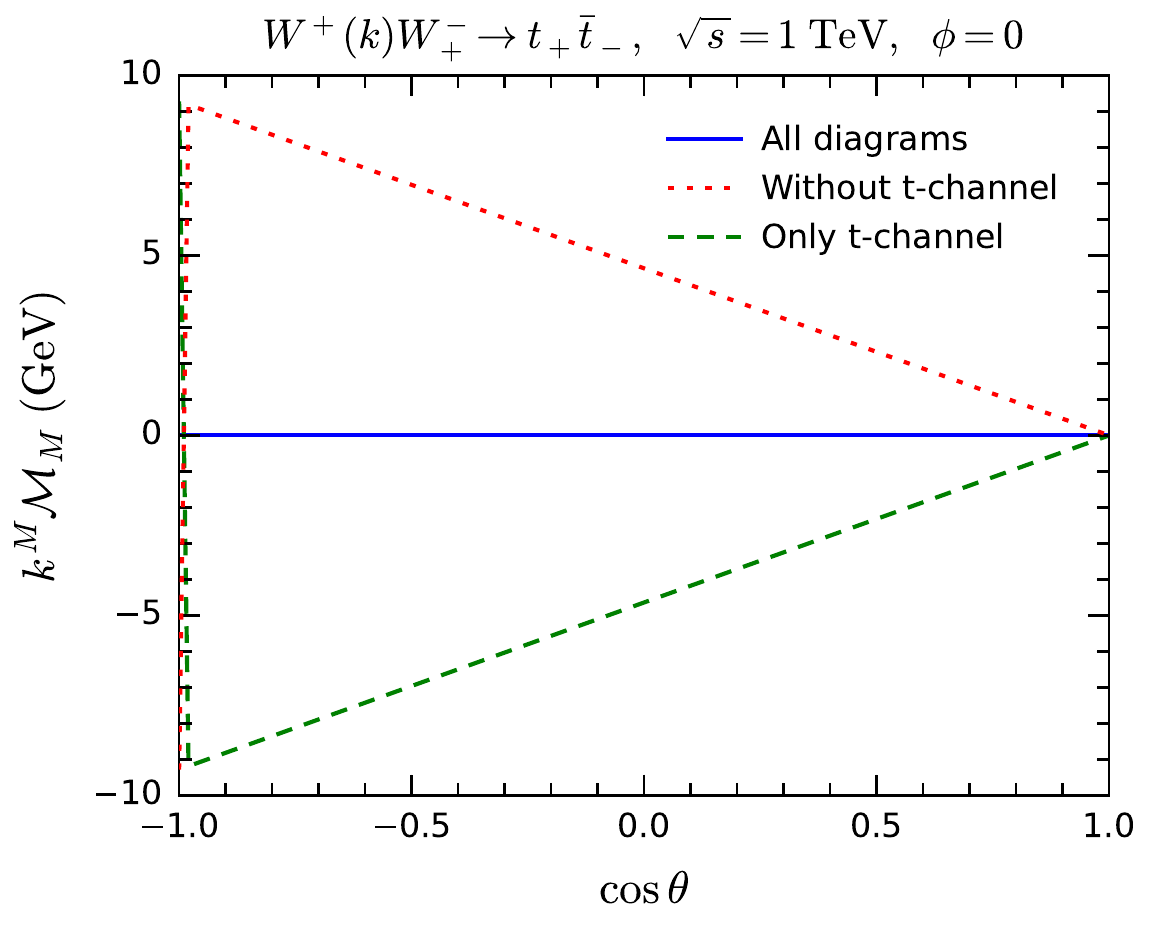}}
\subfigure[]{\includegraphics[height=0.33\textwidth]{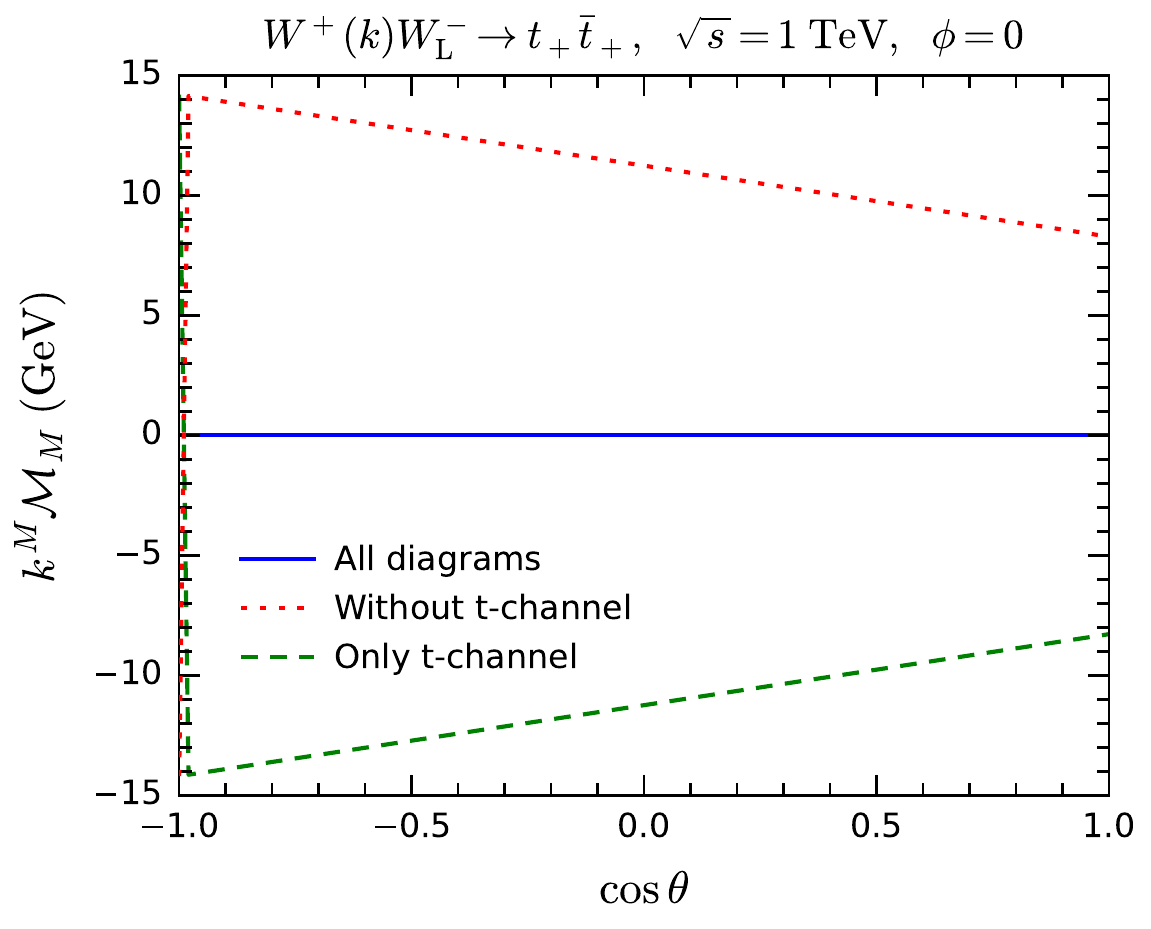}}
\subfigure[]{\includegraphics[height=0.33\textwidth]{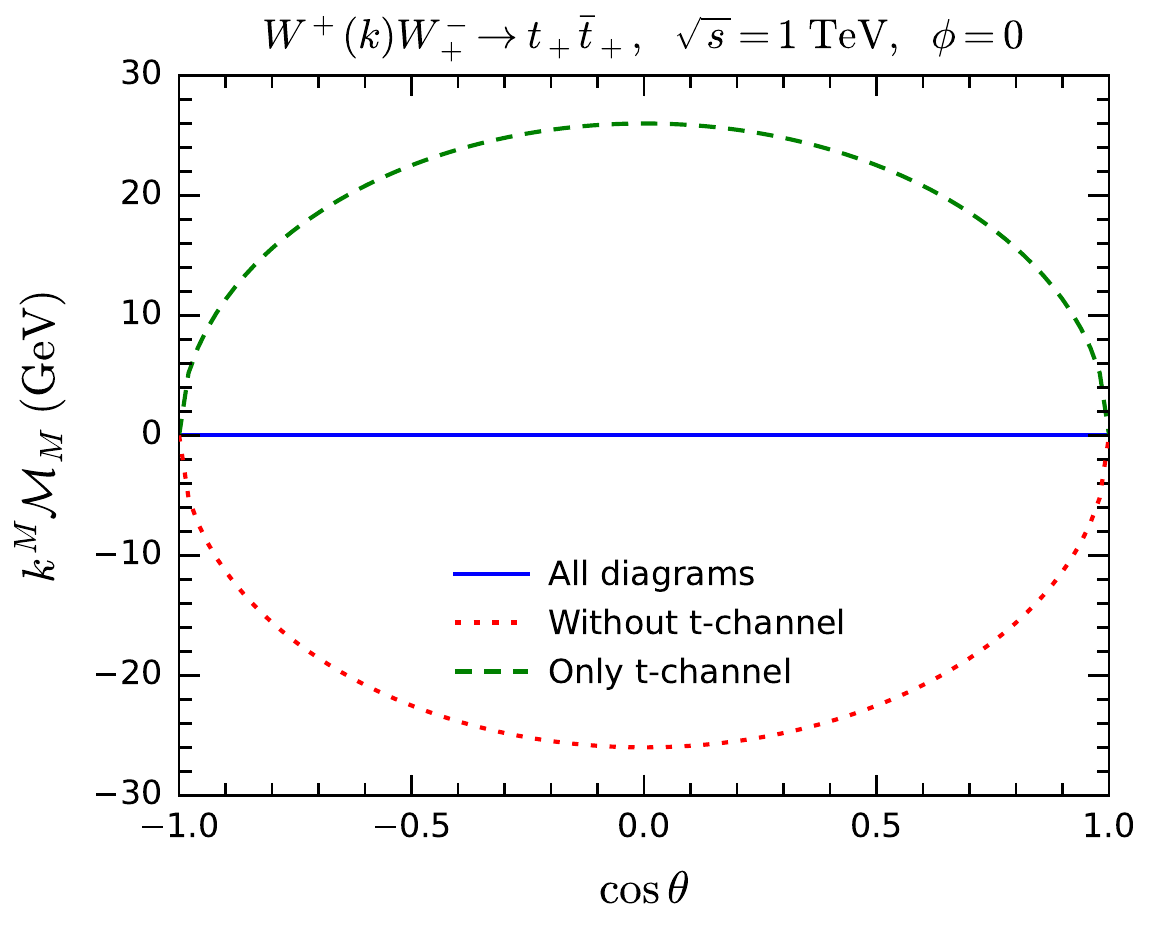}}
\caption{Testing the MWI by computing $k^M\mathcal M_M$ for $W^+W^-\rightarrow t\bar t$.}
\label{fig:MWI-ww2tt}
\end{figure}

In Fig.~\ref{fig:MWI-ww2tt}, we show $k^M\mathcal M_M$ for $W^+ W^-\rightarrow t \bar t$ as a function of $\cos\theta$, where $\theta$ is the scattering angle in the center-of-mass (CM) system. The CM energy and azimuthal angle are fixed as $\sqrt{s} = 1~\mathrm{TeV}$ and $\phi = 0$, respectively. In Figs.~\ref{fig:MWI-ww2tt}(a) and \ref{fig:MWI-ww2tt}(b), the polarization vector of $W^+$ is replaced by $k^M$, with the helicities of $t\bar t$ fixed as $+-$ and the helicity of $W^-$ set to be $0$ (longitudinal polarization) and $+$. In Figs.~\ref{fig:MWI-ww2tt}(c) and \ref{fig:MWI-ww2tt}(d), the setup is similar, except the helicities of $t\bar t$ are fixed as $++$. The blue solid lines represent the results including all tree-level diagrams, demonstrating the MWI $k^M\mathcal M_M = 0$. The red dotted and green dashed lines correspond to the results excluding the $t$-channel diagram and those solely involving the $t$-channel diagram, respectively. These contributions are precisely opposite to each other, indicating a large and exact cancellation between different diagrams, as expected.

\begin{figure}[!t]
\centering
\subfigure[]{\includegraphics[height=0.33\textwidth]{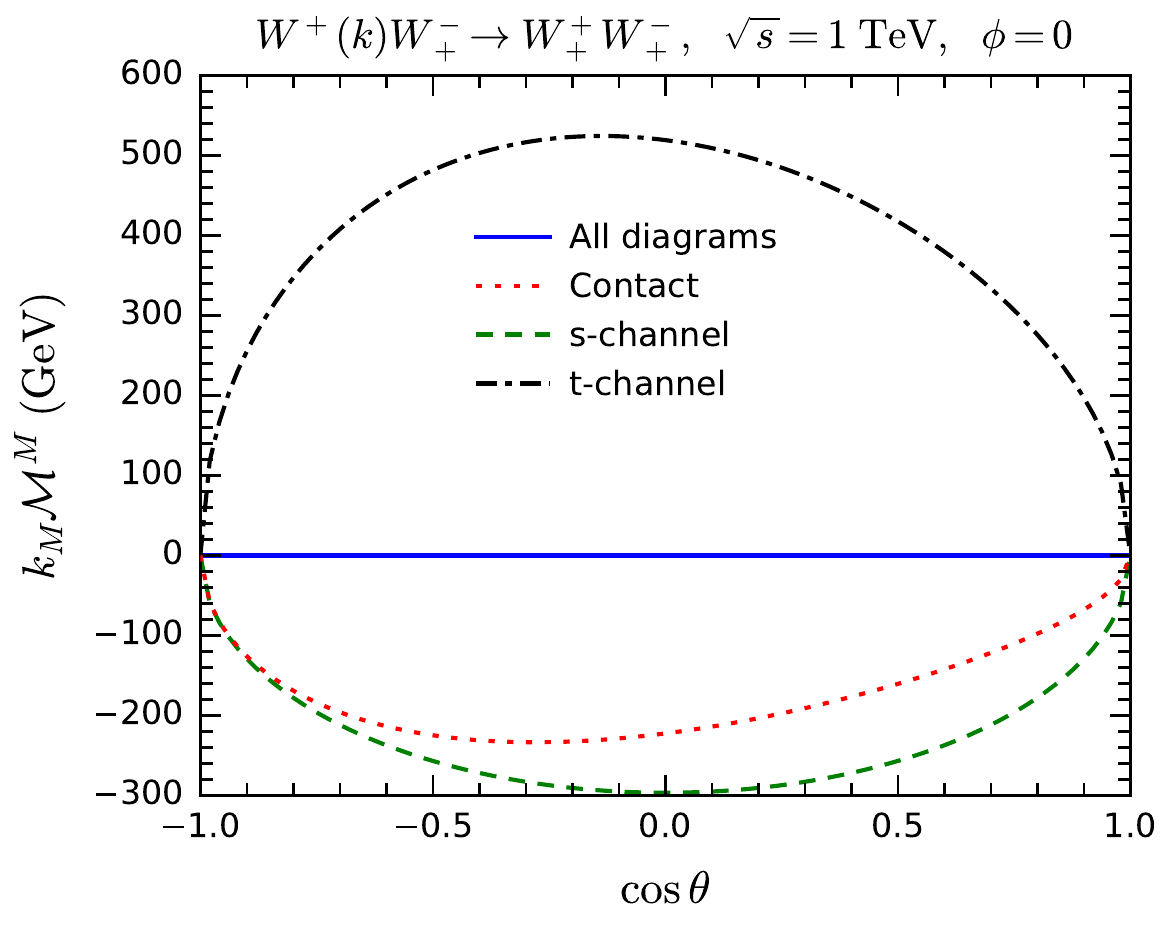}}
%\hspace{.01\textwidth}
\subfigure[]{\includegraphics[height=0.33\textwidth]{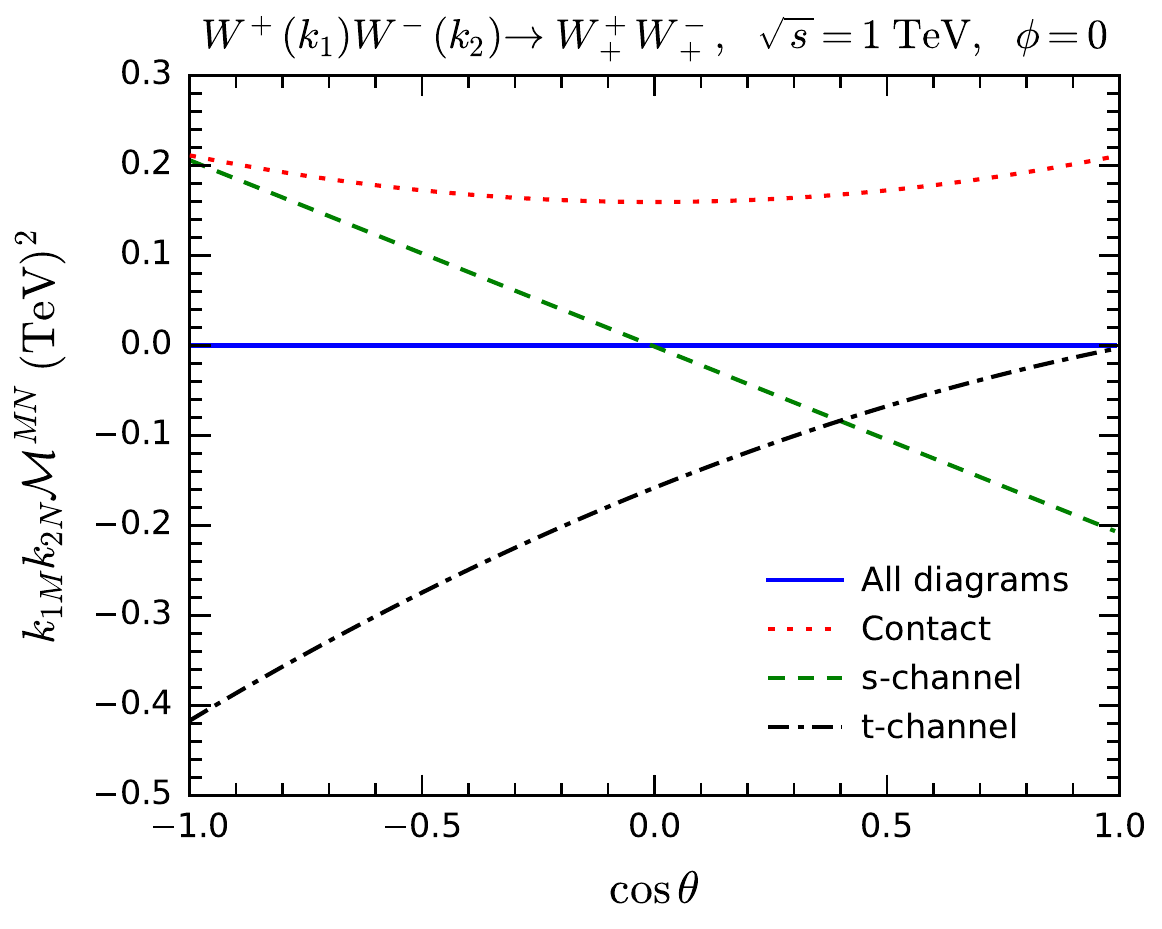}}
\subfigure[]{\includegraphics[height=0.33\textwidth]{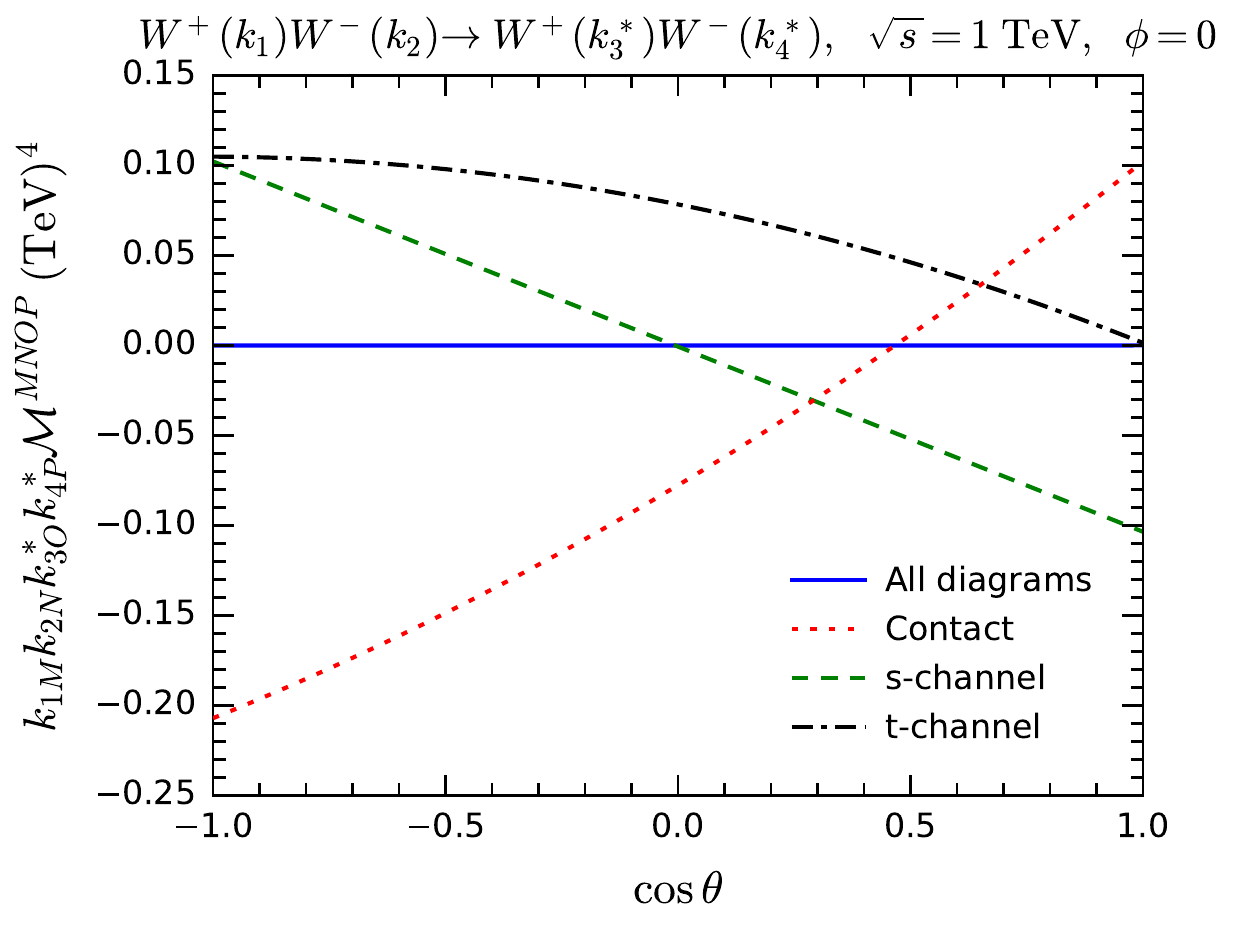}}
\caption{Testing the MWI by computing $k_M \mathcal M^M$ (a), $k_{1M} k_{2N} \mathcal M^{MN}$ (b), and $k_{1M} k_{2N} k^*_{3O} k^*_{4P} \mathcal M^{MNOP}$ for $W^+W^-\rightarrow W^+W^-$.}
\label{fig:MWI-ww2ww}
\end{figure}

In Fig.~\ref{fig:MWI-ww2ww}, we present the results for $W^+W^-\rightarrow W^+W^-$, where one, two, and four polarization vectors of $W$ bosons are replaced by one, two, and four five-component momenta, respectively. The helicities of the remaining $W$ bosons are uniformly set to $+$.
The blue lines represent the sums of all tree-level diagrams, whereas the red dotted, green dashed, and black dot-dashed lines correspond to the results for the contact, $s$-channel, and $t$-channel diagrams, respectively.
Once again, we observe that the MWI is satisfied when all tree-level diagrams are included, confirming the gauge symmetry.  

\section{Testing Gauge Symmetry through Anomalous Couplings}
\label{sec:anom_coup}

In the previous section, we confirmed that the gauge symmetry represented by the MWI ensures precise cancellation among various diagrams when one or multiple polarization vectors $\epsilon^M(k)$ are replaced by five-component momenta $k^M$.  In this section, we take a different approach by modifying certain couplings to examine how the MWI is affected. % An analysis of gauge symmetry implies modify couplings anomalously would result in the violation of  MWI(Eq.(\ref{eq:MWI})).  We will try to confirm it in this section. 
Such a test of gauge symmetry through anomalous couplings can provide deeper insights into how different couplings and parameters are interrelated through gauge symmetry. 

In the GE representation, any vertex involving massive vector bosons contains multiple couplings. This allows us not only to modify the couplings of individual vertices as a whole but also to adjust specific parameters within those couplings.
The latter will be studied in Subsection~\ref{sec:3-point} with three-point vertices, whereas the former will be discussed in Subsection~\ref{sec:4-point} with four-point vertices.

\subsection{Three-point Vertices}
\label{sec:3-point}

In this subsection, we examine gauge symmetry through the analysis of three-point vertices, specifically focusing on the $VVh$, $ff'V$, and $VVV$ vertices within the GE representation. 

%In GE representation,   a vertex involving massive vector boson has more than one parameter.  For example, $V^{MN}_{WWh}$ has the pure gauge component $V^{\mu\nu}_{wwh}$ with coupling $g_{wwh}m_W$, but also  gauge-Goldstone components $V^{M4}_{WWh}$ with coupling $\frac{1}{2}g_{w\varphi h}$ and Goldstone-Goldstone component $V_{wwh}^{44h}$ with coupling $-ig\frac{m_h}{m_W}$(or $-i\frac{\lambda_{\phi\phi h}}{2}v$). So we have $V^{MN}_{WWh}(g_{wwh}, g_{\phi wh}, \lambda_{\phi\phi h}; m_h, m_W)$.  %I have purposely wrote the parameter $g$ in pure gauge and gauge-Goldstone couplings differently because there is no guarantee of their equality without gauge symmetry, especially under quantum corrections.

%In the SM, those parameters are not independent, but are instead related to each other by the following relations protected by gauge symmetry:
%\begin{equation}\label{eq:hVV_coup}
%    g_{wwh}=g_{w\phi h} = g \ \ \ \ \lambda_{\phi\phi h}=g^2\frac{m_h}{2m_W^2}, 
%\end{equation}
%leaving only two independent parameters, of which we choose couplings $g, \lambda_{\phi\phi h}$.  On the other hand,  if  one of  couplings deviates from the SM value, so that Eq.(\ref{eq:hVV_coup}) is no longer valid, then gauge symmetry will also be broken.  For example, if $g_{w\phi h}=g+\delta g_{w\phi }$,  then we have $k^M\mathcal M_M\neq 0$ for processes involving $hWW$ vertex.   The patterns of gauge symmetry and its breaking  also apply to other vertices involving vector bosons, such as $VVV$ and $ffV$.

\subsubsection*{$VVh$ vertex}

The first type of vertices we study is the $VVh$ vertices, which include the $WWh$ and $ZZh$ vertices.  As discussed in Sec.~\ref{sec:GE_int}, a $VVh$ vertex in the GE representation involves four sub-vertices: $VVh$, $V\varphi h$, $V\varphi h$, and $\varphi\varphi h$, with corresponding couplings $g_{hVV}$, $g_{V\varphi h}$, $g_{\varphi V h}$, and $\lambda_{\varphi\varphi h}$.
For the $WWh$ vertex in the SM, the couplings are related to each other by the following relations:
\begin{equation}\label{eq:hVV_coup}
WWh:  \quad  g = 2 g_{W\varphi h} = 2 g_{\varphi W h} =\frac{g_{WWh}}{m_W},\quad
\lambda_{\varphi\varphi h}=\frac{g m_h^2}{2m_W}.
\end{equation}

We intend to test the MWI by modifying these couplings by
\begin{equation}\label{eq:hVV_coup_anom}
\begin{aligned}
g_{WWh} &=gm_W(1+\delta^1_{WWh}),\quad
g_{W\varphi h}=\frac{g}{2}(1+\delta^2_{WWh}),
\\
g_{\varphi W h} &= \frac{g}{2} (1+\delta^3_{WWh}), \quad
\lambda_{\varphi\varphi h}=\frac{gm^2_h}{2m_W}(1+\delta^4_{WWh}).
\end{aligned}
\end{equation}
The SM corresponds to $\delta^1_{WWh}=\delta^2_{WWh}=\delta^3_{WWh}=\delta^4_{WWh}=0$.  For the $ZZh$ vertex, we only need to replace $m_W$ with $m_Z$ in Eqs.~\eqref{eq:hVV_coup} and \eqref{eq:hVV_coup_anom}. To facilitate a clearer comparison with the gauge representation, we leave $g_{WWh}$ unchanged, ensuring $\delta_{WWh}^1=0$ at all times. Additionally, we enforce $\delta^2_{WWh}=\delta^3_{WWh}\equiv \delta^{23}_{WWh}$ when modifying the couplings, as the two sub-vertices $V\varphi h$ and $\varphi V h$ belong to the same type.

\begin{figure}[!t]
    \centering
    \subfigure[]{ \includegraphics[height=0.33\textwidth]{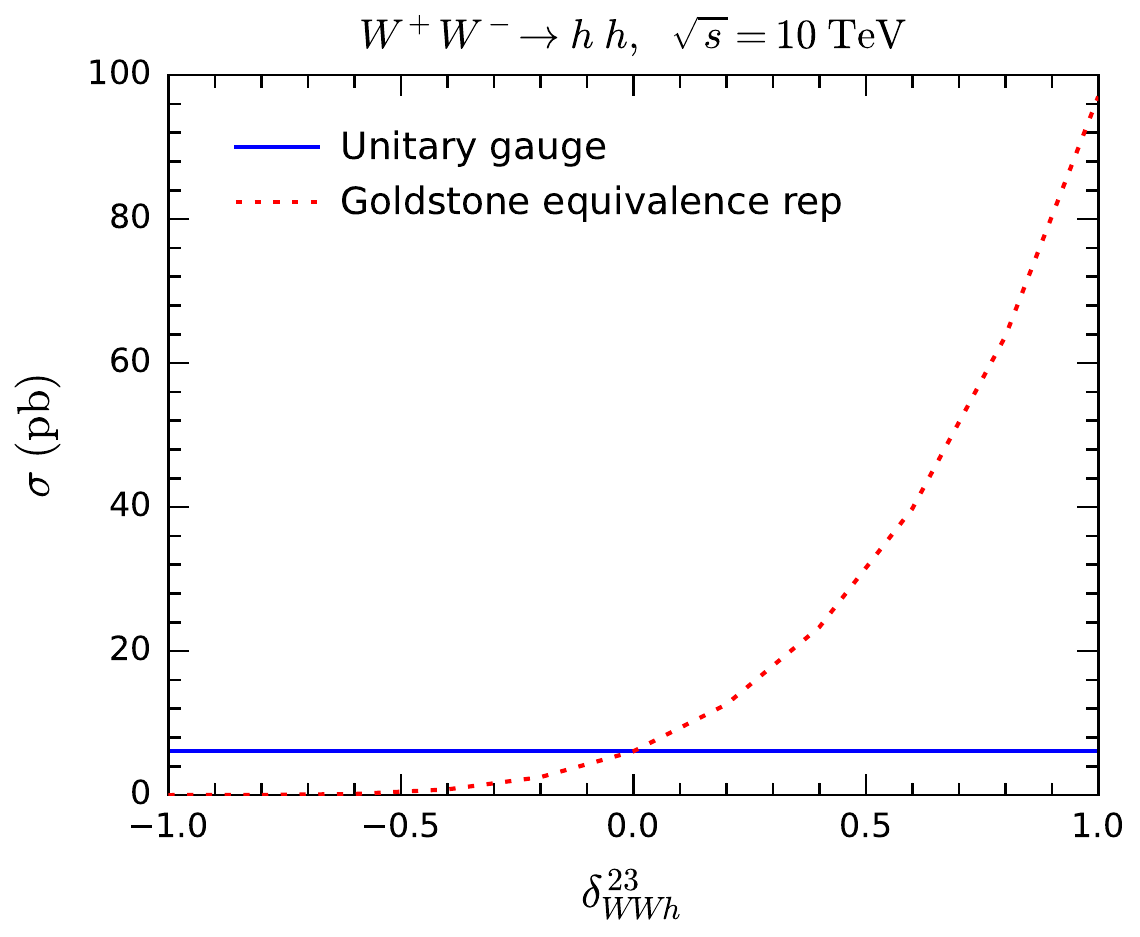}}
    \subfigure[]{ \includegraphics[height=0.33\textwidth]{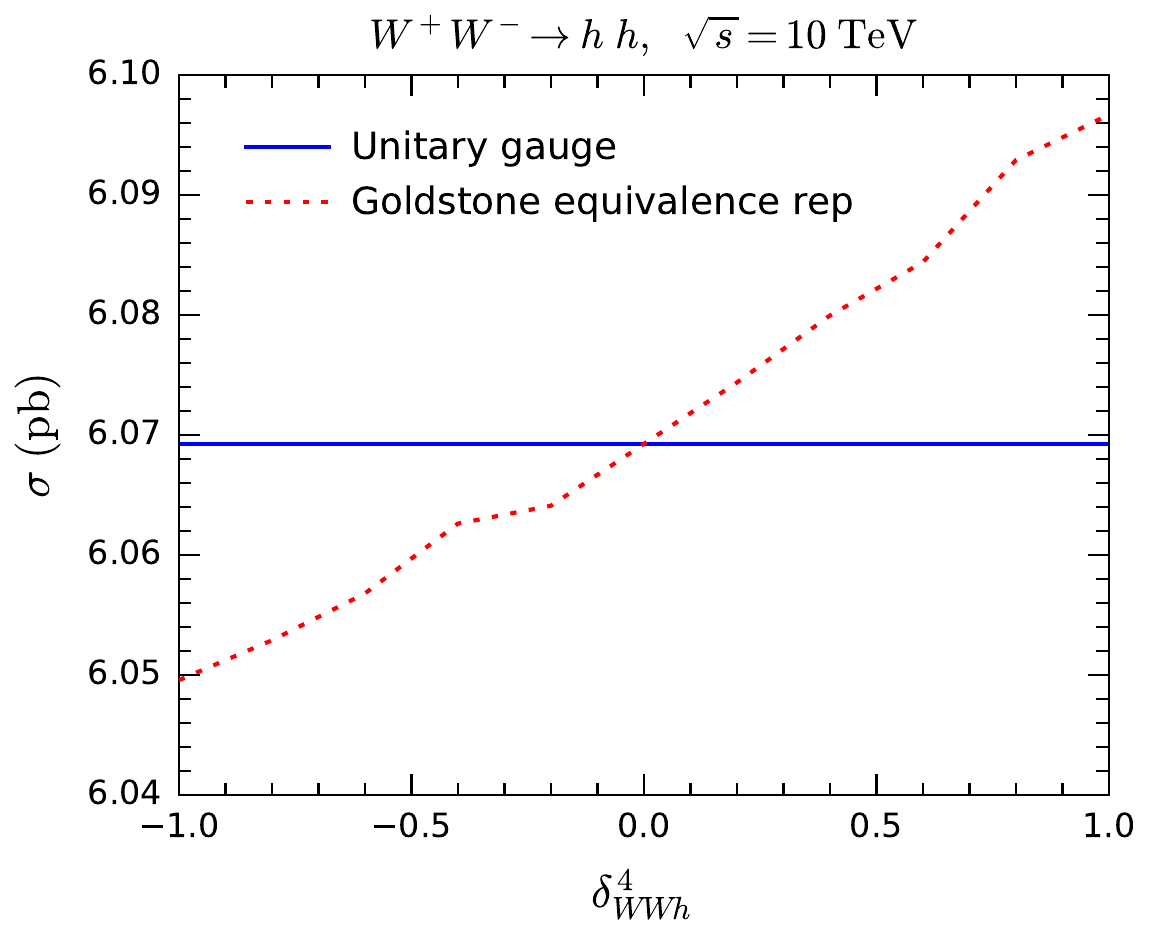}}
    \subfigure[]{ \includegraphics[height=0.33\textwidth]{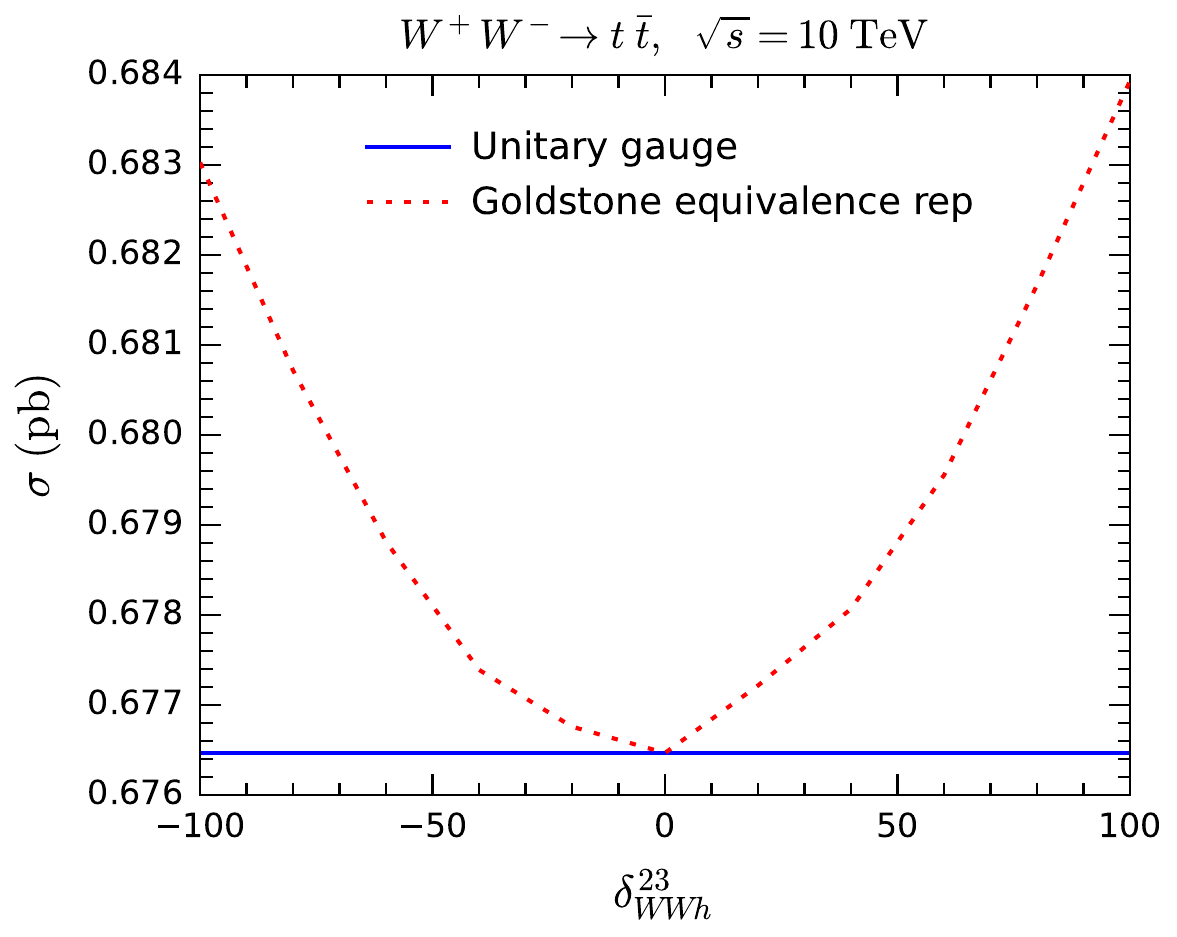}}
    \subfigure[]{ \includegraphics[height=0.33\textwidth]{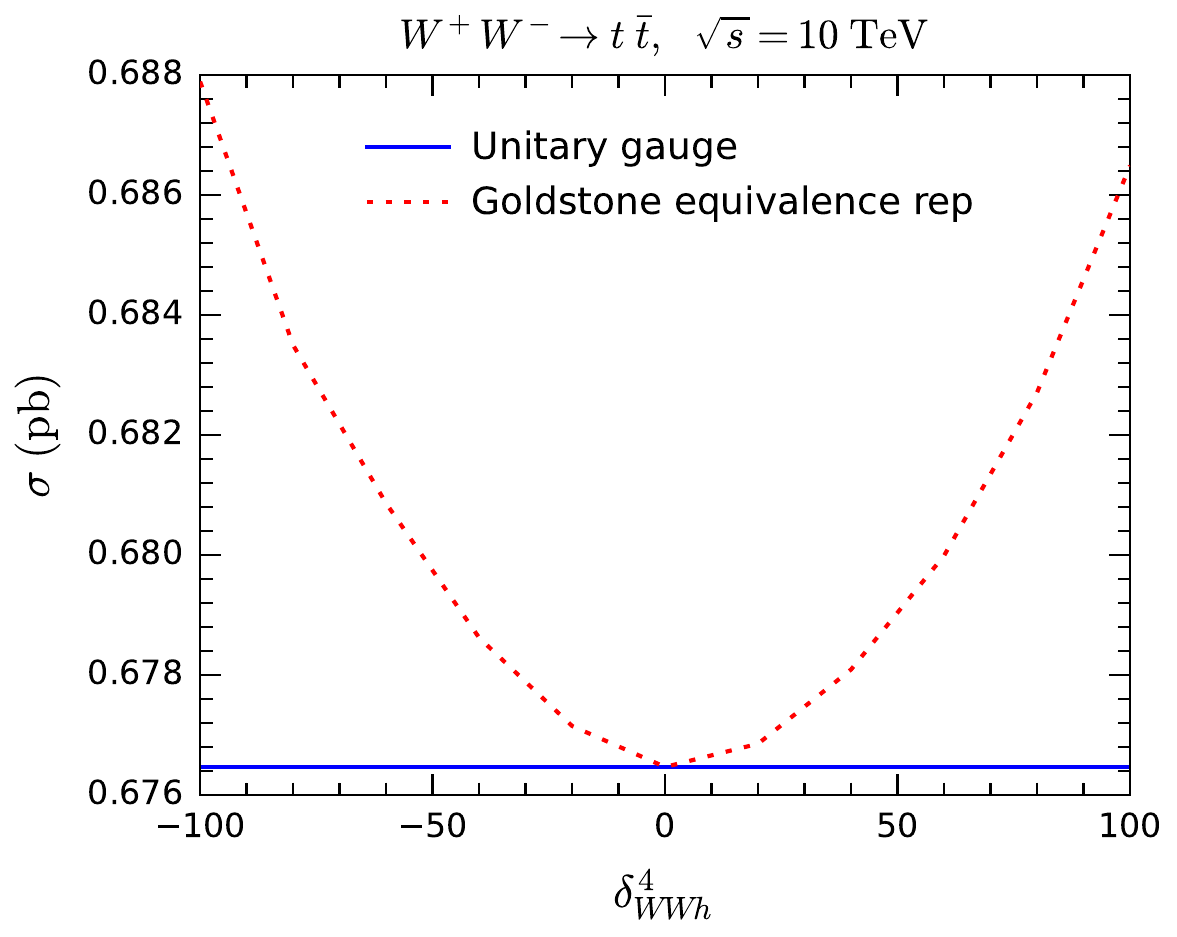}}
    \caption{Cross sections of $WW\rightarrow hh$ (upper panels) and $WW\rightarrow t \bar t$ (lower panels) at $\sqrt{s} = 10~\mathrm{TeV}$ with varying $\delta^{23}_{WWh}$ and $\delta^4_{WWh}$ in the unitary gauge (blue solid lines) and GE representation (red dotted lines).  }
    \label{fig:wwh_gauge }
\end{figure}

The processes we choose to test gauge symmetry with the $WWh$ vertex are $W^+ W^-\rightarrow t\bar t$ and $W^+ W^-\rightarrow hh$.   We first compare the cross sections with respect to $\delta_{WWh}^{23,4}$ in both the unitary gauge and the GE representation,  the results are shown in Fig.~\ref{fig:wwh_gauge }. As expected, when  $\delta^{23}_{WWh}$ and $\delta^4_{WWh}$ are nonzero, the cross sections in the GE representation deviate from those in the unitary gauge, which remain unchanged and retain the SM values. The larger the values of $\delta^{23}_{WWh}$ and $\delta^4_{WWh}$, the larger the deviations.   Compared with $WW\rightarrow hh$, $WW\rightarrow t \bar t$ exhibits significantly lower sensitivity to $\delta^{23}_{WWh}$ and $\delta^4_{WWh}$. We must plot $\delta^{23}_{WWh}$ and $\delta^4_{WWh}$  over the range $[-100, 100]$ for $WW\rightarrow t \bar t$ but only $[-1, 1]$  for $WW\rightarrow hh$. This is not surprising, given that the $WWh$ vertex contributes to three channels of $WW\rightarrow hh$ but only one in $WW\rightarrow t\bar t$.   Another notable observation is that the sensitivity of the cross section is much higher to $\delta^{23}_{WWh}$ than to $\delta^4_{WWh}$ for $WW\rightarrow h h$, but the sensitivities are approximately the same for $WW\rightarrow t\bar t$.

\begin{figure}[!t]
    \centering
     \subfigure[]{\includegraphics[height=0.33\textwidth]{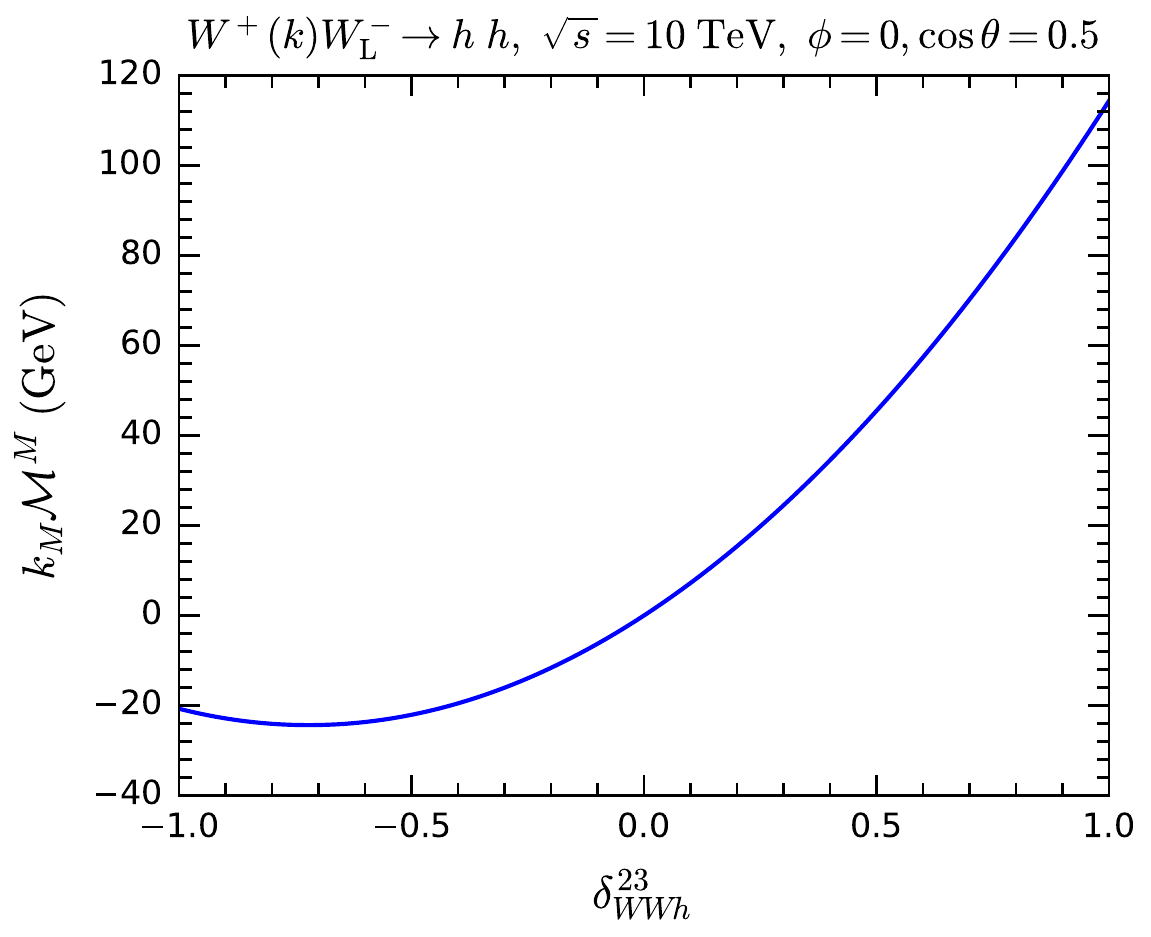}}
    \subfigure[]{\includegraphics[height=0.33\textwidth]{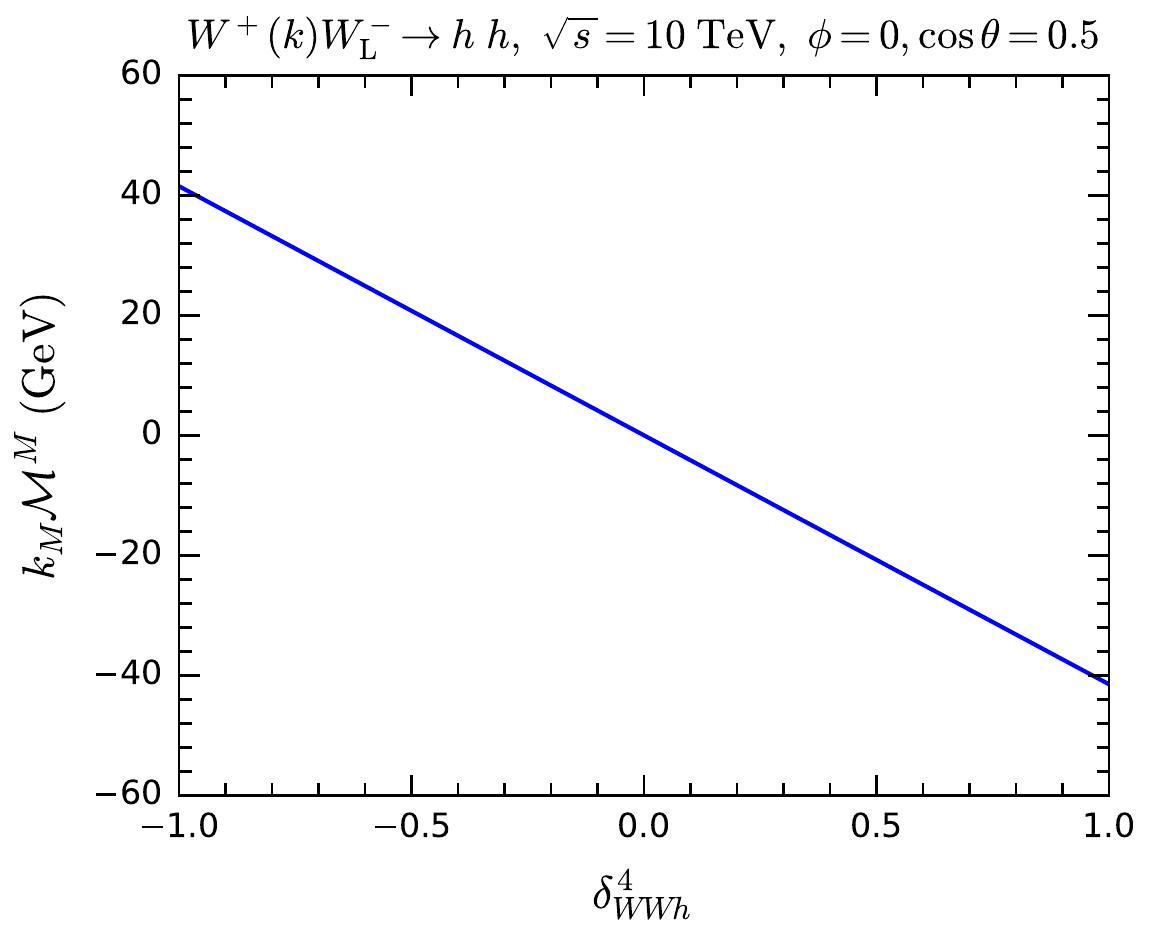}}
     \subfigure[]{\includegraphics[height=0.33\textwidth]{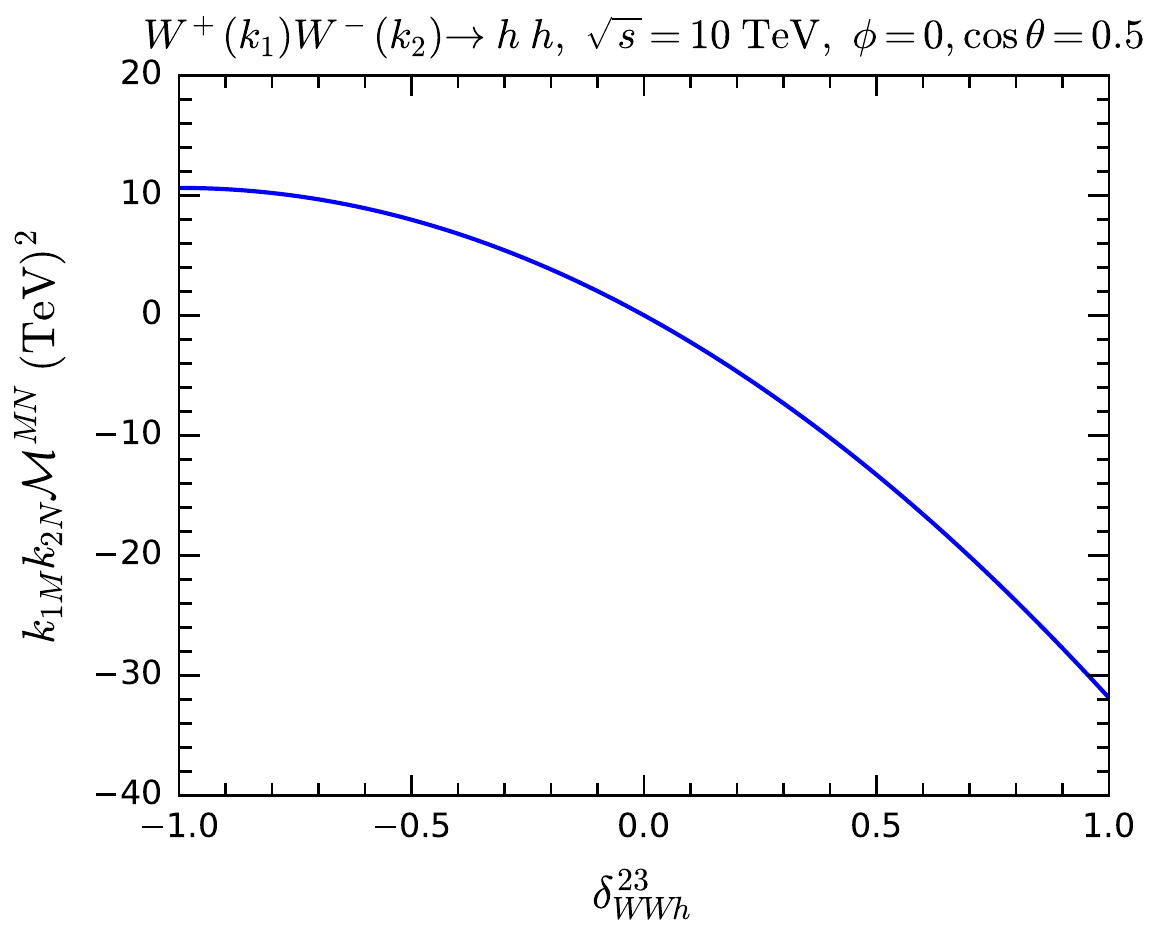}}
     \subfigure[]{\includegraphics[height=0.33\textwidth]{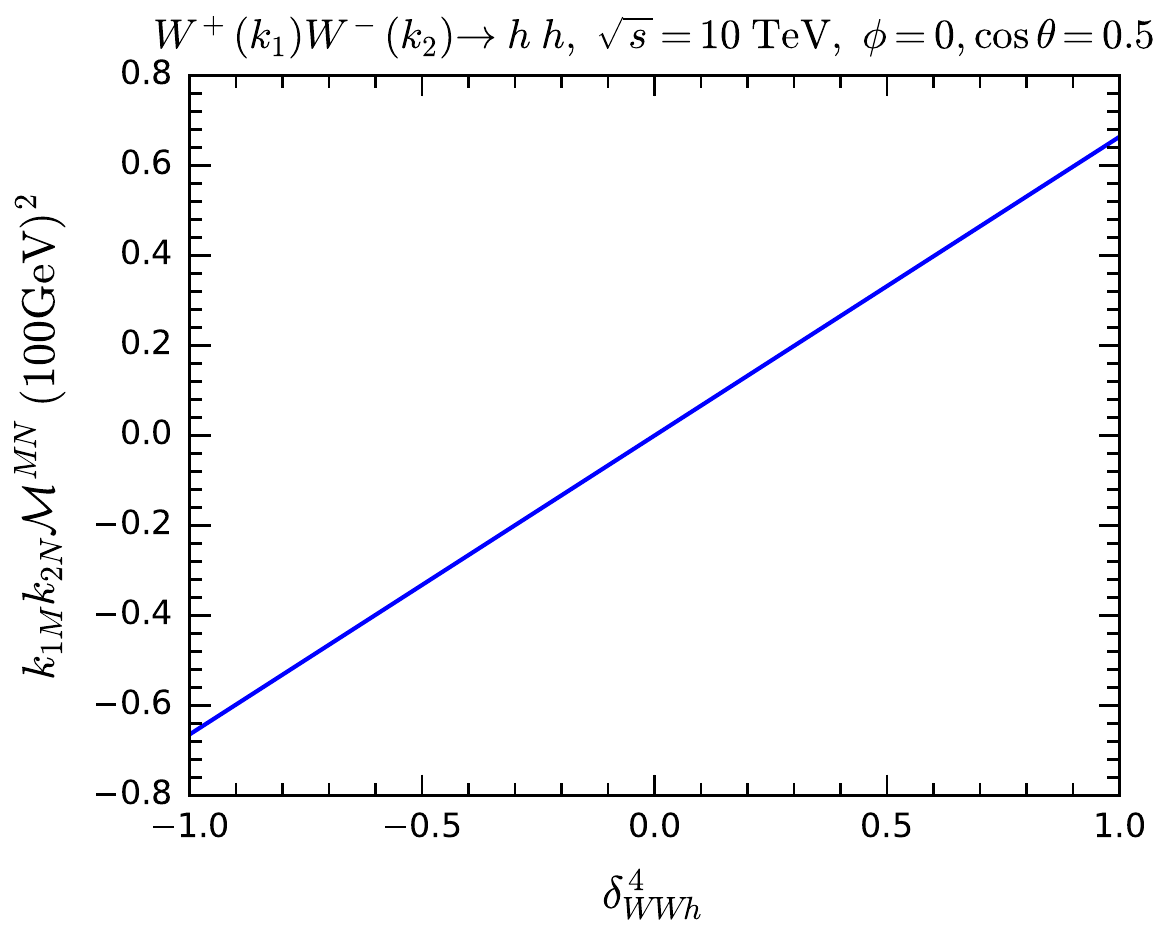}}
    \caption{$k_M\mathcal M^M$ and $k_{1M} k_{2N} \mathcal M^{MN}$ as functions of $\delta^{23}_{WWh}$ and $\delta^4_{WWh}$ for $WW\rightarrow hh$ with one or two $W$ polarization vectors are replaced by five-component momenta.}
    \label{fig:WI-ww2hh}
\end{figure}

\begin{figure}[!t]
    \centering
     \subfigure[]{\includegraphics[height=0.33\textwidth]{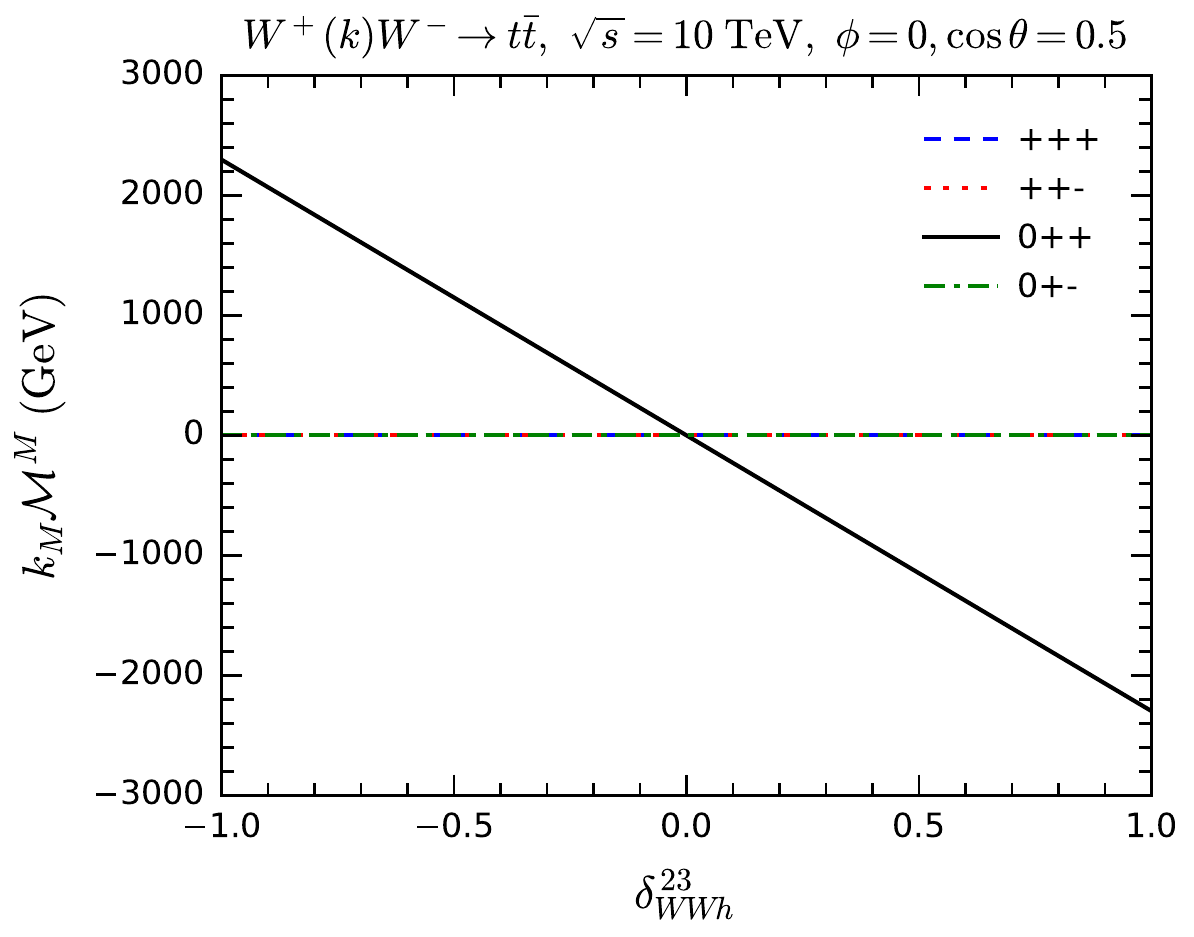}}
    \subfigure[]{\includegraphics[height=0.33\textwidth]{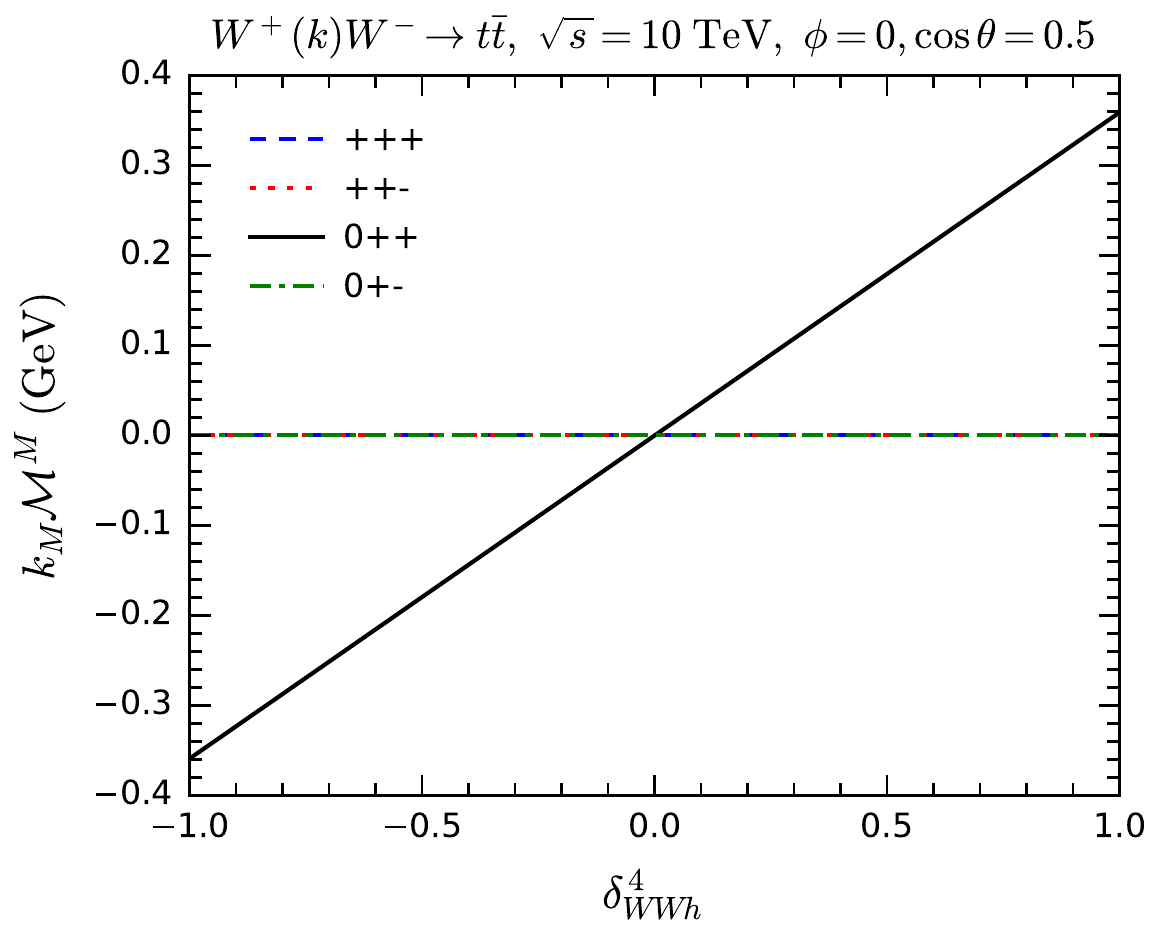}}
\hspace{0.5\textwidth}
     \subfigure[]{\includegraphics[height=0.33\textwidth]{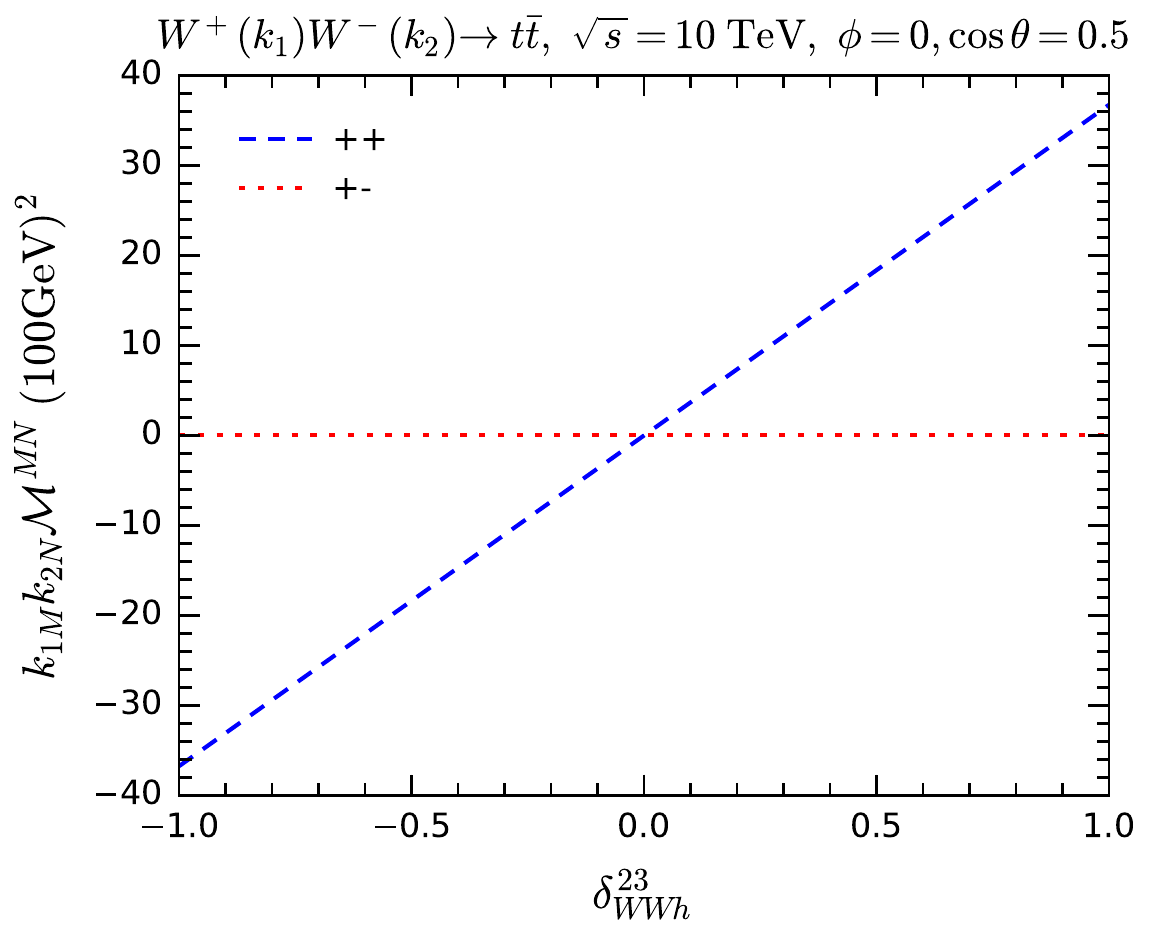}}
\subfigure[]{\includegraphics[height=0.33\textwidth]{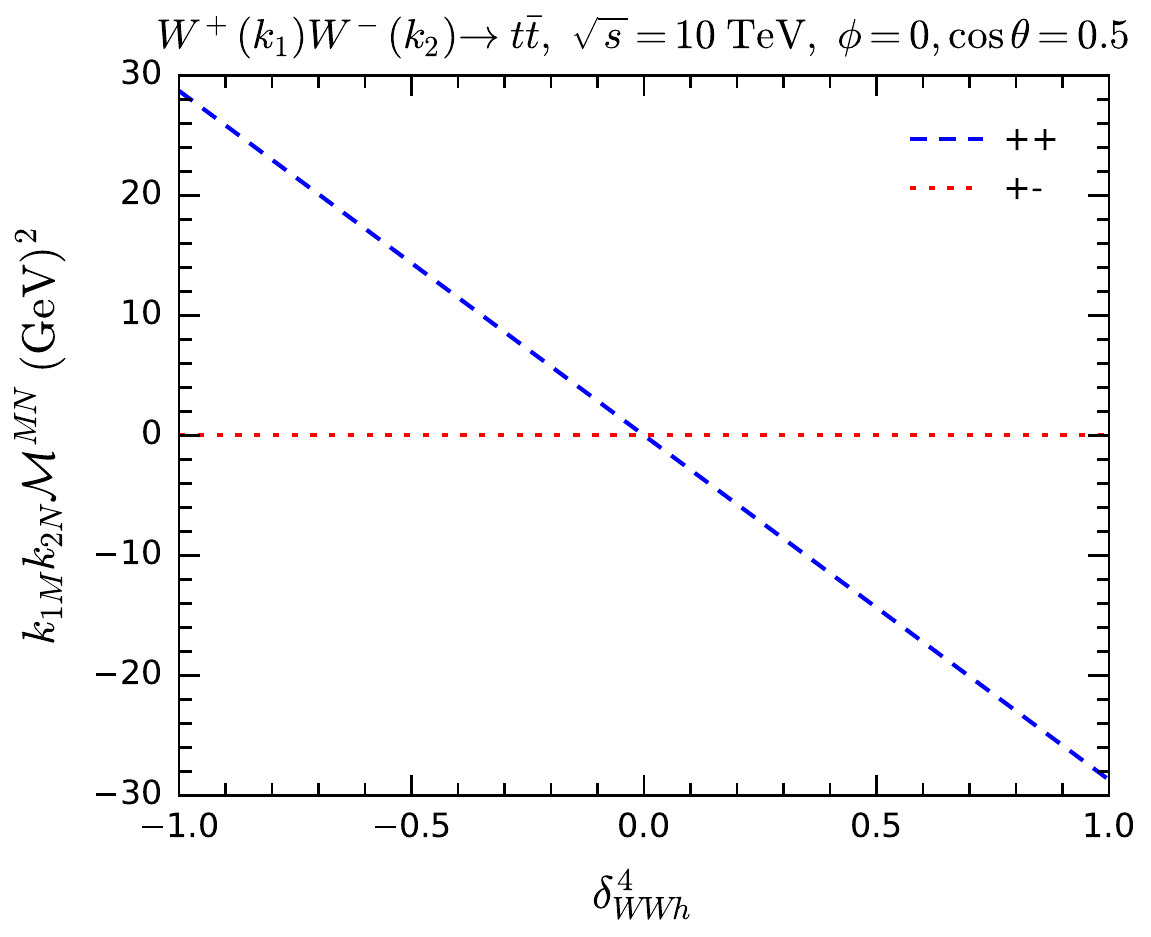}}
    \caption{$k_M\mathcal M^M$ and $k_{1M} k_{2N} \mathcal M^{MN}$ as functions of $\delta^{23}_{WWh}$ and $\delta^4_{WWh}$ for $WW\rightarrow t\bar t$ with one or two $W$ polarization vectors replaced by five-component momenta. The results for various helicity combinations are shown.}
    \label{fig:WI-ww2tt}
\end{figure}

Next, we compute $k_M\mathcal M^M$ or $k_{1M} k_{2N} \mathcal M^{MN}$ for these processes as a way of testing gauge symmetry by replacing the polarization vectors of one or two $W$ bosons with five-component momenta. The results are shown in Fig.~\ref{fig:WI-ww2hh} for $WW\rightarrow hh$ and Fig.~\ref{fig:WI-ww2tt} for $WW\rightarrow t\bar t$. For $WW\rightarrow t\bar t$, we analyze various helicity combinations. In certain helicity configurations, the results consistently vanish due to the violation of angular momentum conservation. Apart from these cases, the anomalous couplings lead to violations of the MWI. For $WW\rightarrow t\bar t$, the deviations for $\delta^{23}_{WWh}$ are larger than those for $\delta^4_{WWh}$. 

%\JM{We  compare the cross section/amplitudes with different gauges: GEG v.s unitarity gauge (UG), when modifying $\delta_2/\delta_3$ and $\delta_4$ respectively.   As shown in Fig.(), we found....}

%We also analyze $hZZ$ with the processes  $ZZ\rightarrow ZZ$ and $ZZ\rightarrow hh$. The method is the same as $hww$ except we change $w$ to $z$. In Fig.(), we show all the results. 

\subsubsection*{$ff'V$ vertex}

The $ff'V$ vertex has two sub-vertices: $ff'V$ and $ff'\varphi$. The $ff'V$ coupling can be parameterized as $-i \gamma^\mu (g_\mathrm{L}  P_\mathrm{L} + g_\mathrm{R} P_\mathrm{R})$, whereas  the $ff'\varphi$ coupling is $y_\mathrm{L} P_\mathrm{L} + y_\mathrm{R} P_\mathrm{R}$.  Similar to the $VVh$ case, gauge couplings and Yukawa couplings are not independent, as the latter can be expressed in terms of the former.    

For a charged current vertex, such as $u\bar d W$, we have 
\begin{equation}
    g_\mathrm{R}=0,\quad
    g_\mathrm{L}=\frac{g}{\sqrt{2}},\quad
    y_\mathrm{L}=\frac{g m_d}{\sqrt 2 m_W},\quad
    y_\mathrm{R}=-\frac{g m_u}{\sqrt 2 m_W}.
\end{equation}
%such as $u\bar d W$,  $g_R=0$, $g_L=\frac{g}{\sqrt{2}}$; while the Yukawa couplings $y_L=y_d, y_R=-y_u$ are related to the gauge coupling $g$ by $y_{d,u}=g\frac{m_{d,u}}{\sqrt{2}m_W}$.  
For a neutral current vertex, such as $u\bar u Z$, we have 
\begin{equation}
    g_\mathrm{R} = -\frac{Q_f g s_\mathrm{W}^2}{c_\mathrm{W}},\quad
    g_\mathrm{L} = g_\mathrm{R} + \frac{g}{2c_\mathrm{W}},\quad
    y_\mathrm{L}=-y_\mathrm{R}=\frac{g m_u}{2 m_W},
\end{equation}
with $s_\mathrm{W} \equiv \sin \theta_\mathrm{W}$ and $c_\mathrm{W} \equiv \cos \theta_\mathrm{W}$.
%$g_L-g_R=\frac{g}{2c_w},g_R=-Q_f\frac{g}{c_w}s_w^2$, $y_L=-y_R=\frac{y_u}{\sqrt{2}}=g\frac{m_u}{2m_W}$.  The full SM fermion couplings are summarized in cite(). 
To sum up, the $ff'\varphi$ couplings are Yukawa couplings $y_f$, which are related to the fermion masses $m_f$ by  $y_f = g m_f/(\sqrt{2}m_W)$. This relation is  protected by gauge symmetry, and therefore, the breaking of gauge symmetry will be reflected in the violation of this relation. 

Similar to $VVh$,  we test gauge symmetry by modifying the $ff'V$ couplings as follows:  
\begin{equation}
    y_{f}= \frac{g m_f}{\sqrt{2}m_W} (1+\delta_f).
\end{equation}
The process we choose to examine gauge symmetry for the $ff'V$ vertex is $WW\rightarrow t\bar t$. This process is ideal because its $t$-channel diagram includes the $t\bar b W$ vertex, whereas its $s$-channel diagram includes the $t\bar t Z$ vertex, allowing us to examine both charged and neutral currents.

\begin{figure}[!t]
    \centering
     \subfigure[]{\includegraphics[height=0.33\textwidth]{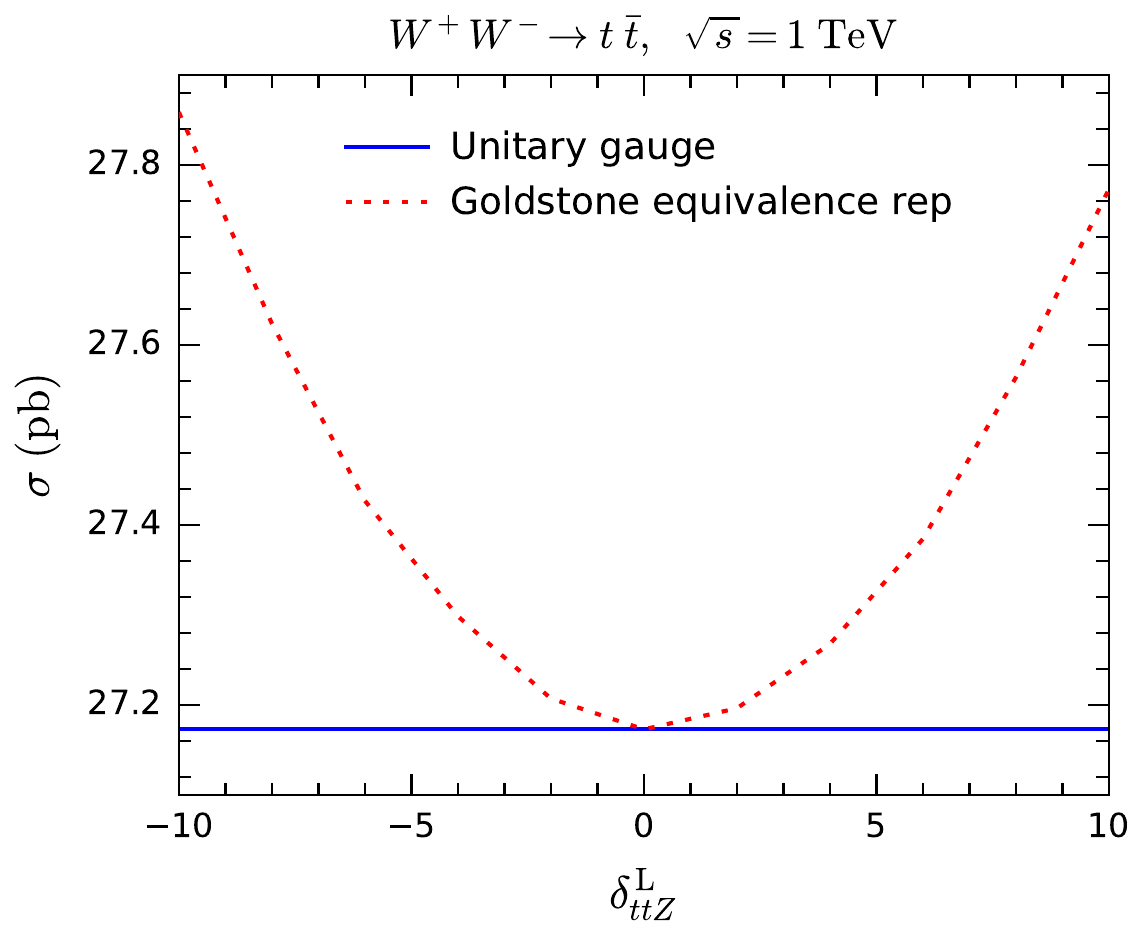}}
    \subfigure[]{\includegraphics[height=0.33\textwidth]{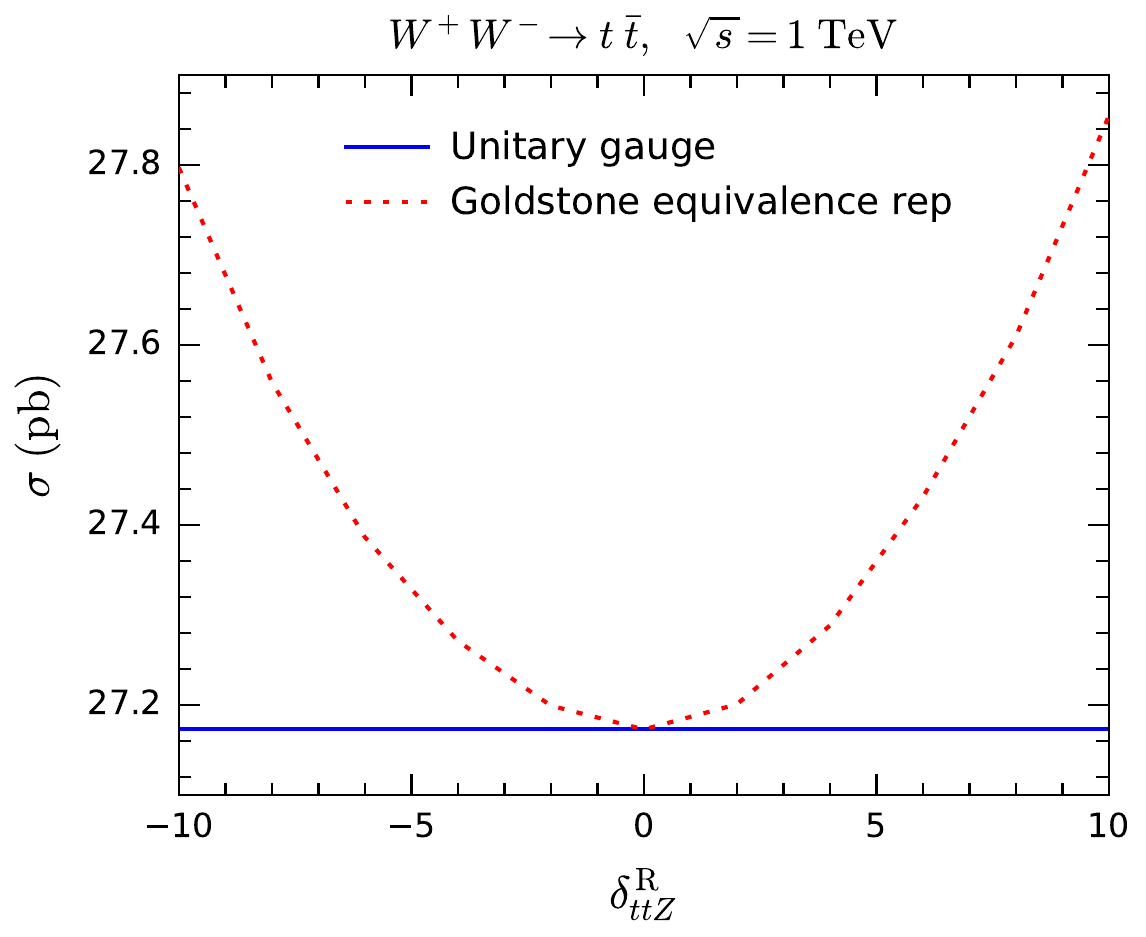}}
%\hspace{0.5\textwidth}
     \subfigure[]{\includegraphics[height=0.33\textwidth]{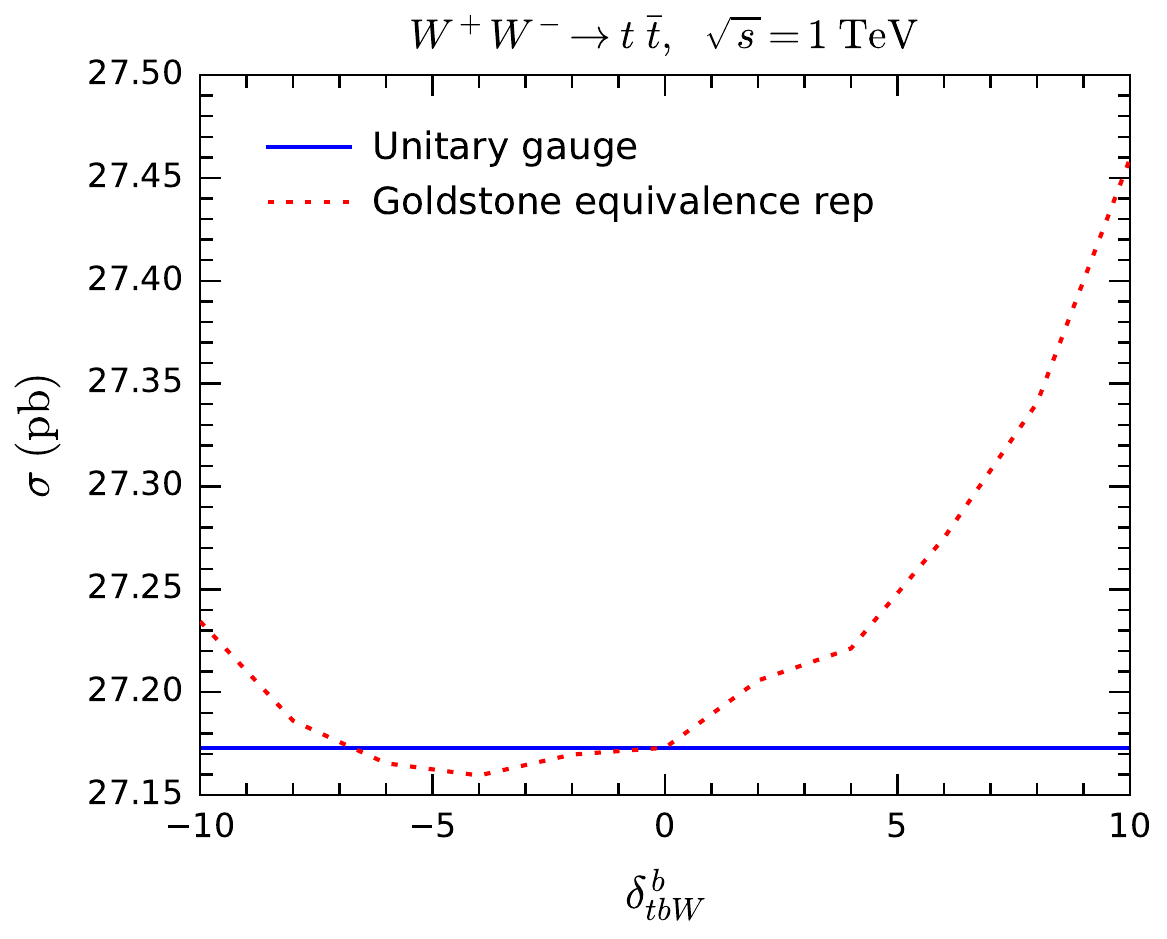}}
\subfigure[]{\includegraphics[height=0.33\textwidth]{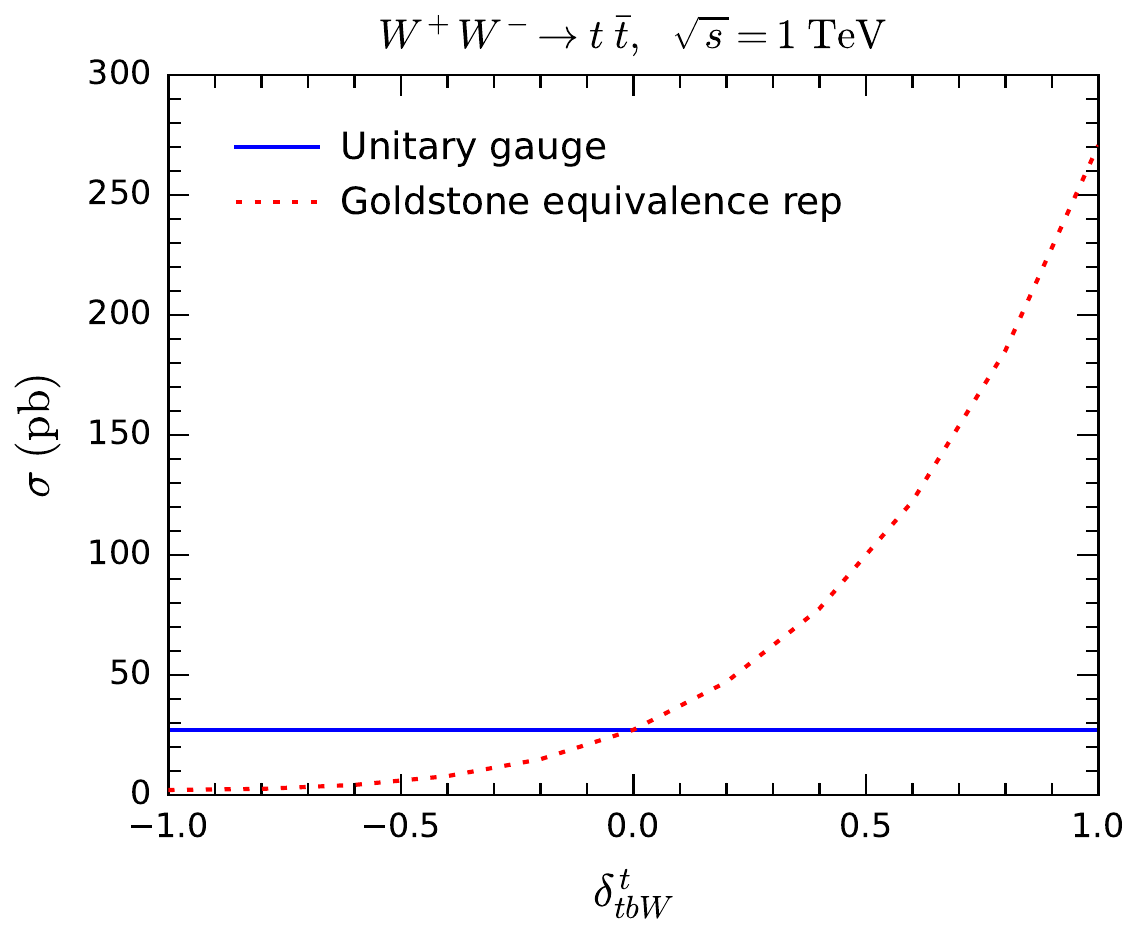}}
    \caption{Cross sections of $WW\rightarrow t\bar{t}$  in the unitary gauge and GE representation as functions of the anomalous couplings of $ttZ$ (upper panels) and $tbW$ (lower panels).}
    \label{fig:ffV_ww2tt_gauge}
\end{figure}

\begin{figure}[!t]
    \centering
     \subfigure[]{\includegraphics[height=0.33\textwidth]{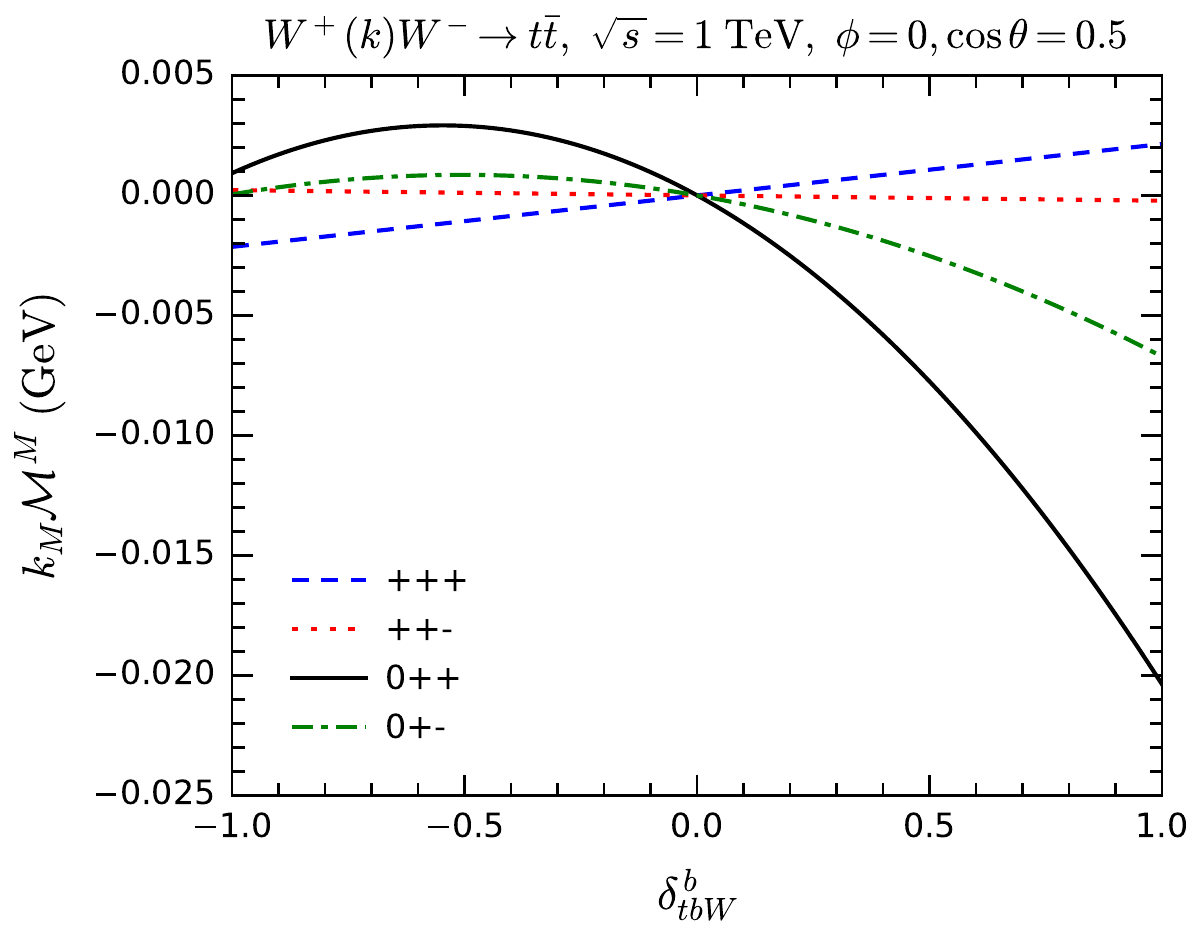}}
    \subfigure[]{\includegraphics[height=0.33\textwidth]{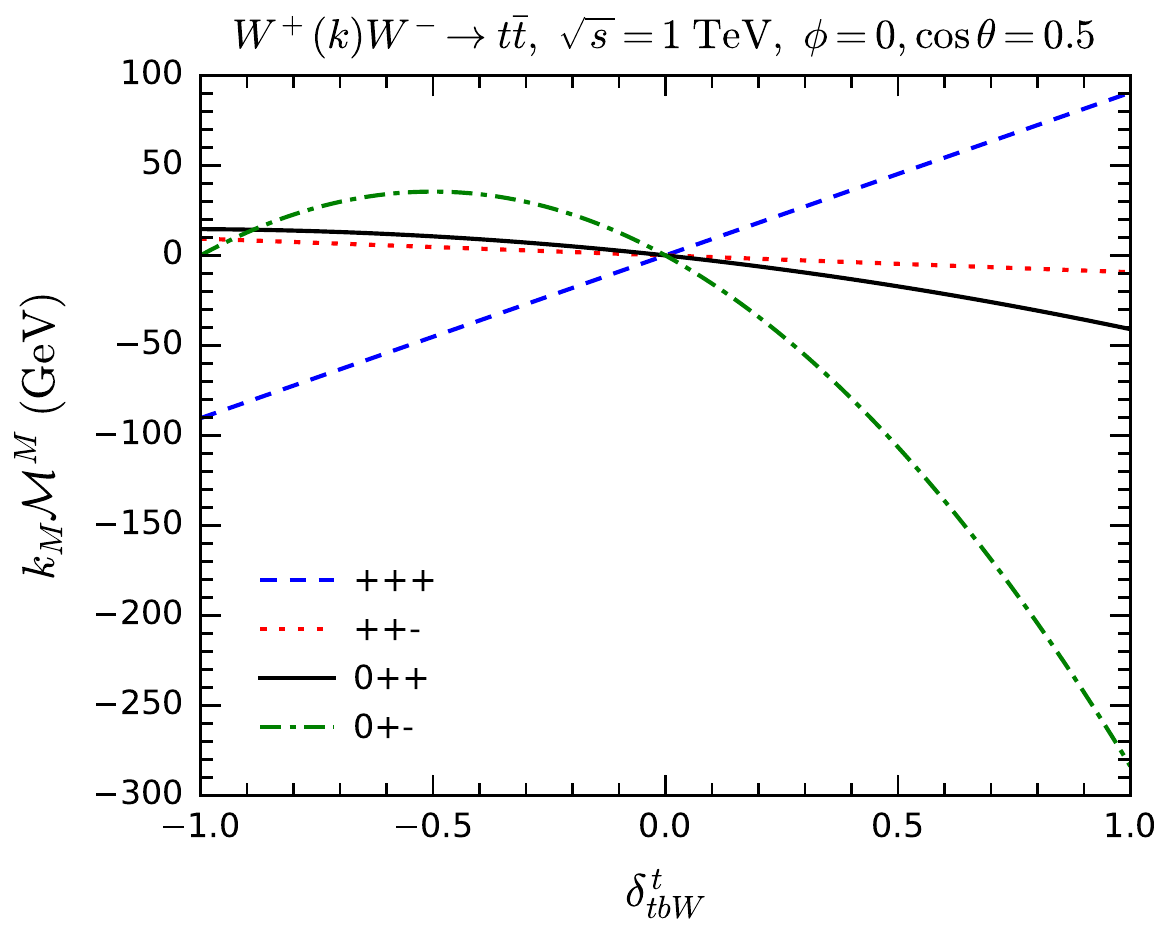}}
%\hspace{0.5\textwidth}
     \subfigure[]{\includegraphics[height=0.33\textwidth]{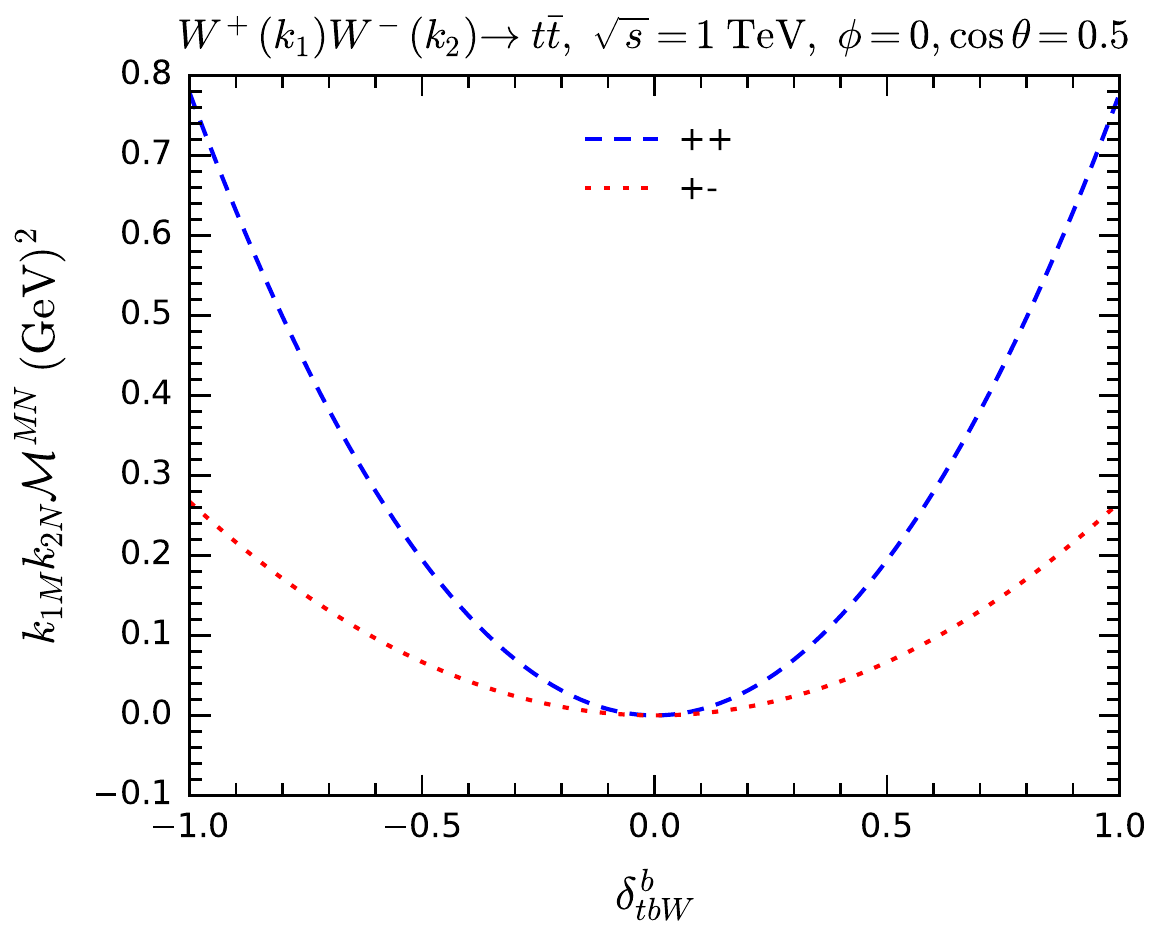}}
    \subfigure[]{\includegraphics[height=0.33\textwidth]{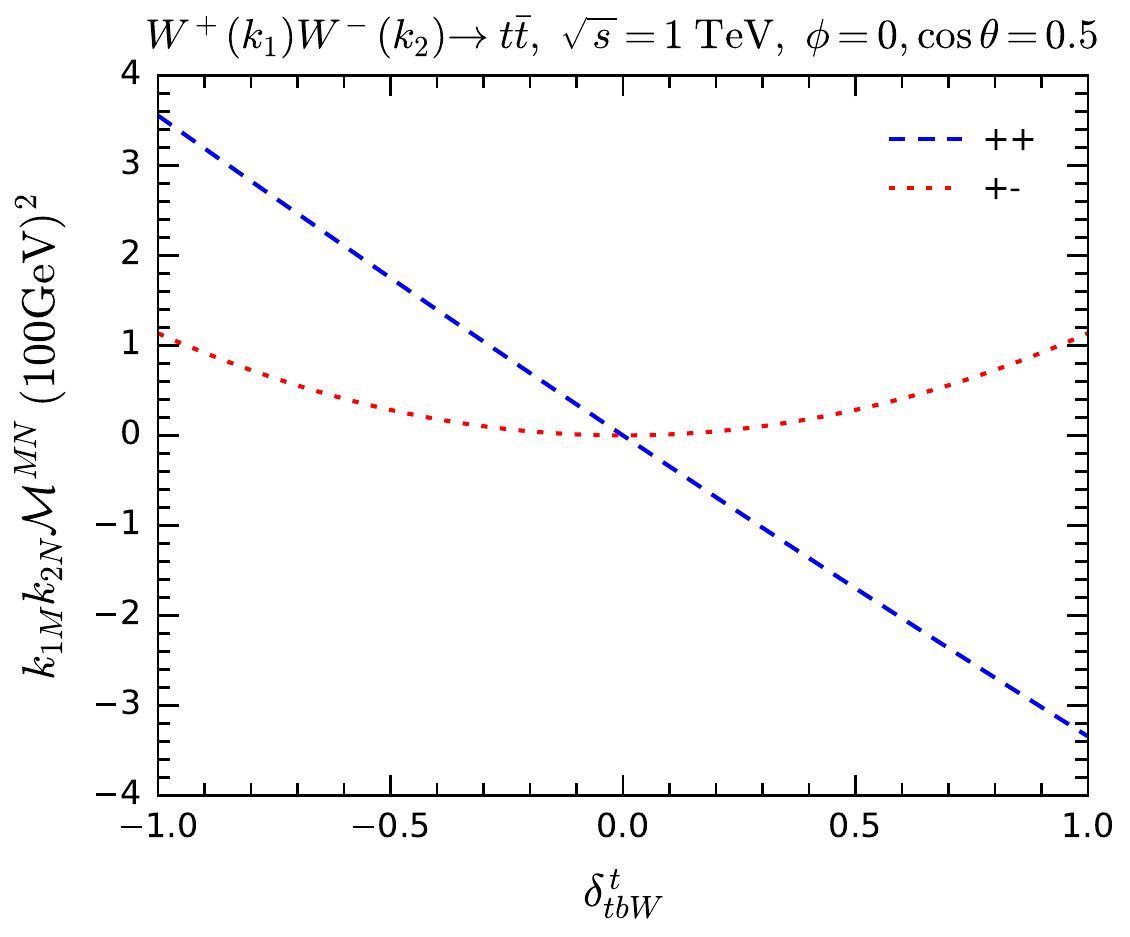}}
    \caption{Testing the MWI by computing $k_M\mathcal M^M$ (upper panels) and $k_{1M} k_{2N}\mathcal M^{MN}$ (lower panels) with anomalous $tbW$ couplings for the process $WW\rightarrow t\bar t$. Various helicity combinations are shown.}
    \label{fig:tbW_ww2tt_anom}
\end{figure}

\begin{figure}[!t]
    \centering
    \subfigure[]{\includegraphics[height=0.33\textwidth]{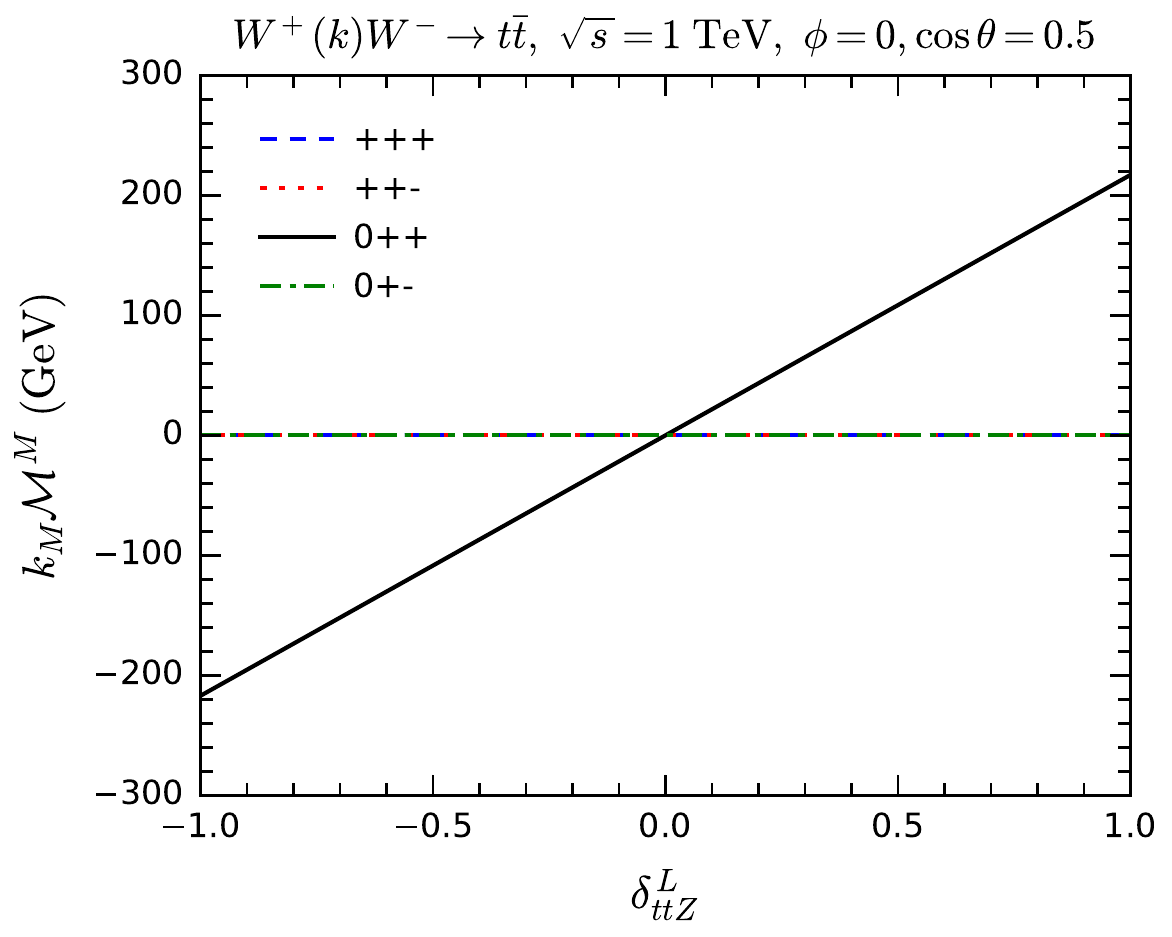}}
\subfigure[]{\includegraphics[height=0.33\textwidth]{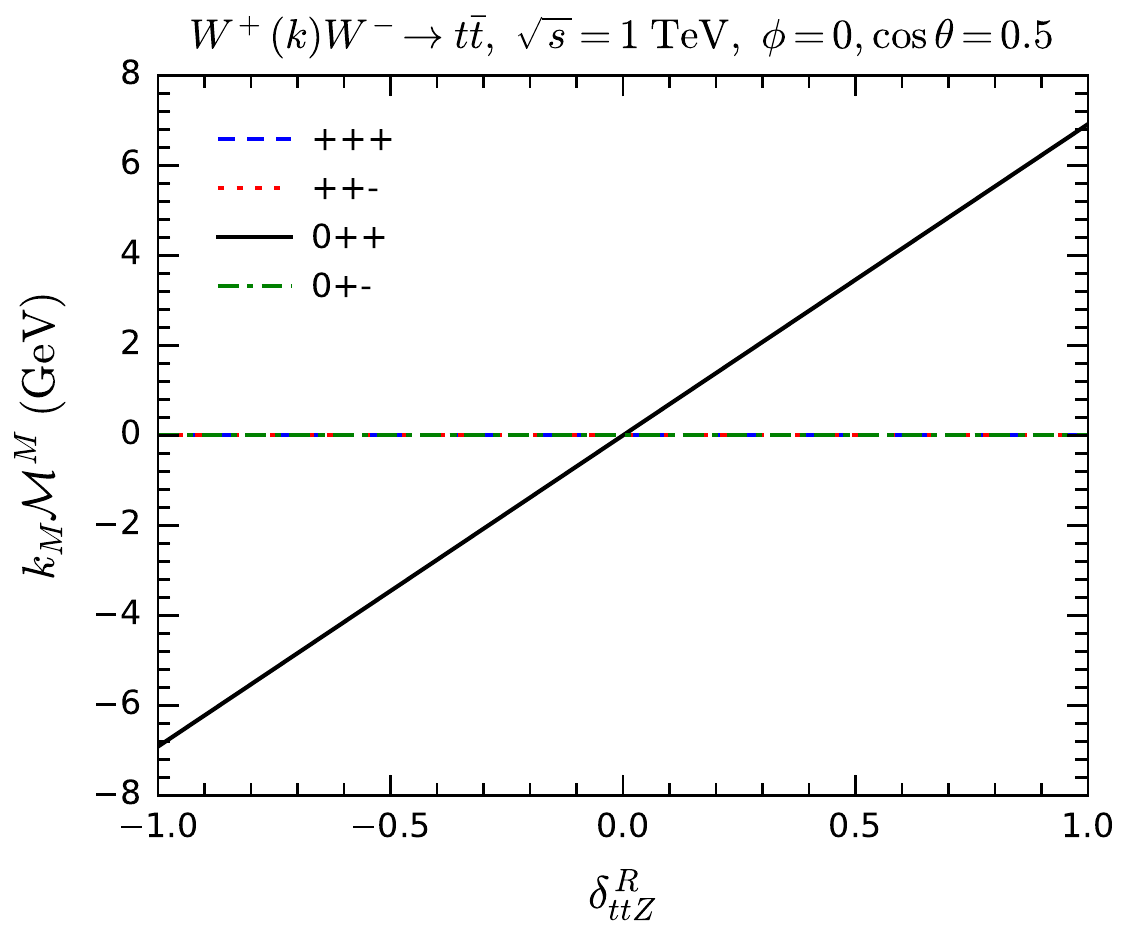}}
%\hspace{0.5\textwidth}
    \subfigure[]{\includegraphics[height=0.33\textwidth]{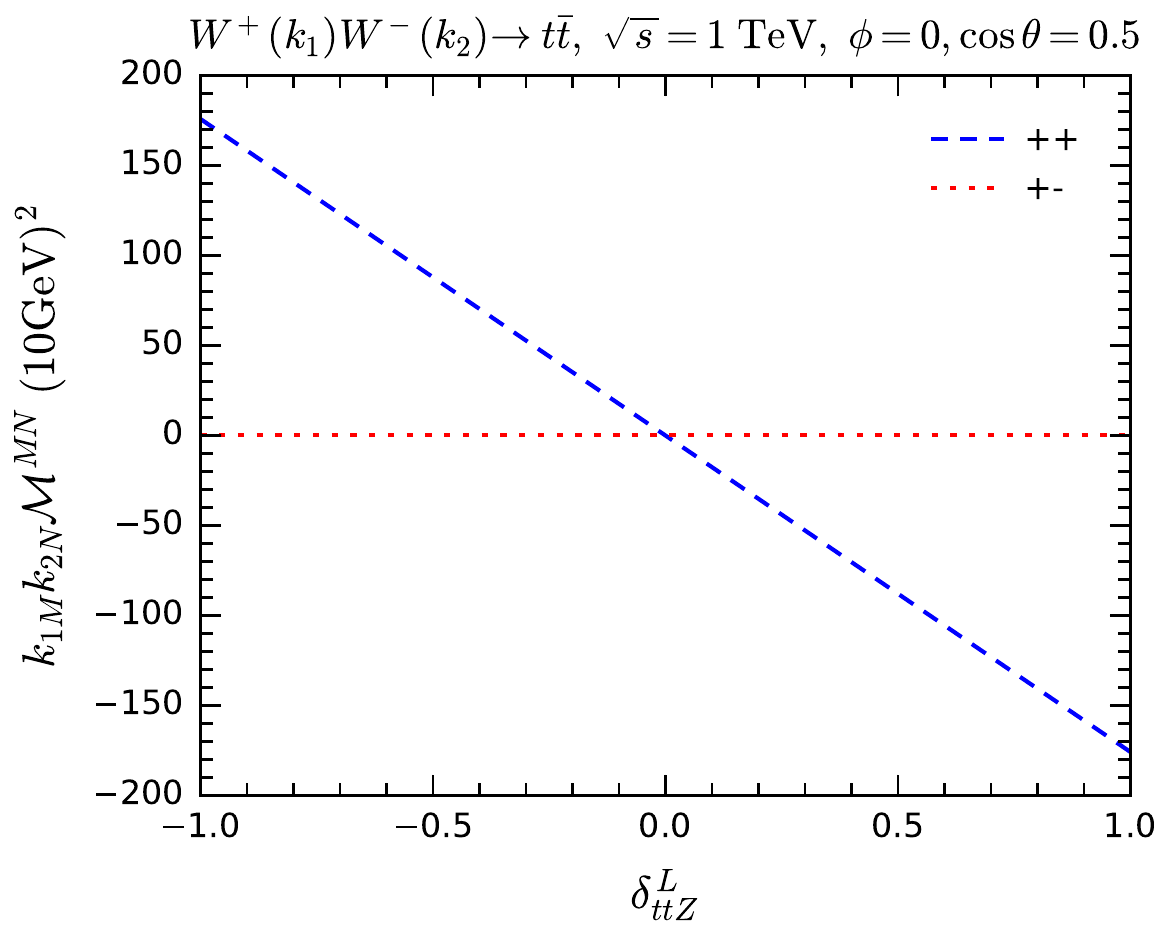}}
    \subfigure[]{\includegraphics[height=0.33\textwidth]{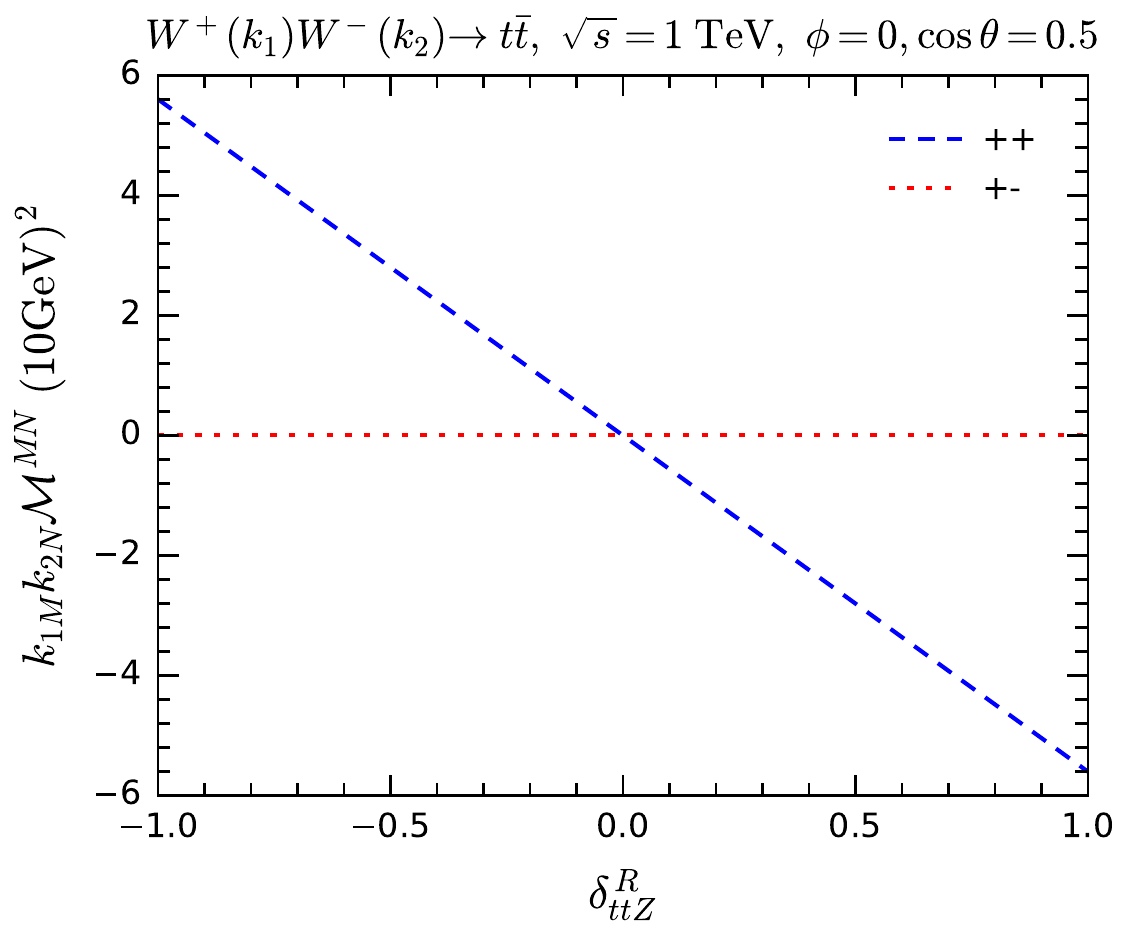}}
    \caption{Testing the MWI by computing $k_M\mathcal M^M$ (upper panels) and $k_{1M} k_{2N}\mathcal M^{MN}$ (lower panels) with anomalous $ttZ$ couplings for the process $WW\rightarrow t\bar t$. Various helicity combinations are shown.}
    \label{fig:ttZ_ww2tt_anom}
\end{figure}

In Fig.~\ref{fig:ffV_ww2tt_gauge},  we demonstrate how the $WW\rightarrow t\bar{t}$  cross sections are changed when the couplings $y_{ttZ}^{\mathrm{L}/\mathrm{R}}$  and $y_{tbW}^{b/t}$ are modified in both the unitary gauge and GE representation.  Similarly, we show how $k_M\mathcal M^M$ or $k_{1M} k_{2N}\mathcal M^{MN}$ changes with the anomalous $tbW$ couplings in Fig.~\ref{fig:tbW_ww2tt_anom}, and with the anomalous $ttZ$ coupling in Fig.~\ref{fig:ttZ_ww2tt_anom}. As the figures show, gauge symmetry is broken when one of the components of $tbW/ttZ$ couplings is modified, both in terms of cross sections and the MWI. For $ttZ$, the sensitivity of $k_M\mathcal M^M$ and $k_{1M} k_{2N}\mathcal M^{MN}$ to $\delta_{ttZ}^\mathrm{L}$ is much higher than that to $\delta_{ttZ}^\mathrm{R}$. This is expected, as the SM $ttZ$ vertex is dominated by the left-handed interaction. For $tbW$, the sensitivity to $\delta_{tbW}^t$ is much higher than that to $\delta_{tbW}^b$. The reason is also not difficult to understand: the top Yukawa coupling is much larger than the bottom Yukawa coupling.

%\JM{reduce MWI figures of  $tbW$ and $ttZ$ into one. }

\subsubsection*{$VVV$ vertex}

%\JM{WWZ /A has 3 types of couplings:wwz, w$\phi$z, w$\phi\phi$ -- change respectively}

In the SM, there are two $VVV$ vertices, $W^+W^+Z$ and $W^+W^-A$, either of which has three types of sub-vertices: $VVV$, $V\varphi V$, and $\varphi\varphi V$. There is no $\varphi\varphi\varphi$-type vertex.  

For the $WWZ$ vertex, the couplings are related to each other by
\begin{eqnarray}
    g_{WWZ} &\equiv& g c_\mathrm{W},\quad
    g_{\varphi\varphi Z}=\frac{gc_{2\mathrm{W}}}{2c_\mathrm{W}},\quad
    g_{\varphi W\varphi}=\frac{g}{2},\quad
    g_{W\varphi\varphi}=\frac{g}{2},
    \nonumber\\
    g_{WW\varphi} &=&0,\quad
    g_{W\varphi Z}=\frac{e s_\mathrm{W} m_W}{c_\mathrm{W}},\quad
    g_{\varphi WZ}=-\frac{e s_\mathrm{W} m_W}{c_\mathrm{W}},
\end{eqnarray}
with $c_{2\mathrm{W}}\equiv \cos 2\theta_\mathrm{W}$ and $e = g s_\mathrm{W}$.
%Finally, its $vv\phi$ type sub-vertex has 3 components, denoted as $w^+w^-\phi^0$ with coupling $g_{ww\phi}=0$, $w^+\phi^-z$ with coupling $g_{w\phi z}=\frac{e}{c_W}m_W$ and $\phi^+w^-z$ with coupling $g_{\phi wz}=-\frac{e}{c_W}m_W$ ($e=g\sin\theta_W$).  
Following the analysis for the $VVh$ and $ff'V$ vertices, we modify the couplings to test gauge symmetry as follows:
\begin{eqnarray}\label{eq:wwz_anom}
    g_{\varphi\varphi Z}&=& \frac{gc_{2\mathrm{W}}}{2c_\mathrm{W}}(1+\delta_{WWZ}^{234}),\quad
    g_{\varphi W\varphi}=g_{W\varphi\varphi} = \frac{g}{2}(1+\delta_{WWZ}^{234}),
    \nonumber \\
    g_{W\varphi Z}&=& \frac{e s_\mathrm{W} m_W}{c_\mathrm{W}}(1+\delta_{WWZ}^{567})=-g_{\varphi WZ}.
\end{eqnarray}
%Similarly we also modify couplings of $WWA$ in the same way as $WWZ$ in Eq.(\ref{eq:wwz_anom}). 

\begin{figure}[!t]
    \centering
     \subfigure[]{\includegraphics[height=0.33\textwidth]{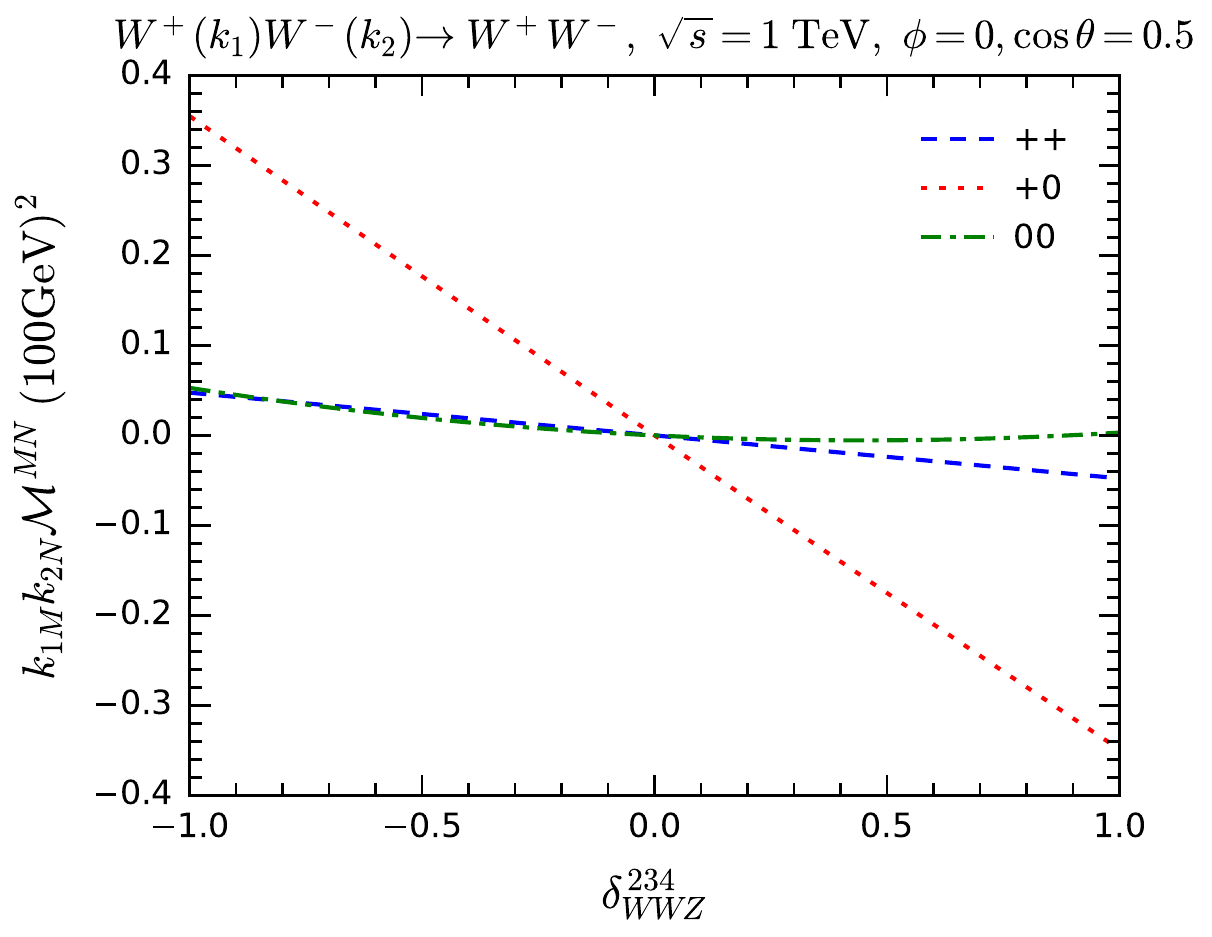}}
    \subfigure[]{\includegraphics[height=0.33\textwidth]{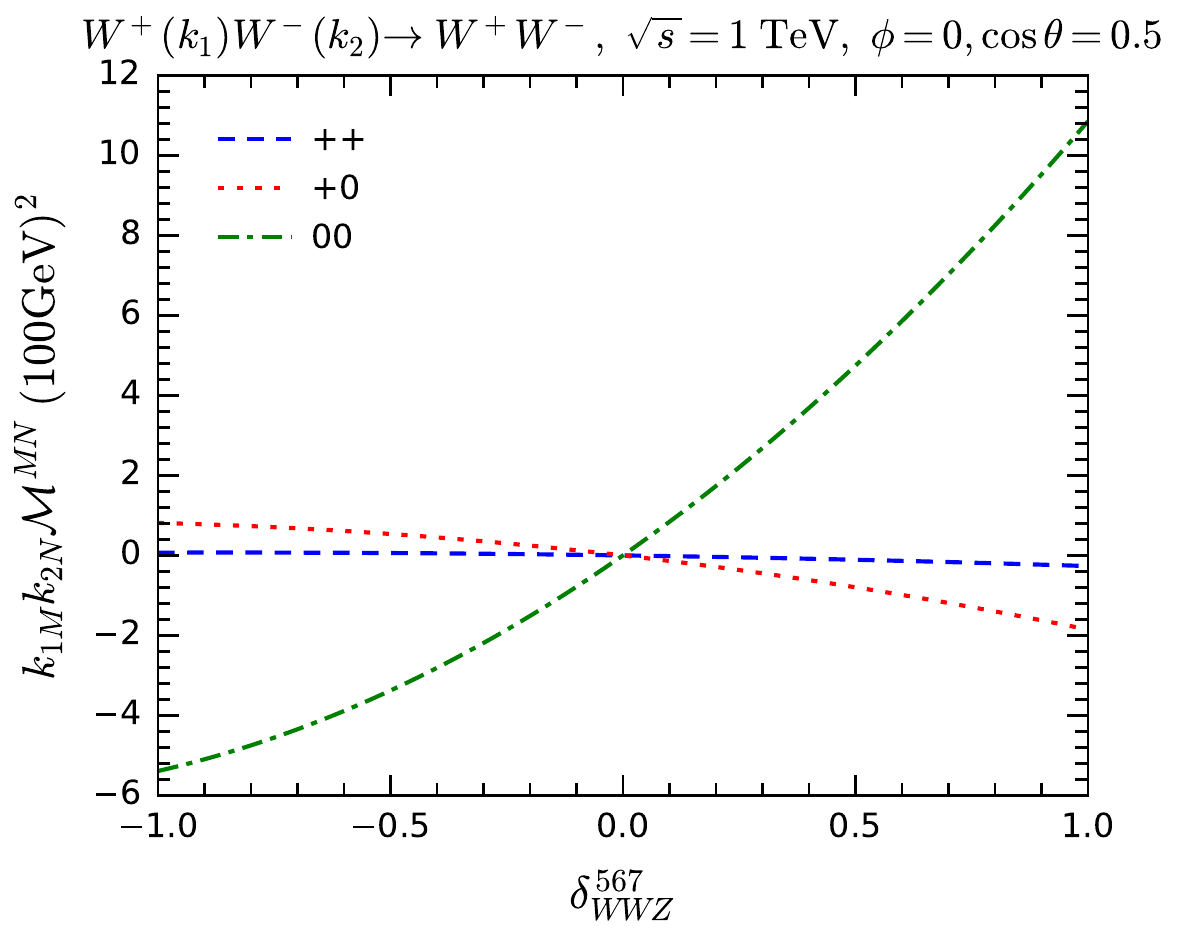}}
\hspace{0.5\textwidth}
     \subfigure[]{\includegraphics[height=0.33\textwidth]{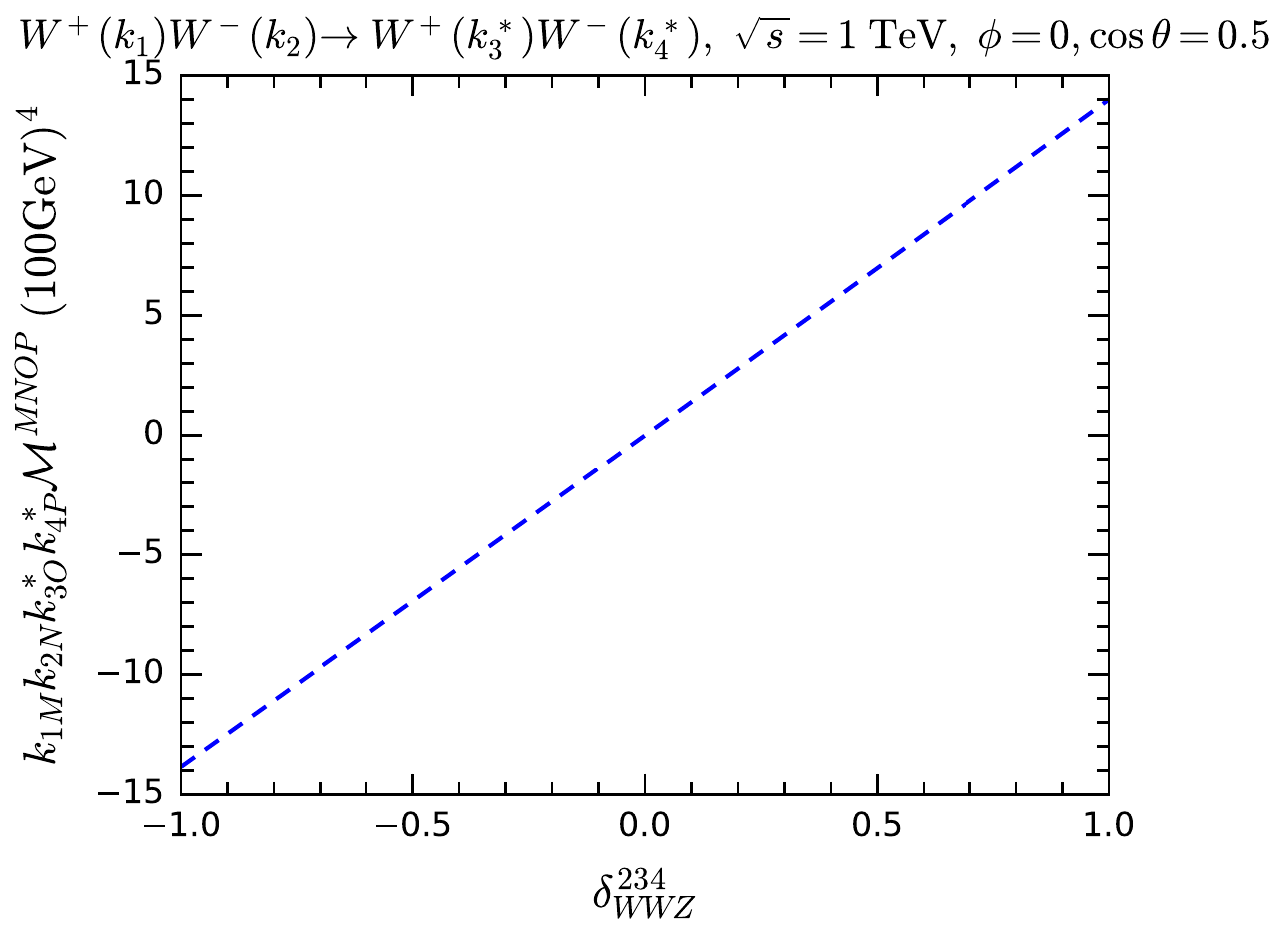}}
\subfigure[]{\includegraphics[height=0.33\textwidth]{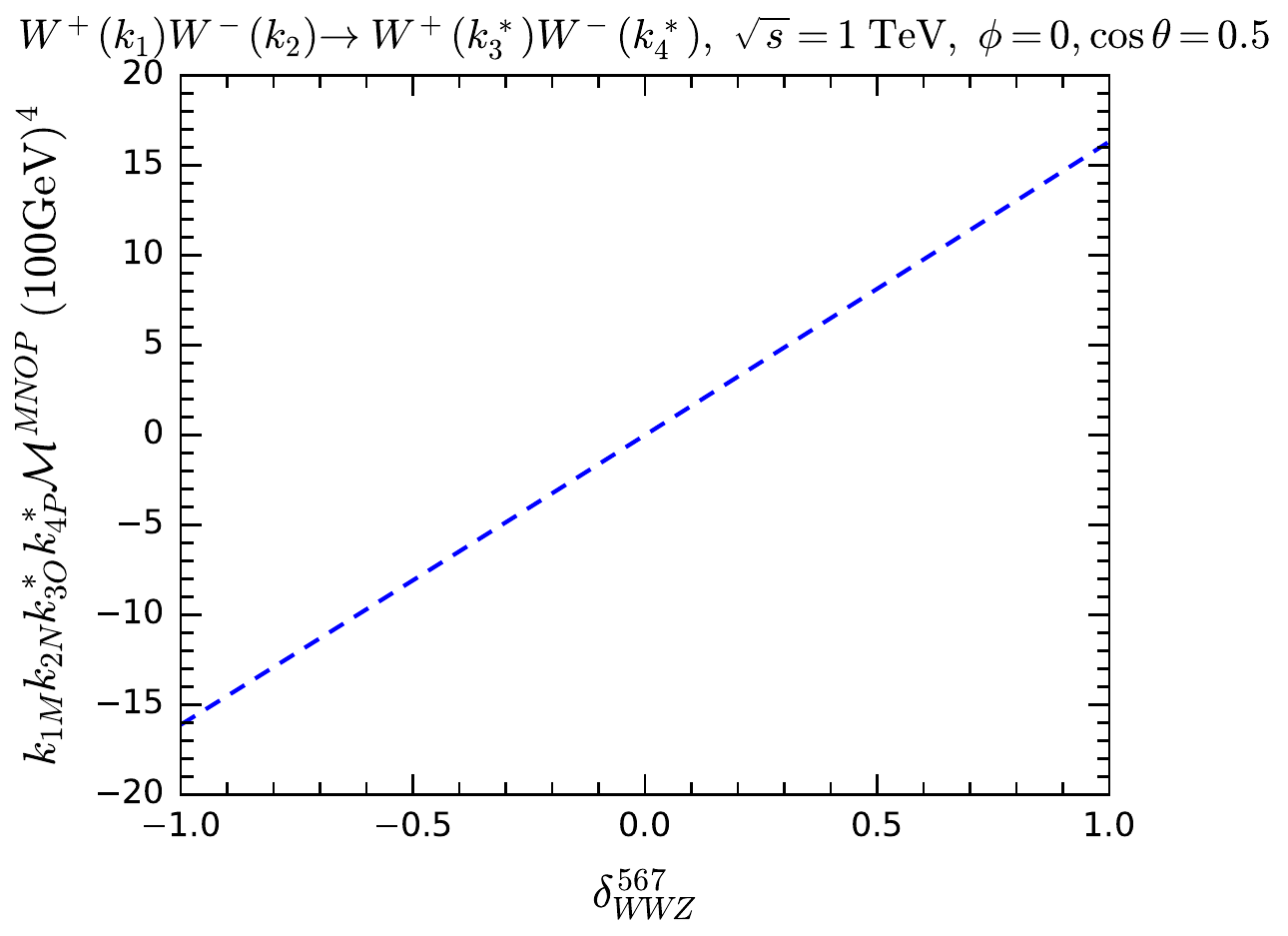}}
    \caption{Testing the MWI by computing $k_{1M} k_{2N} \mathcal M^{MN}$ (upper panels), and $k_{1M} k_{2N} k^*_{3O} k^*_{4P} \mathcal M^{MNOP}$ (lower panels) with anomalous $WWZ$ couplings for the process $W^+ W^+ \rightarrow W^+ W^-$. In the upper panels, various helicity combinations are shown.}
    \label{fig:wwz_ww2ww_anom}
\end{figure}

\begin{figure}[!t]
    \centering
     \subfigure[]{\includegraphics[height=0.33\textwidth]{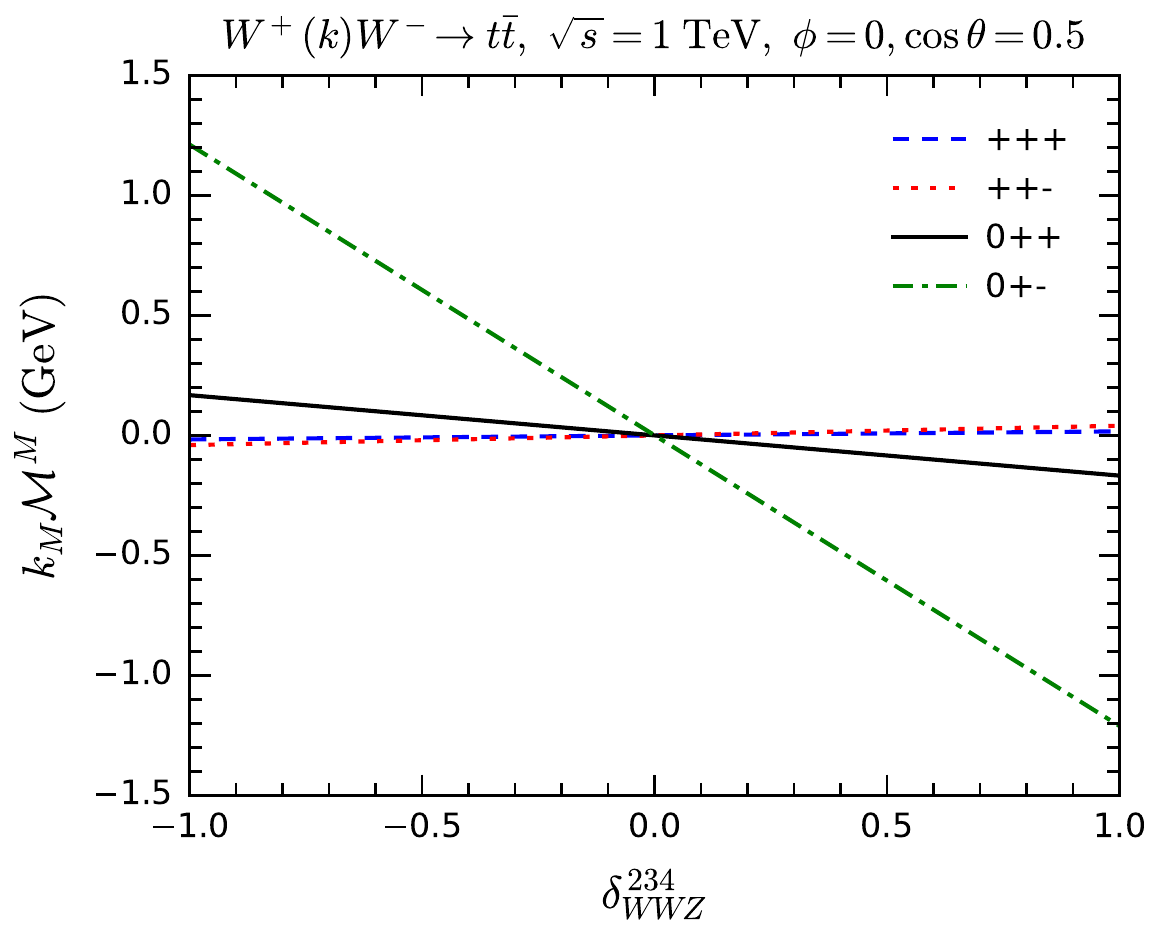}}
\subfigure[]{\includegraphics[height=0.33\textwidth]{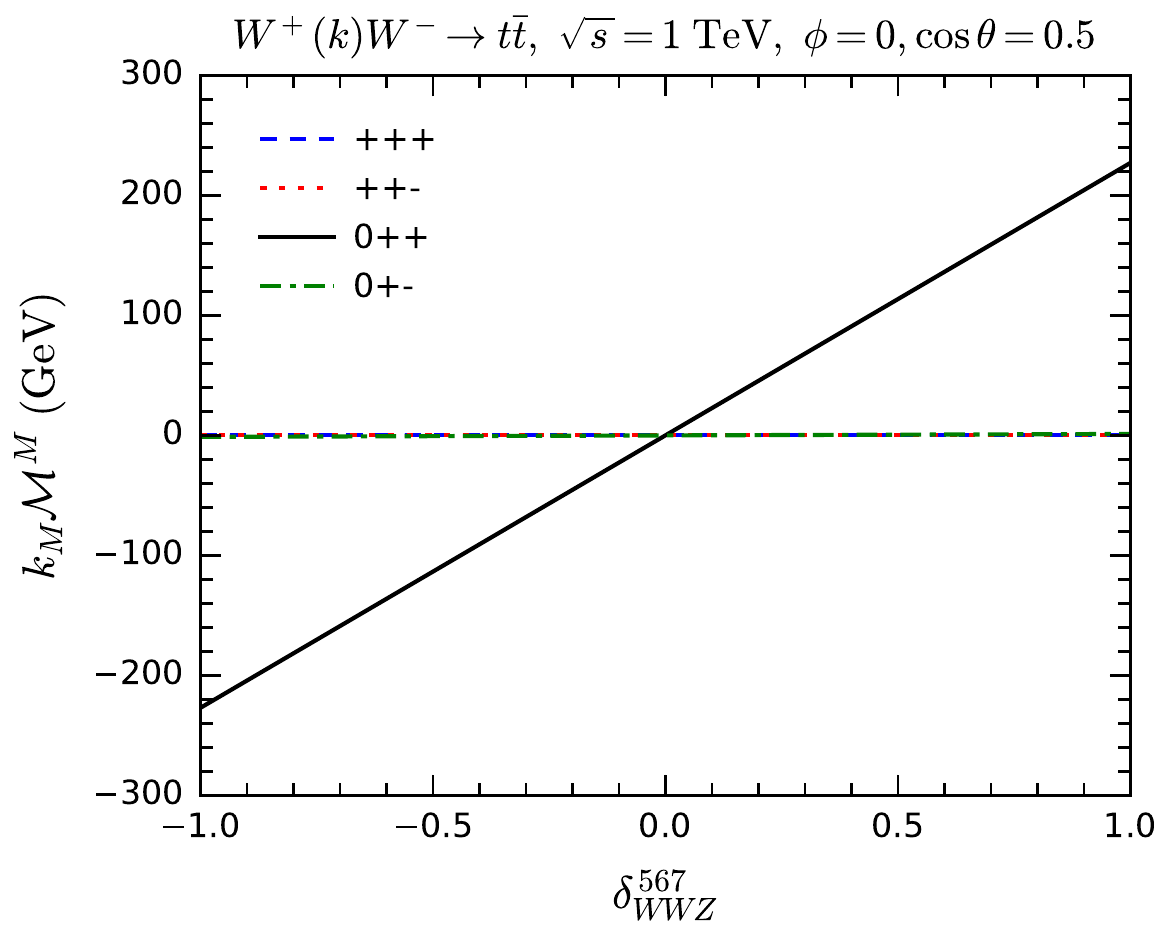}}
     \subfigure[]{\includegraphics[height=0.33\textwidth]{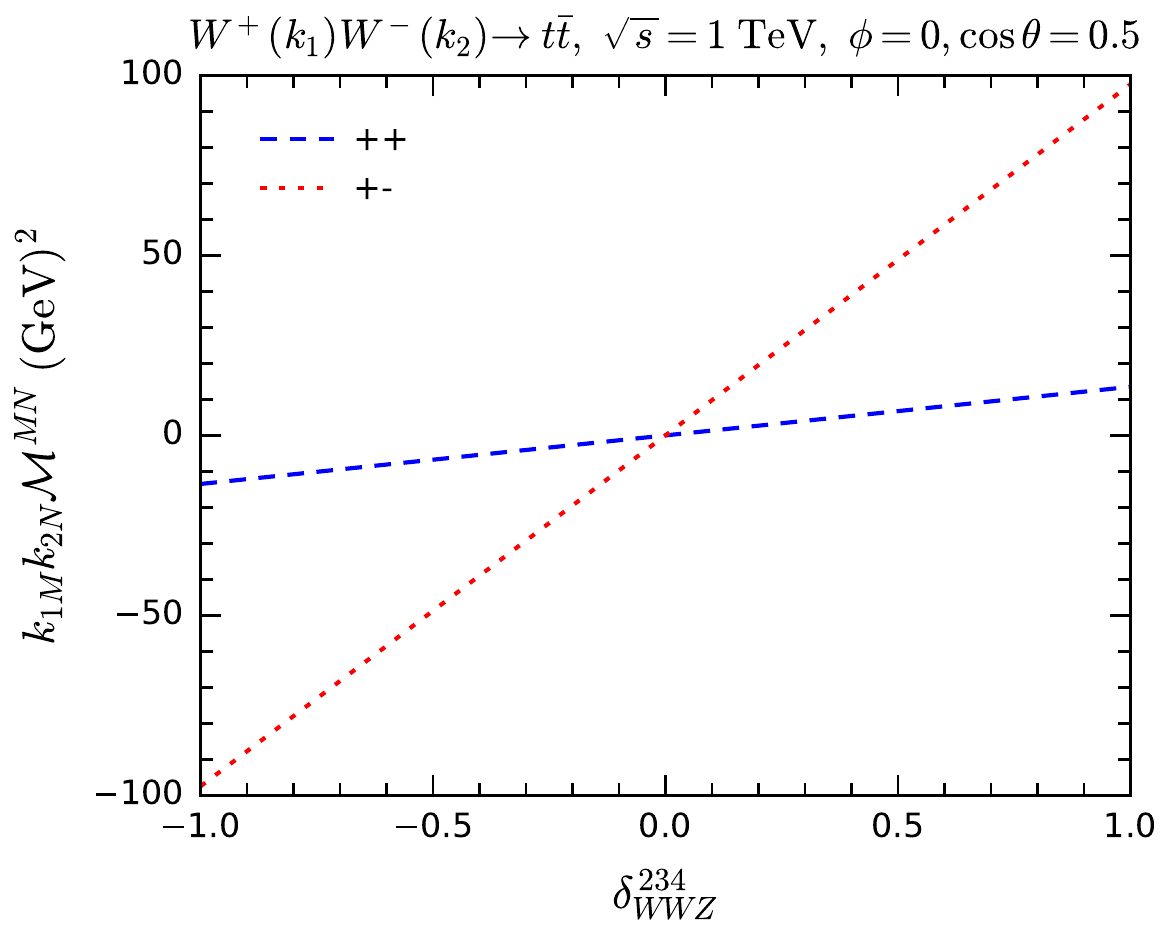}}
    \subfigure[]{\includegraphics[height=0.33\textwidth]{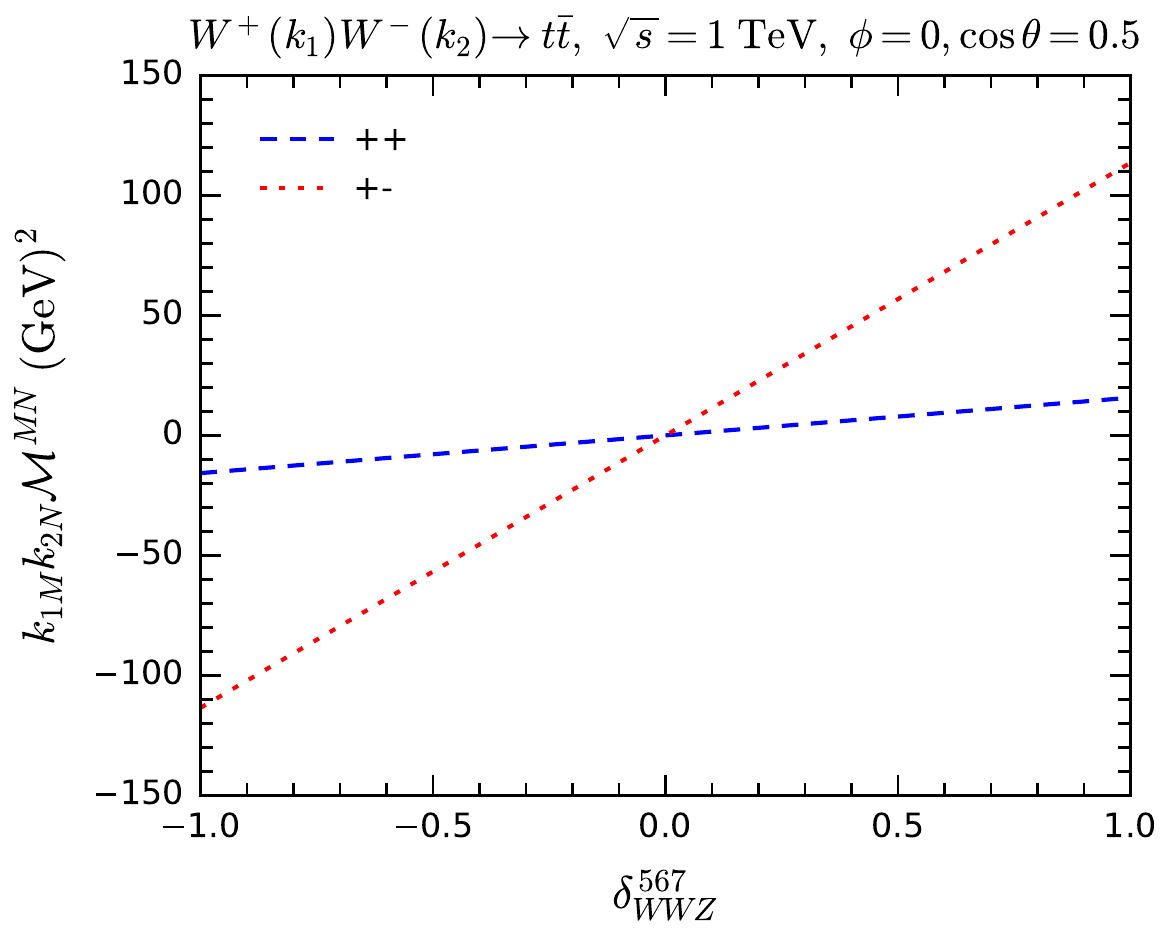}}
    \caption{Testing the MWI by computing $k_M\mathcal M^M$ (upper panels) and $k_{1M} k_{2N} \mathcal M^{MN}$ (lower panels) with anomalous $WWZ$ couplings for the the process $WW\rightarrow t\bar t$.  Various helicity combinations are shown.}
    \label{fig:wwz_ww2tt_anom}
\end{figure}

We examine the gauge symmetry of the $WWZ$ vertex with both the $W^+ W^+ \rightarrow W^+ W^-$ and $WW\rightarrow t\bar t$ processes.  In Figs.~\ref{fig:wwz_ww2ww_anom} and \ref{fig:wwz_ww2tt_anom}, we test the MWI by modifying $\delta_{WWZ}^{234}$ and  $\delta_{WWZ}^{567}$ for $W^+ W^+ \rightarrow W^+ W^-$ and $WW\rightarrow t\bar t$, respectively. In both cases, the sensitivity of the violation of the MWI to $\delta_{WWZ}^{567}$ is higher than that to $\delta_{WWZ}^{234}$, sometimes significantly so. This is somewhat surprising, as the $VV\varphi$-type vertex that $\delta_{WWZ}^{567}$ modifies is suppressed by $e = g s_\mathrm{W}$. However, they are typically proportional to $m_W$, providing an enhancement that effectively counterbalances the suppressive influence of the weak mixing angle $\theta_\mathrm{W}$.

%%\begin{figure}[!t]
%    \centering
%     \subfigure[]{\includegraphics[width=0.48\textwidth]{fig/Xsec_WW2tt_ywwz_delta1_100.pdf}}
%     \subfigure[]{\includegraphics[width=0.48\textwidth]{fig/Xsec_WW2tt_ywwz_delta2_100.pdf}}
%    \subfigure[]{\includegraphics[width=0.48\textwidth]{fig/Xsec_WW2tt_ywwa_delta1.pdf}}
%     \subfigure[]{\includegraphics[width=0.48\textwidth]{fig/Xsec_WW2tt_ywwa_delta2.pdf}}
%    \caption{Caption}
%    \label{fig:VVV_ww2tt_gauge}
%\end{figure}

\subsection{Four-point Vertices}
\label{sec:4-point}

Gauge symmetry is reflected not only in the couplings within a single vertex but also in the relations among different vertices.  For example, in the $WW\rightarrow W W$ process, where vertices, such as $WWWW$, $WWZ$, $WW\gamma$, and $WWh$, are involved, the MWI requires precise relations among their couplings. Any deviation from the SM values breaks gauge symmetry, similar to how it violates unitarity in the gauge representation. We modify the overall couplings of the vertices to test the MWI. 
The processes we choose are $W^+ W^-\rightarrow W^+ W^-$, $W^+ W^-\rightarrow hh$, and $W^+ W^-\rightarrow t\bar t$. 

%Indeed, the analogue to the gauge representation implies that there will also be large cancellation for even slight deviation of couplings in the Goldstone representation.

%\JM{change 3-point and 4-point vertices as wholes, similar to gauge rep -- processes:$WW\rightarrow t\bar t$, $WW\rightarrow hh$,$WW\rightarrow WW$}

%\noindent $W^+ W^-\rightarrow W^+ W^-$

%\JM{vary $WWWW$,  $WWh$, $WWZ/A$ respectively}

\begin{figure}[!t]
    \centering
    \subfigure[]{\includegraphics[height=0.33\linewidth]{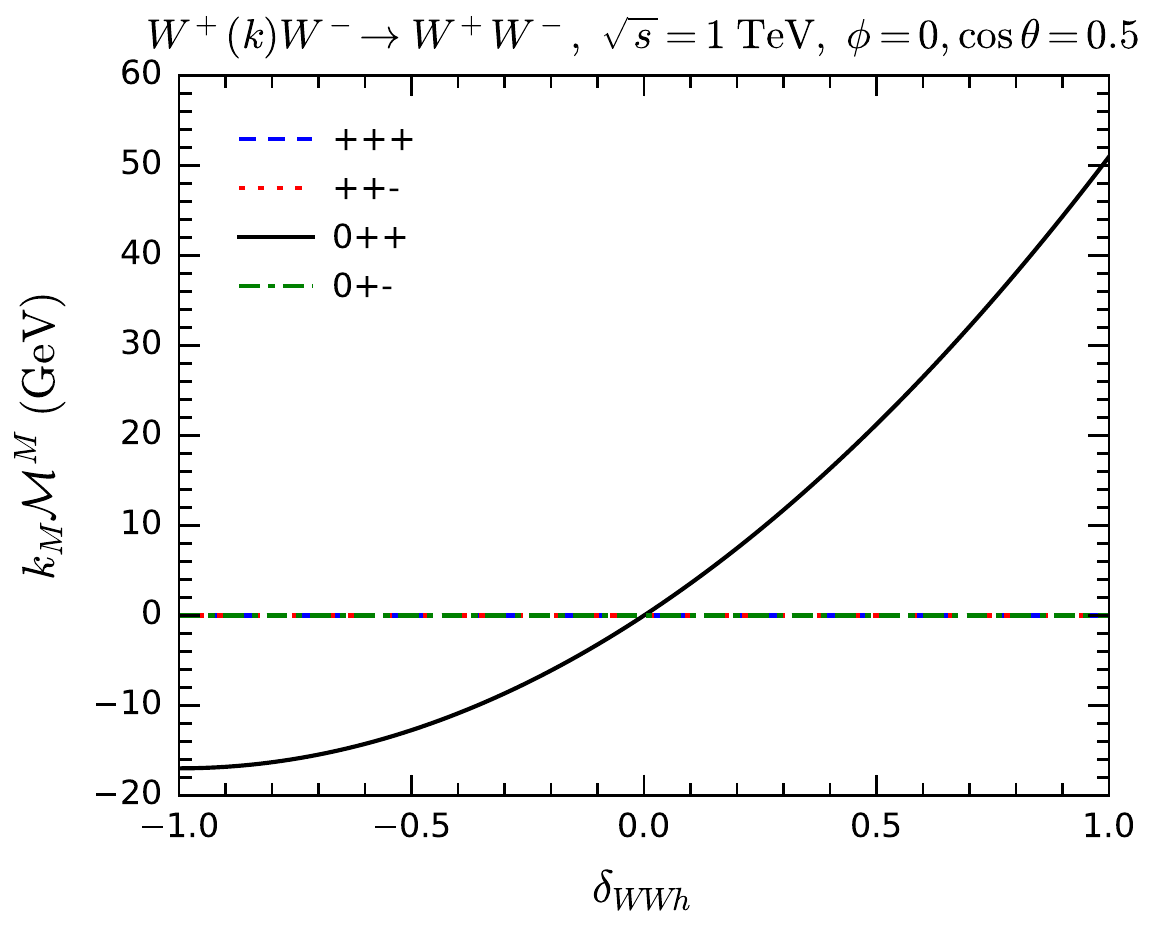}}
    \subfigure[]{\includegraphics[height=0.33\linewidth]{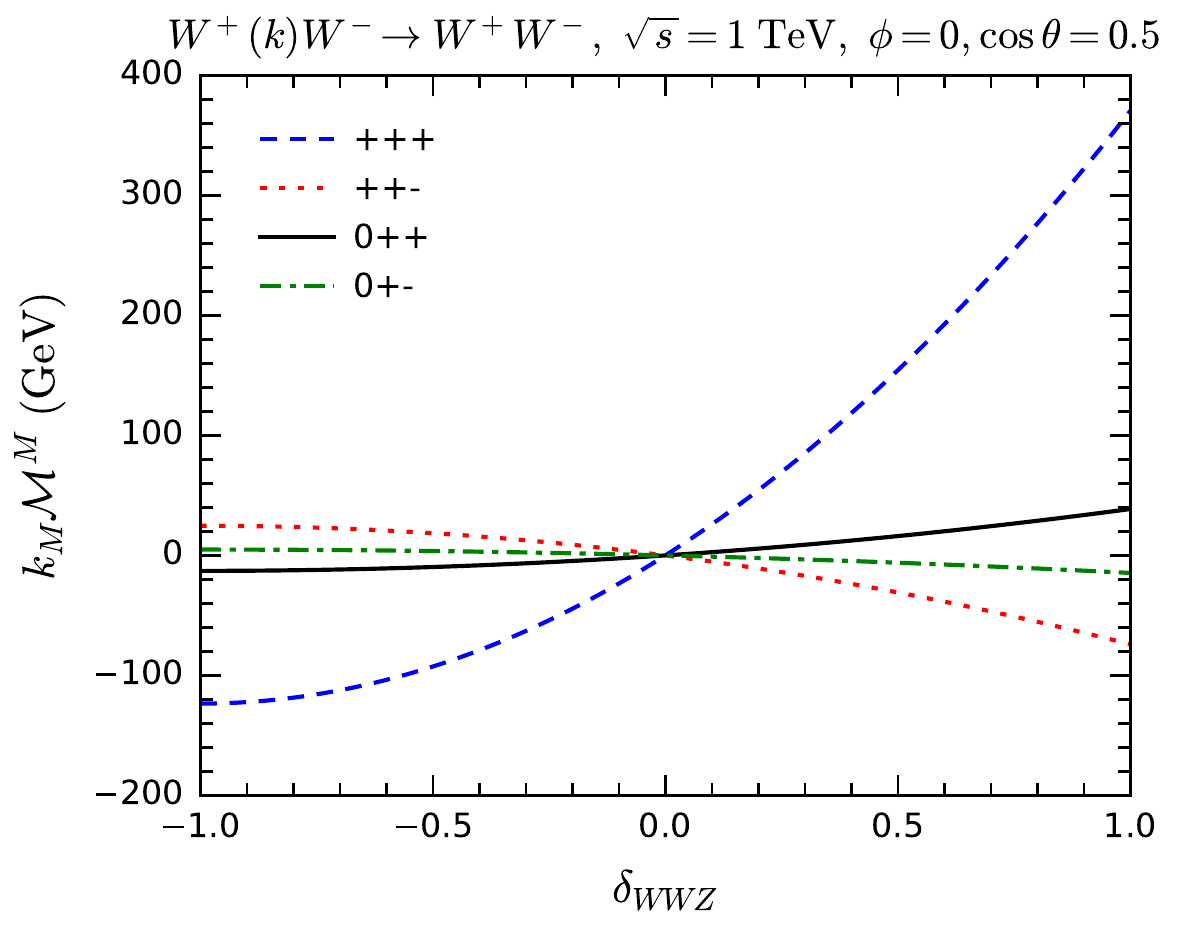}}
    \subfigure[]{\includegraphics[height=0.33\linewidth]{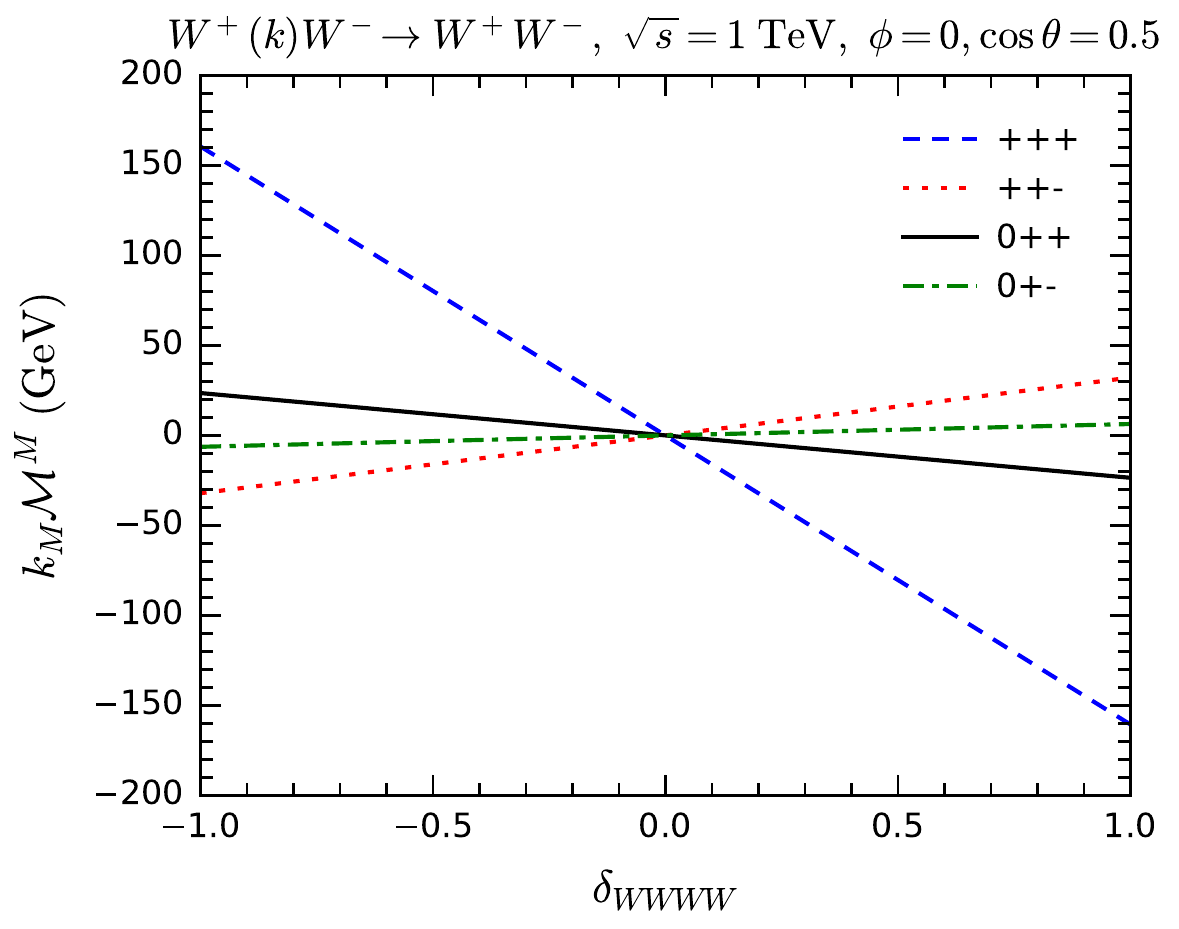}}
    \caption{$k_M\mathcal M^M$ for $W^+ W^-\rightarrow W^+ W^-$ as a function of anomalous overall couplings $\delta_{WWh}$ (a), $\delta_{WWZ}$ (b), and $\delta_{WWWW}$ (c). Various helicity combinations are shown.}
    \label{fig:4pt_ww2ww}
\end{figure}

For $W^+ W^-\rightarrow W^+ W^-$, we separately vary the overall couplings of the $WWh$, $WWZ$, and $WWWW$ vertices and compute $k_M \mathcal M^M$. In Fig.~\ref{fig:4pt_ww2ww}, we show the results, listing different helicity combinations in the process. As expected, we can observe the violation of the MWI when the anomalous couplings are nonzero.  This violation is most sensitive to $\delta_{WWZ}$ and $\delta_{WWWW}$. 

One interesting observation is that, in most helicity combinations, $k_M \mathcal M^M$ remains zero or close to zero.  This reminds us to be careful in selecting helicities when testing the gauge symmetry of massive amplitudes.  The exact vanishing of $k_M \mathcal M^M$ in the presence of an anomalous $\delta_{WWh}$ appears to be due to angular momentum conservation: the amplitude vanishes when angular momentum conservation is violated; hence, only those helicity combinations that conserve angular momentum can yield nonzero values.
%The reason for $k_M \mathcal M^M$ being significantly large for certain helicity combinations with $\delta_{WWZ}$ and $\delta_{WWWW}$ is less straightforward and requires further investigation.

% It's also interesting to know the reason for this phenomenon. Generally, $k^M\mathcal M_M$ equals or approximately equals to 0, is because the amplitude generated by the anomalous coupling gives 0 or close to 0. But why?

%\noindent $WW\rightarrow h h$

%\JM{vertices: $hWW$, $WWhh$, $hhh$;  vary $g_{WWhh}$ $hWW$ and $hhh$ respectively.  }

\begin{figure}[!t]
    \centering
    \subfigure[]{\includegraphics[height=0.33\linewidth]{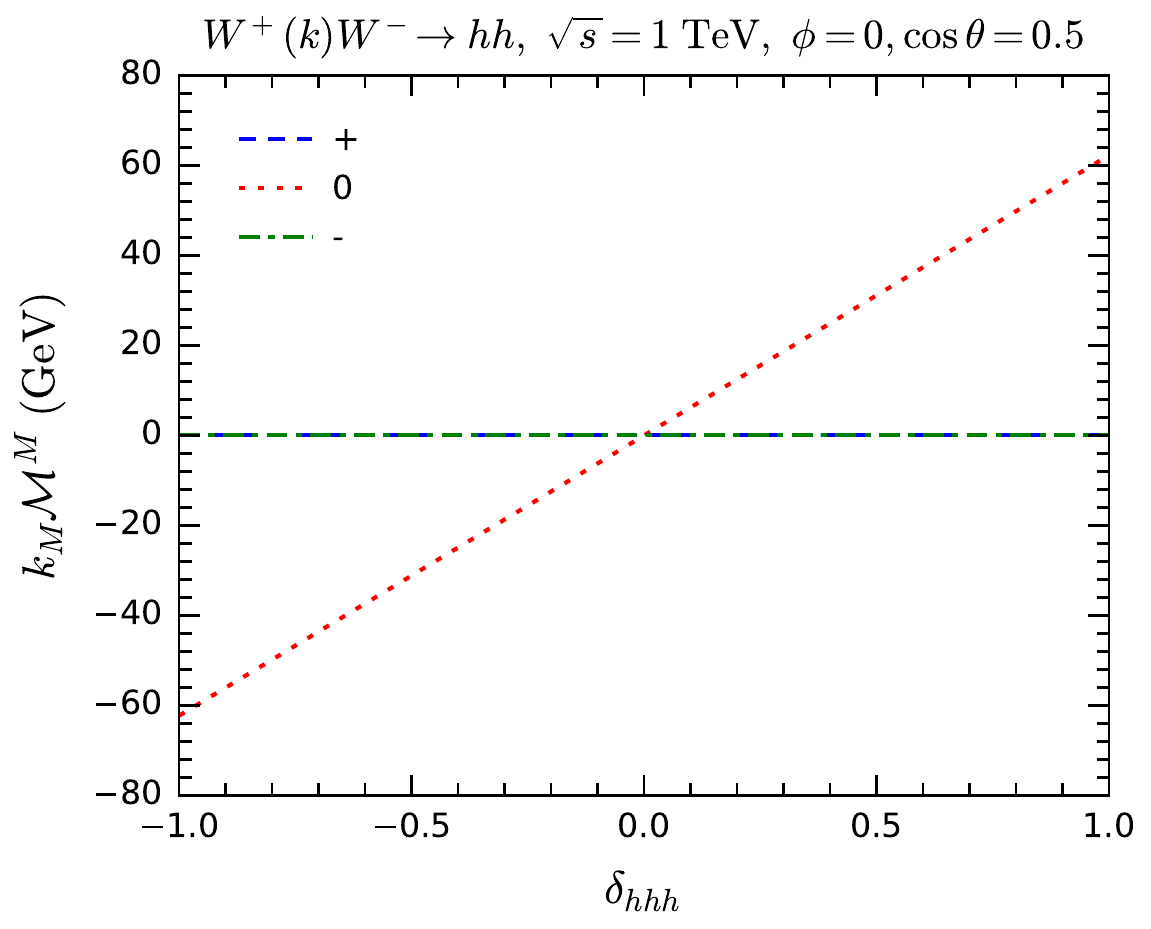}}
    \subfigure[]{\includegraphics[height=0.33\linewidth]{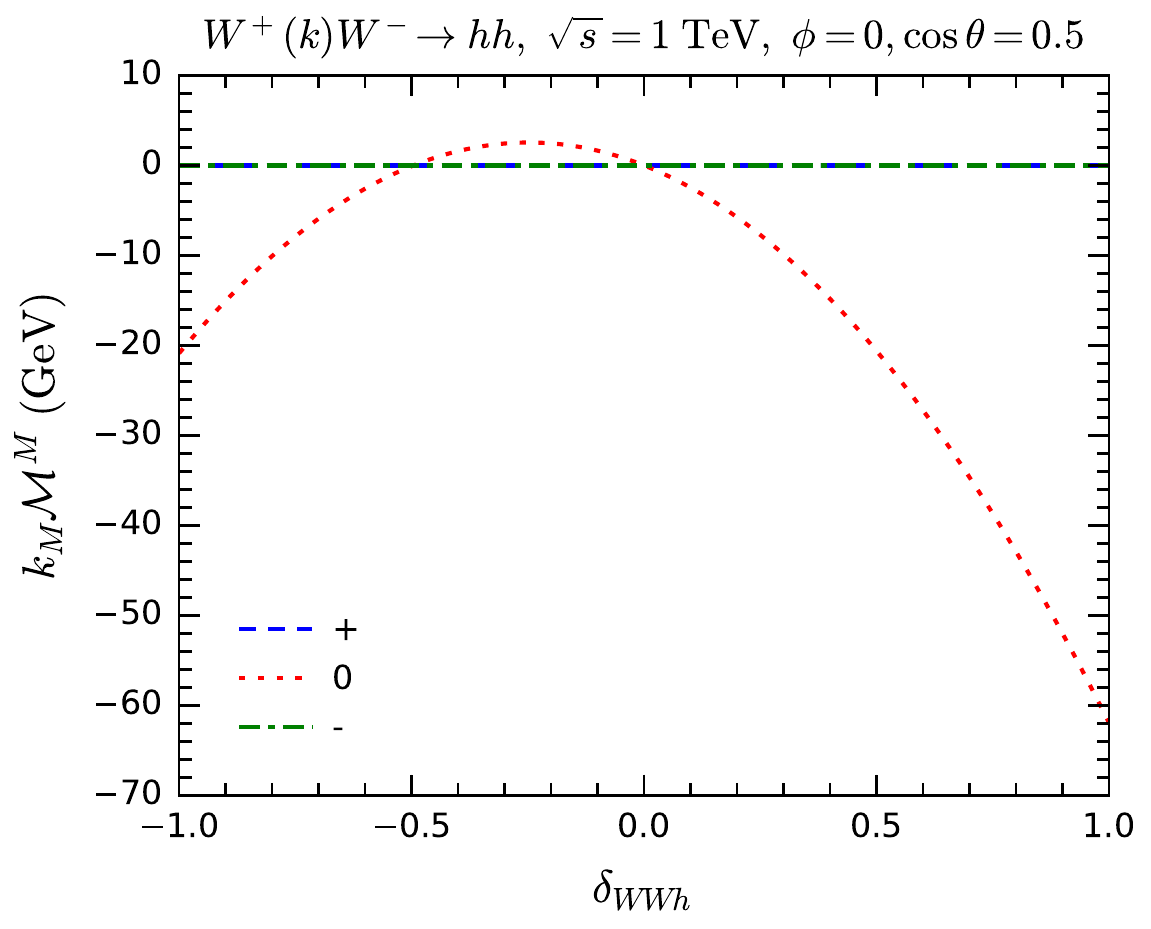}}
    \subfigure[]{\includegraphics[height=0.33\linewidth]{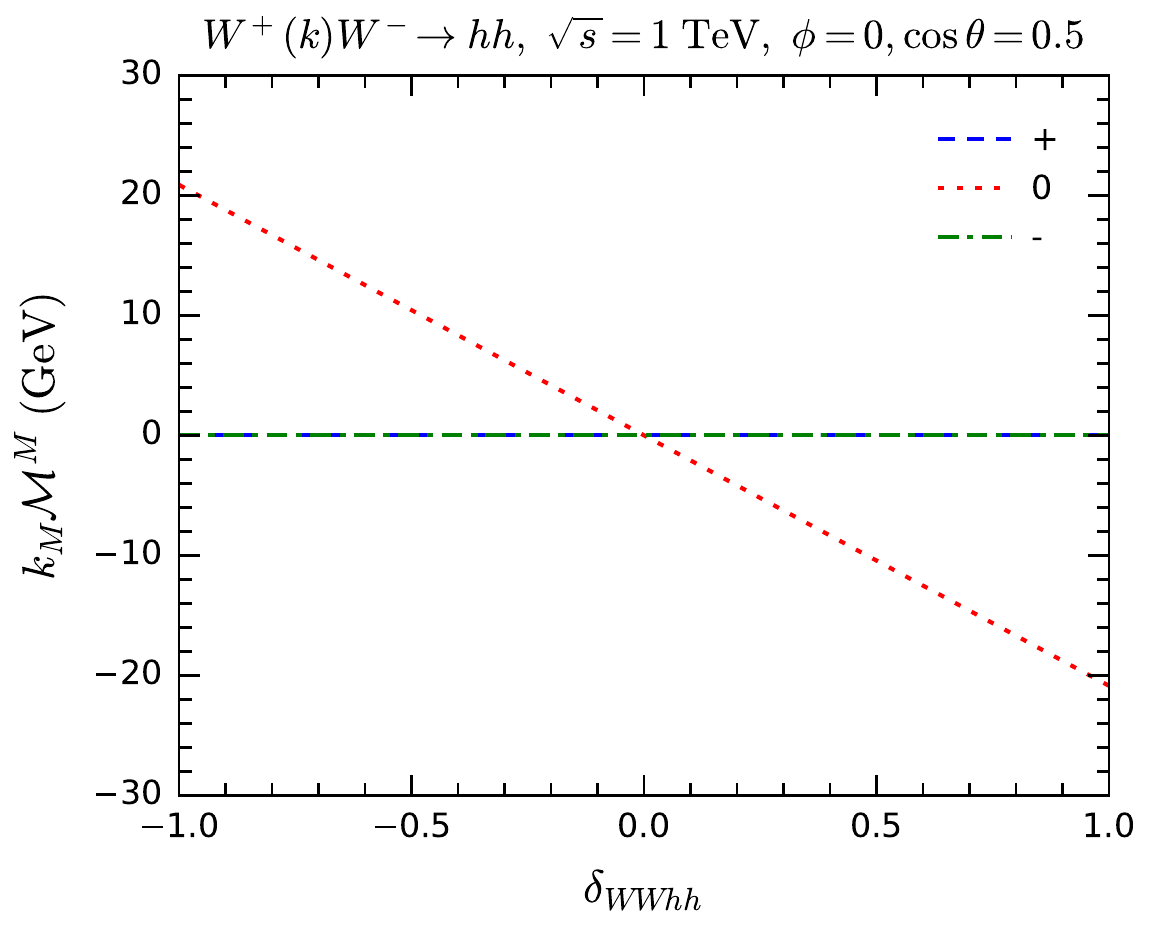}}
    \caption{$k_M\mathcal M^M$ for $WW\rightarrow hh$ as a function of anomalous overall couplings $\delta_{hhh}$ (a), $\delta_{WWh}$ (b), and $\delta_{WWhh}$ (c). Various helicity configurations are shown. }
    \label{fig:4pt_ww2hh}
\end{figure}

For $WW\rightarrow hh$, we separately modify the couplings of the $hhh$, $WWh$, and $WWhh$ vertices, and  compute $k_M\mathcal M^M$.   The results are shown in Fig.~\ref{fig:4pt_ww2hh}. 

%\noindent $WW\rightarrow t\bar t$

%\JM{vertices: $hWW$,$A/ZWW$,  $ht\bar t$, $Z/A t\bar t$, $Wtb$; 
%vary $hWW$, $htt$ and $A/ZWW$ respectively, fix other}

\begin{figure}[!t]
    \centering
    \subfigure[]{\includegraphics[height=0.33\linewidth]{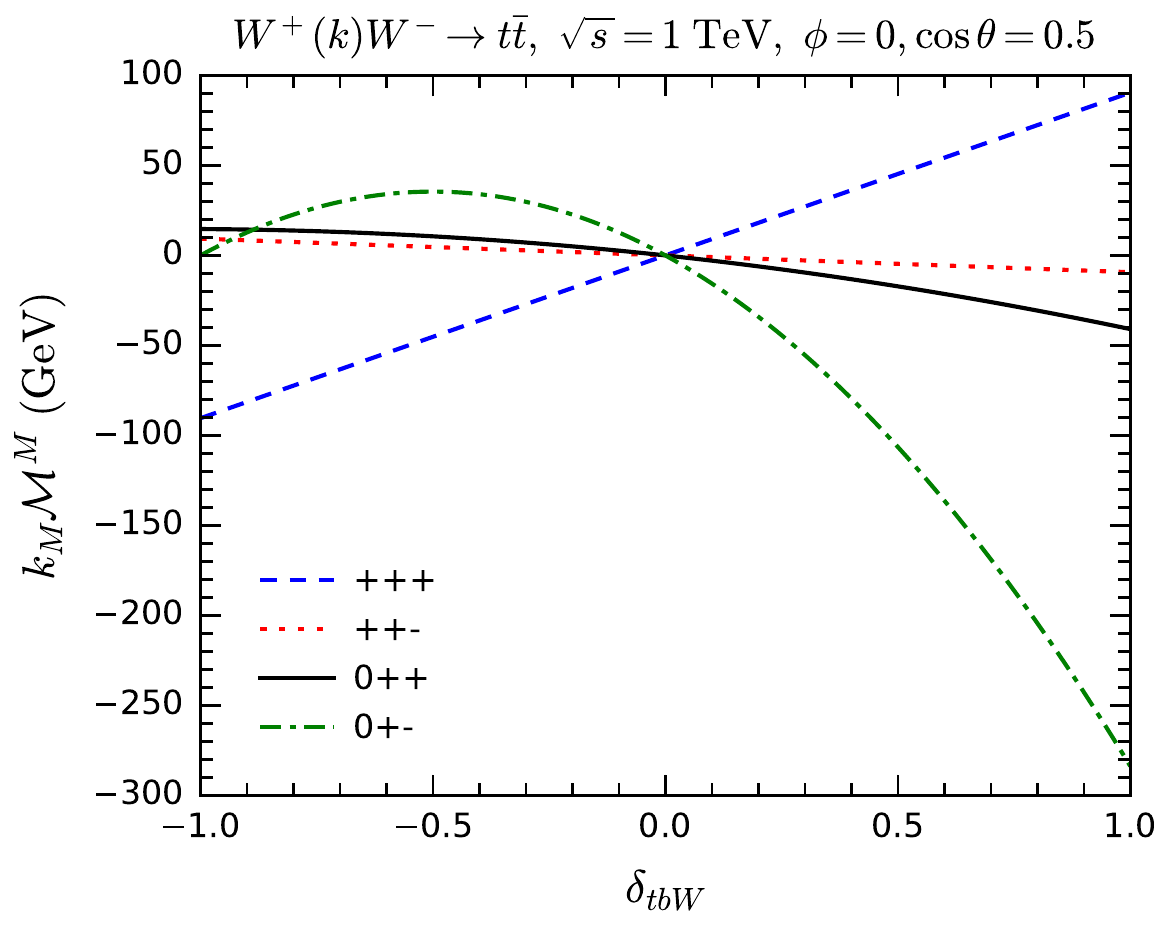}}
    \subfigure[]{\includegraphics[height=0.33\linewidth]{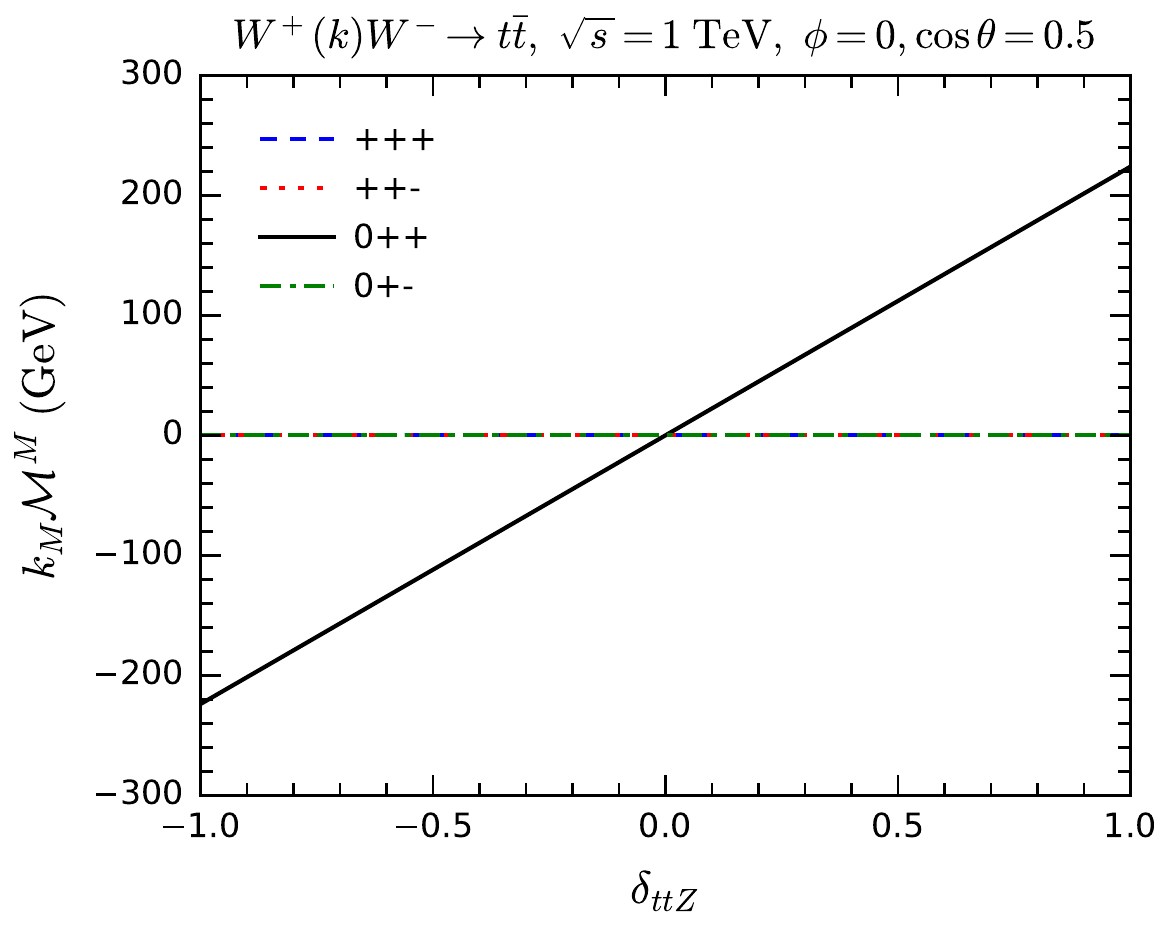}}
     \subfigure[]{\includegraphics[height=0.33\linewidth]{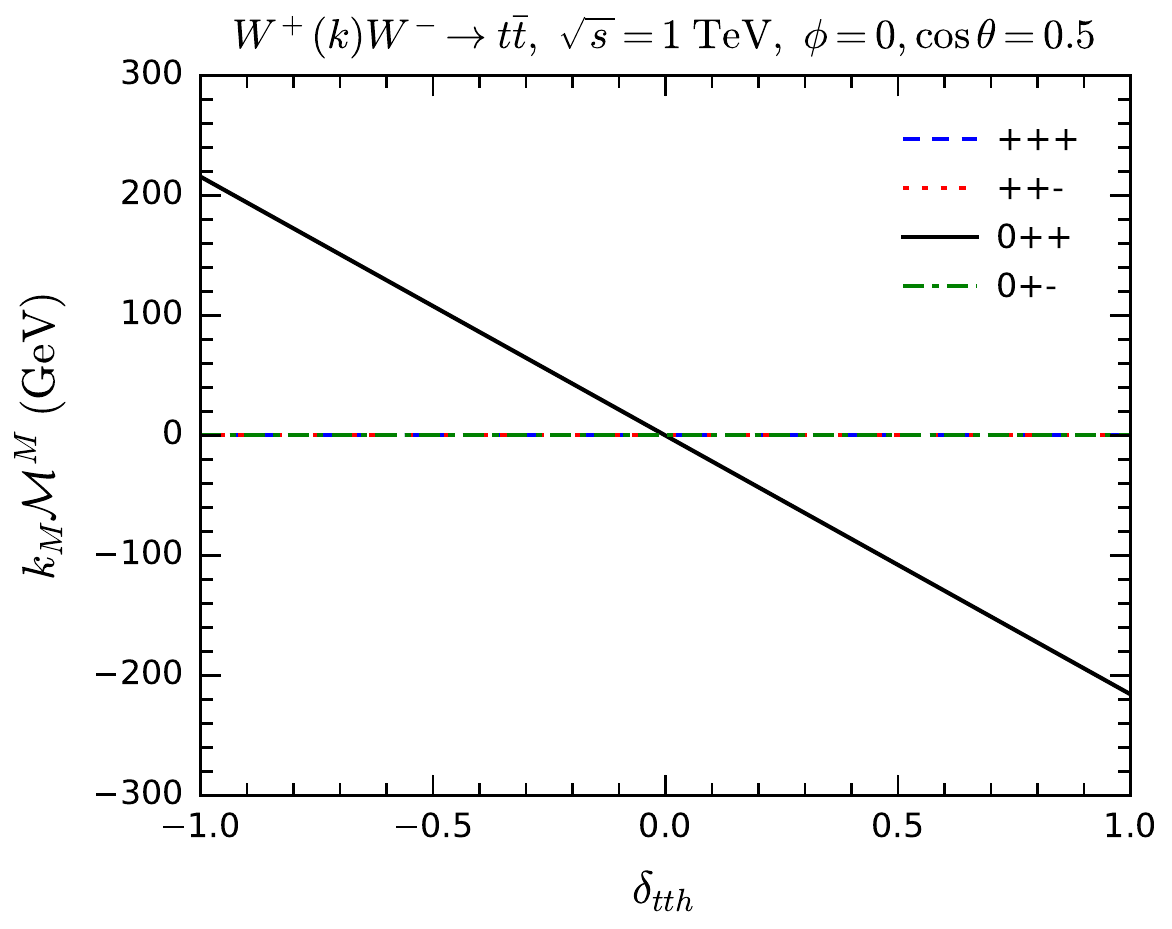}}
    \caption{$k_M\mathcal M^M$ for $WW\rightarrow t\bar t$ as a function of anomalous overall couplings $\delta_{tbW}$~(a), $\delta_{ttZ}$~(b), and $\delta_{tth}$~(c). Various helicity combinations are shown.}
    \label{fig:4pt_ww2tt}
\end{figure}

For $WW\rightarrow t\bar t$, we separately modify the couplings of the $tbW$, $ttZ$, and $tth$ vertices.  The results are shown in Fig.~\ref{fig:4pt_ww2tt}. 

The overall pattern is similar for all three processes. Angular conservation forbids some helicity combinations, giving $k_M\mathcal M^M = 0$ automatically, regardless of how the couplings of individual vertices are modified. However, this does not imply that the MWI is valid regardless of the couplings, as there are always some helicity combinations in which a modification of one coupling results in the violation of the MWI.

\section{Anomalous Couplings and SMEFT}

%\JM{Anomalous coupling: $\Delta y_t -- c_{t\phi}$, process: $WW\rightarrow t\bar t$, $WW\rightarrow t\bar t h$; $\Delta \lambda_h \rightarrow c_6$, process: $WW\rightarrow  hh$, $WW\rightarrow WWh/hhh$ }

Thus far, the anomalous couplings we considered are only instruments to demonstrate  the gauge symmetry of related amplitudes. They do not have intrinsic, gauge-invariant, physical meanings.  However, there is a scenario in which anomalous couplings can be physical: when those couplings are part of a set of anomalous SM couplings from new physics that respects gauge symmetry.  An example that we explore in this section is the SMEFT, which is an effective field theory respecting the SM gauge symmetry. In particular, we discuss two dim-6 SMEFT operators~\cite{Grzadkowski:2010es,Dedes:2017zog,Dedes:2019uzs,Barger:2023wbg}: $\mathcal O_6=C_6(\Phi^\dagger \Phi)^3$ and $\mathcal{O}_{t\Phi}=C_{t\Phi}(\Phi^\dagger \Phi)(\bar Q_\mathrm{L} t_\mathrm{R}\tilde \Phi) + \mathrm{H.c.}$, where $\Phi$ is the SM Higgs doublet, and $\tilde\Phi \equiv i\sigma^2 \Phi^*$.  We study how the gauge symmetry is restored in the presence of these operators when the three-point Higgs self-coupling $\lambda_{hhh}$ and the top Yukawa coupling $y_{tth}$ are modified in the $WW\rightarrow h h$ and $WW\rightarrow t\bar t$ processes, respectively.

\subsubsection*{$WW\rightarrow h h$}

Adding an SMEFT operator $\mathcal{O}_6=C_6(\Phi^\dag \Phi)^3$ into the SM Lagrangian, the $hhh$ and $hh\varphi^+\varphi^-$ couplings are  modified to
\begin{equation}
\lambda_{hhh} = \frac{3m_h^2}{v} +  6C_6v^3,\quad
\lambda_{hh\varphi^+\varphi^-} = \frac{m_h^2}{v^2} + 6C_6v^2.
\end{equation}
%\JM{It's consistent as along as the the $\mp$ is the same for both couplings. }
As the two couplings are modified by the same operator, their deviations from the SM values are related. This relation is further protected  by gauge symmetry. Conversely, if we modify one of the couplings alone, say $\lambda_{hhh}=\lambda_{hhh}^\mathrm{SM}+\delta_{hhh}$, the gauge symmetry will be broken. To restore gauge symmetry, the other coupling $\lambda_{hh\varphi^+\varphi^-}$ has to be modified by adding
\begin{equation}\label{eq:delta_c6}
\delta_{hh\varphi^+\varphi^-} = \frac{\delta_{hhh}}{v}.
\end{equation}
Any deviation from the above relation will break gauge symmetry for a process involving both $hhh$ and $hhVV$ couplings, which can be tested with the MWI. 

\begin{figure}[!t]
    \centering
    \includegraphics[width=0.4\linewidth]{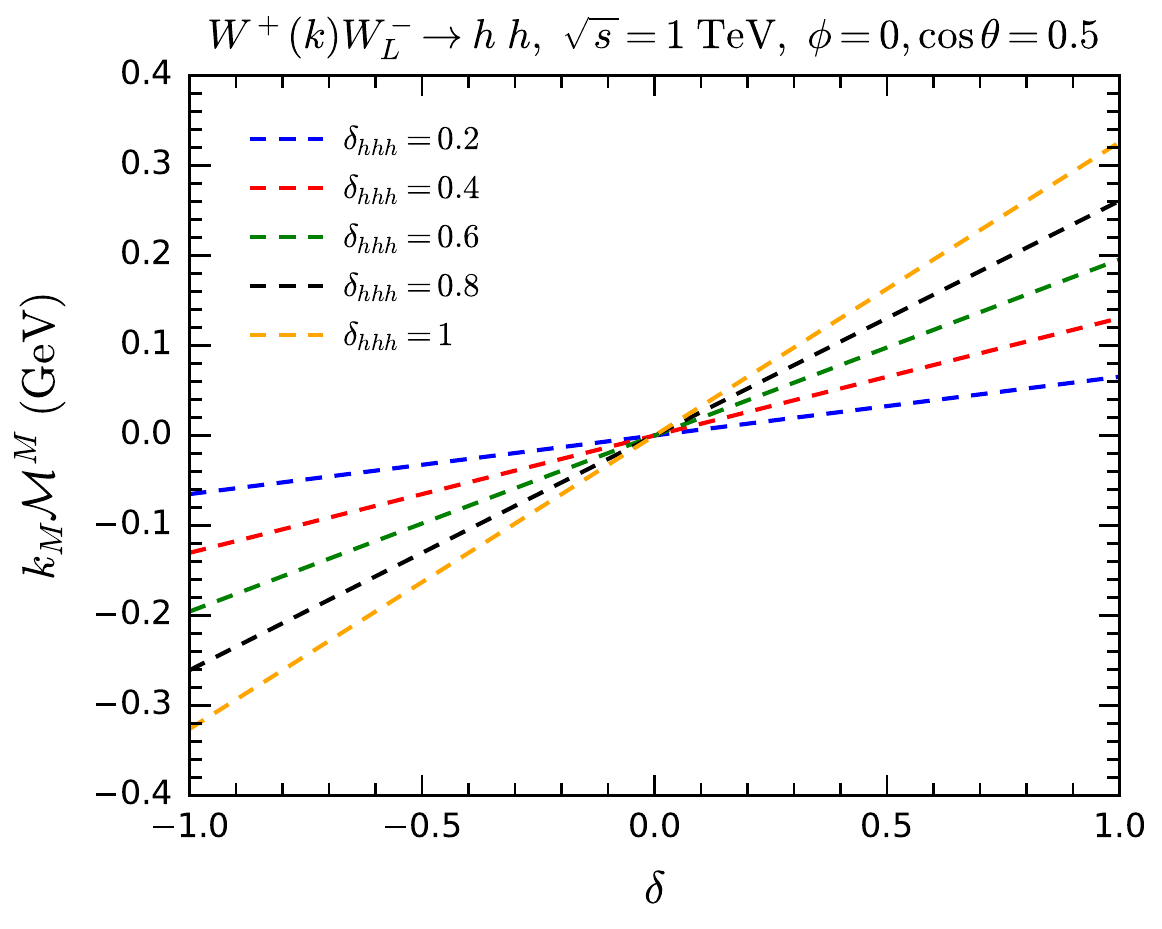}
    \caption{$k_M\mathcal M^M$ for $WW\rightarrow hh$ as a function of $\delta$ with different values of $\delta_{hhh}$, serving as a test for gauge symmetry in the presence of the SMEFT operator $\mathcal{O}_6$.}
    \label{fig:smeft_ww2hh}
\end{figure}

To test the relation~\eqref{eq:delta_c6}, we introduce a deviation parameter $\delta$ as follows:
\begin{equation}\label{eq:C6_delta}
\delta_{hhh} = \delta_{hh\varphi^+\varphi^-}  v  (1+\delta).
\end{equation}
Thus, $\delta = 0$ corresponds to the relation~\eqref{eq:delta_c6}, which respects gauge symmetry.
We consider $\delta_{hhh}$ to be $0.2$, $0.4$, $0.8$, and  $1$. For every $\delta_{hhh}$ value, we compute $k_M\mathcal M^M$ as a function of $\delta$ for the $WW\rightarrow h h$ process with one $W$ polarization vector replaced by the five-component momentum.
The results are summarized in Fig.~\ref{fig:smeft_ww2hh}. As shown in the figure, $k_M\mathcal M^M = 0$ only appears when $\delta =0$, i.e., when the relation~\eqref{eq:delta_c6} holds. This confirms our argument that, with an anomalous coupling of $\lambda_{hhh}$, gauge symmetry can be restored by adding a proper anomalous coupling of $\lambda_{hh\varphi\varphi}$.

\subsubsection*{$WW\rightarrow t\bar t$}

We then focus on the gauge symmetry of the $WW\rightarrow t\bar t$ process with the incorporation of the SMEFT operator $\mathcal{O}_{t\Phi}=C_{t\Phi}(\Phi^\dag \Phi)(\bar Q_\mathrm{L} t_\mathrm{R}\tilde \Phi) + \mathrm{H.c.}$. The couplings of $tth$ and $tt\varphi\varphi$ are defined as
\begin{eqnarray}
 \mathcal L_{tth}&=&-\bar t(y_{tth} +i y^5_{tth}\gamma^5)t h,
 \\
 \mathcal L_{tt\varphi\varphi}&=& -\bar t(y_{tt\varphi\varphi}+i y^5_{tt\varphi\varphi}\gamma^5)t \varphi^+\varphi^- .
\end{eqnarray}
In the SM, the values of these couplings are
\begin{equation}
y_{tth}=\frac{m_t}{v},\quad
y^5_{tth} = y_{tt\varphi\varphi} = y^5_{tt\varphi\varphi} = 0. 
\end{equation}
After the addition of $\mathcal{O}_{t\varPhi}$, the couplings are modified to
\begin{eqnarray}
y_{tth} &=& \frac{m_t}{v} - \frac{v^2}{\sqrt{2}} \operatorname{Re}(C_{t\Phi}),\quad
y^5_{tth}=-\frac{v^2}{\sqrt{2}} \operatorname{Im}(C_{t\Phi}),
\\
y_{tt\varphi\varphi} &=& -\frac{v^2}{\sqrt{2}}  \operatorname{Re}(C_{t\Phi}),\quad
y^5_{tt\varphi\varphi}= - \frac{v^2}{\sqrt{2}}  \operatorname{Im}(C_{t\Phi}).
\end{eqnarray}
Based on the above equations, we can obtain the relations between the modifications to these couplings as
\begin{equation}
\delta y_{tth} = \delta y_{tt\varphi\varphi} v,\quad
\delta y^5_{tth} = \delta y^5_{tt\varphi\varphi} v.
\end{equation}

If expressed in left-handed and right-handed  couplings, the Lagrangian terms for the $\bar tth$ and $tt\varphi\varphi$ couplings become
\begin{eqnarray}
    \mathcal{L}_{tth}&=&-\bar t(y_\mathrm{L} P_\mathrm{L} + y_\mathrm{R} P_\mathrm{R})t h , \nonumber\\ 
    \mathcal{L}_{tt\varphi\varphi} &=& -\bar t(g_\mathrm{L}P_\mathrm{L} + g_\mathrm{R} P_\mathrm{R})t\varphi^+\varphi^-.
\end{eqnarray}
The modifications to $y_\mathrm{L}$, $y_\mathrm{R}$, $g_\mathrm{L}$, and $g_\mathrm{R}$ are related to $C_{t\Phi}$ as
\begin{equation}
\delta y_\mathrm{L} = -\frac{C_{t\Phi}^* v^2}{\sqrt{2}},\quad
\delta y_\mathrm{R} = -\frac{C_{t\Phi} v^2}{\sqrt{2}},\quad
\delta g_\mathrm{L} = -\frac{C_{t\Phi}^* v}{\sqrt{2}},\quad
\delta g_\mathrm{R} = -\frac{C_{t\Phi}v}{\sqrt{2}}.
\end{equation}
We can observe that they are related to each other by
\begin{equation}\label{eq:smeft_wwtt}
    \delta y_\mathrm{L/R}=\delta g_\mathrm{L/R} v .
\end{equation}

\begin{figure}[!t]
    \centering
    \subfigure[]{\includegraphics[height=0.33\linewidth]{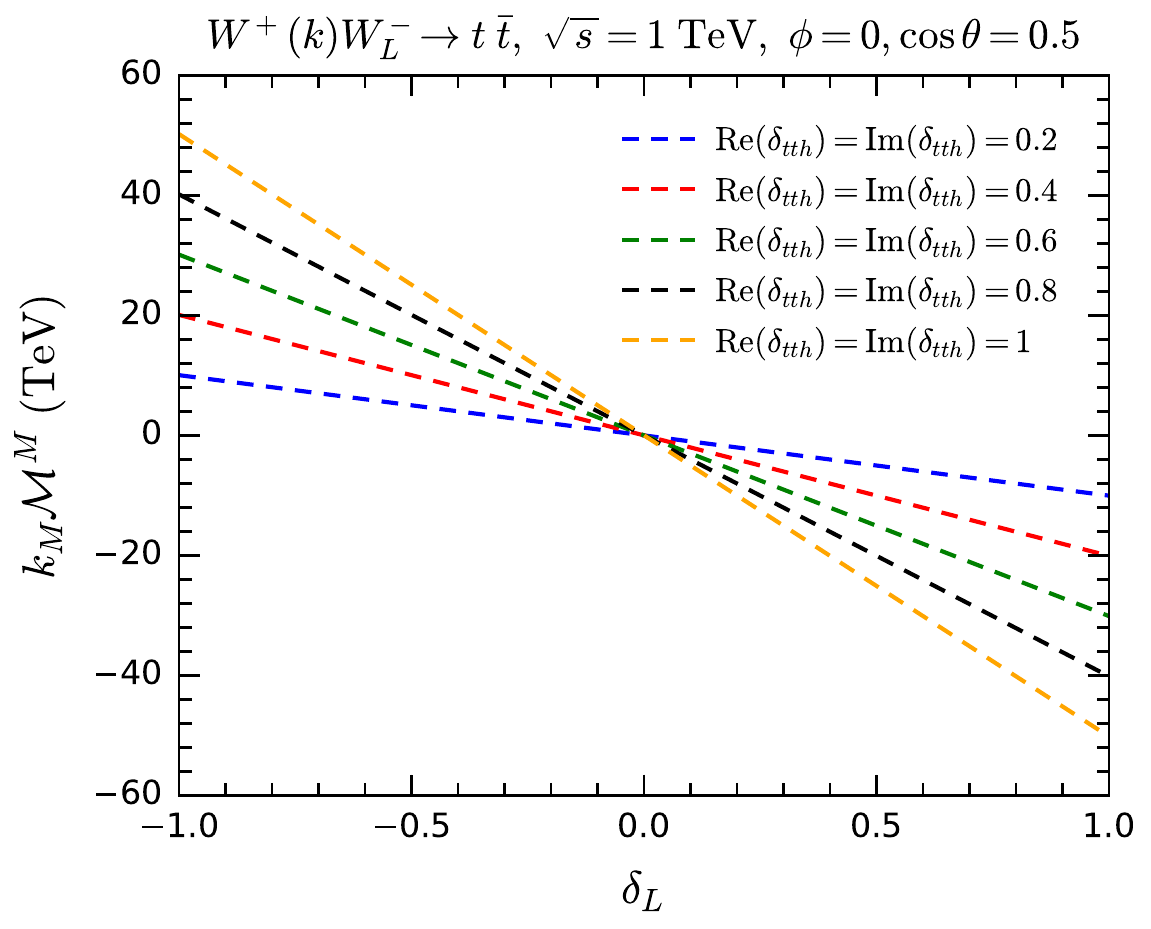}}
    \subfigure[]{\includegraphics[height=0.33\linewidth]{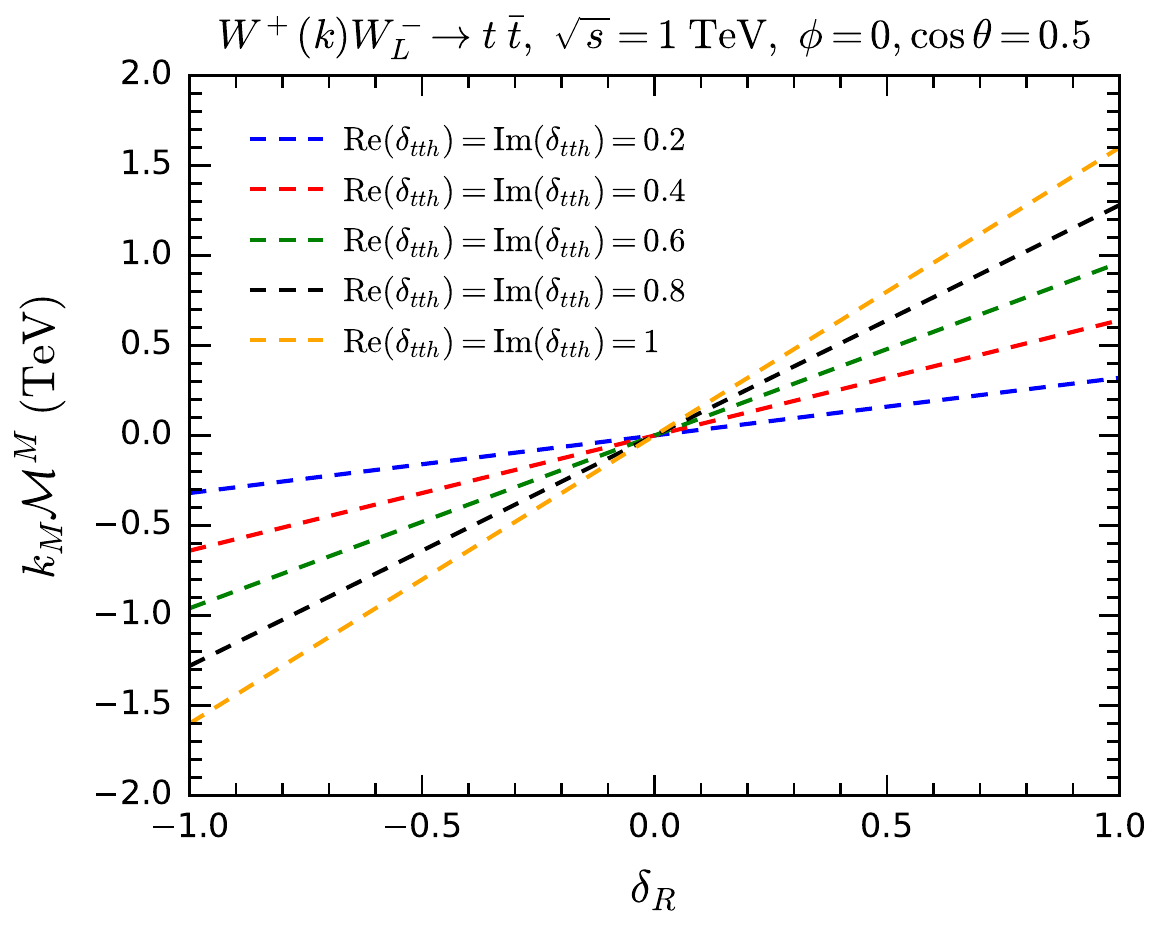}}
    \caption{$k_M\mathcal M^M$ for $WW\rightarrow t\bar{t}$ as a function of $\delta_\mathrm{L}$ (a) and $\delta_\mathrm{R}$ (b) with different values of $\operatorname{Re}(\delta_{tth})$ and $\operatorname{Im}(\delta_{tth})$, serving as a test for gauge symmetry in the presence of the SMEFT operator $\mathcal{O}_{t\Phi}$.}
    \label{fig:smeft_ww2tt}
\end{figure}

Our method of testing the gauge symmetry is similar to that for $WW\rightarrow hh$ with the $\mathcal{O}_6$ operator. 
Two deviation parameters $\delta_\mathrm{L}$ and $\delta_\mathrm{R}$ are introduced as follows:
\begin{equation}\label{eq:smeft_wwtt_delt}
\delta y_\mathrm{L/R} = \delta g_\mathrm{L/R} v (1+\delta_\mathrm{L/R}).
\end{equation}
Therefore, $\delta_\mathrm{L} = \delta_\mathrm{R} = 0$ corresponds to the modifications contributed by the $\mathcal{O}_{t\Phi}$ operator that preserve gauge symmetry.
In Figs.~\ref{fig:smeft_ww2tt}(a) and \ref{fig:smeft_ww2tt}(b), we show $k_M\mathcal M^M$ for $WW\rightarrow t\bar{t}$ as functions of $\delta_\mathrm{L}$ and $\delta_\mathrm{R}$ assuming various values of $\operatorname{Re}(\delta_{tth})$ and $\operatorname{Im}(\delta_{tth})$, where $\delta_{tth}$ is defined as $\delta_{tth} \equiv C_{t\Phi} v^2/\sqrt{2}$.
As expected, gauge symmetry is restored when the condition in Eq.~\eqref{eq:smeft_wwtt} is satisfied. This demonstrates that an anomalous Yukawa coupling can be counterbalanced by a corresponding $ff\varphi\varphi$ contact vertex, ensuring that the $WW\rightarrow t\bar{t}$ amplitude remains gauge invariant.

\section{Summary and Conclusions}

In this study, we demonstrated that the GE representation of EW interactions, which we had a brief introduction first, has gauge symmetry imprinted in the structure of amplitudes, manifesting in the MWI~\eqref{eq:MWI}. This approach to gauge symmetry has been rarely studied before, because it involves the vertices of both gauge bosons and Goldstone bosons simultaneously. We numerically studied this important property of EW interactions in the GE representation in several different aspects, including directly testing the MWI, modifying couplings within vertices, and modifying the overall couplings of vertices. Our main conclusion is that there are precise relations within and without the overall vertices, which guarantee the MWI. Any violation of these relations results in the violation of the MWI and gauge symmetry.

By directly testing gauge symmetry on four-point amplitudes, such as $WW\rightarrow t\bar t$ and $W^+ W^- \rightarrow W^+ W^-$, with different helicity combinations, we observed that the full amplitudes always satisfy the MWI after summing over all tree-level diagrams. This result typically involves large cancellations between individual diagrams, similar to unitarity cancellation in the gauge representation. 

For testing gauge symmetry on three-point vertices with anomalous couplings, we studied the $VVh$, $ff'V$, and $VVV$ vertices.  We observed that there are precise relations governing the couplings of Goldstone and gauge components for these vertices. We then numerically tested if the MWI still holds when some of the couplings are modified so that those relations are violated. We found that any deviation from those relations by anomalous couplings would violate the MWI.   

For testing gauge symmetry on four-point amplitudes with anomalous couplings, we studied $W^+ W^+\rightarrow W^+ W^+$, $WW\rightarrow hh$, and $WW\rightarrow t\bar t$. We focused on how modifying the overall couplings of individual vertices affects gauge symmetry. Our results are similar to those for three-point vertices: gauge symmetry is manifested as precise relations between couplings, and modifying couplings to violate those relations also results in violating the MWI. 

After studying the gauge symmetry of the SM, we studied the effective operators in the SMEFT, specifically the $\mathcal O_6$ operator, which modifies the Higgs self-couplings, and the $\mathcal O_{t\Phi}$ operator, which modifies the top Yukawa coupling. We observed that the operators modify both the couplings mentioned and the related Goldstone couplings, giving relations between the couplings involving $C_6$ and $C_{t\Phi}$, respectively.  Numerically testing the MWI indicates that gauge symmetry is preserved as long as all the related couplings are modified in accordance with those relations. On the other hand, if those relations are violated, gauge symmetry would be broken. 
Similar to the SM, our results for these SMEFT operators demonstrate that not only the gauge components but also the Goldstone components are crucial for maintaining gauge invariance of the theory.  This is also the reason why, gauge cancellation in the unitary gauge, in which there is no Goldstone mode, appears to be ad hoc without an apparent physical mechanism.

We believe that this paper contributes to a deeper understanding of EW interactions and massive gauge theory in general. In addition, it provides a convenient method of checking the self-consistency of EW interactions in the GE representation. 

Thus far, the five-component formalism has been only systematically implemented in HELAS for the tree-level Feynman rules of the SM. Moreover, for the purpose of this study, we included the dim-6 operators $\mathcal O_6$ and $\mathcal O_{t\Phi}$ in HELAS. However, the majority of the SMEFT operators remain to be incorporated. This limits the utility of our method and calls for more progress in this aspect. In Ref.~\cite{Hagiwara:2024xdh}, a general method for automatically implementing the five-component framework has been proposed recently; which, we hope could help apply our method to more general theories.

\begin{acknowledgments}

The authors thank Kaoru Hagiwara for helpful discussions.
Junmou Chen is supported by National Natural Science Foundation of China under Grant No.~12205118.
Wang-Fa Li, Qian-Jiu Wang, and Zhao-Huan Yu are supported by the Guangzhou Science and Technology Planning Project under Grant No.~2024A04J4026.

\end{acknowledgments}

\bibliographystyle{JHEP}
\bibliography{reference-HELAS}

\end{document}